\documentclass[letterpaper, aps, prd, twocolumn, superscriptaddress, showpacs, nofootinbib]{revtex4}
\pdfoutput=1

\usepackage{graphicx}
\usepackage{bm}
\usepackage{amssymb}
\usepackage{amsmath}
\usepackage{amsfonts}
\usepackage{dcolumn}
\usepackage{natbib}
\usepackage{color}
\usepackage{subfigure}

\usepackage[utf8]{inputenc}
\usepackage[T1]{fontenc}

\newcommand{\raiseentry}[1]{\smash{\raise 0.7 em \hbox{#1}}}

\newcommand{\scri}{\ensuremath{\mathcal{J}}}

\newcommand{\tg}{\tilde\gamma}
\newcommand{\tG}{\tilde\Gamma}
\newcommand{\tA}{\tilde A}

\newcommand{\hp}{\hat{\phi}_\kappa}

\newcommand{\p}{\ensuremath{\partial}}
\newcommand{\norm}[1]{\lVert #1 \rVert}

\def\mmsun{M_\odot}

\newenvironment{equationarray*}
{\arraycolsep 0.14 em
\begin{eqnarray*}}
{\end{eqnarray*}}

\newcommand{\code}[1]{{\tt #1}}

\begin{document}

\title{Three-Dimensional General-Relativistic Hydrodynamic Simulations of \\
      Binary Neutron Star Coalescence and Stellar Collapse with Multipatch Grids}

\author{C. Reisswig}
\thanks{Einstein Fellow}
\email{reisswig@tapir.caltech.edu}
\affiliation{TAPIR, MC 350-17, California Institute of Technology, 1200 E California Blvd.,
Pasadena, CA 91125, USA}

\author{R. Haas}
\affiliation{TAPIR, MC 350-17, California Institute of Technology, 1200 E California Blvd.,
Pasadena, CA 91125, USA}

\author{C. D.\ Ott}
\thanks{Alfred P. Sloan Research Fellow}
\affiliation{TAPIR, MC 350-17, California Institute of Technology, 1200 E California Blvd.,
Pasadena, CA 91125, USA}
\affiliation{Kavli Institute for the Physics and
 Mathematics of the Universe (Kavli IPMU), The University of Tokyo, Kashiwa, Japan}

\author{E. Abdikamalov}
\affiliation{TAPIR, MC 350-17, California Institute of Technology, 1200 E California Blvd.,
Pasadena, CA 91125, USA}

\author{P. M\"osta}
\affiliation{TAPIR, MC 350-17, California Institute of Technology, 1200 E California Blvd.,
Pasadena, CA 91125, USA}

\author{D. Pollney}
\affiliation{
  Department of Mathematics,
  Rhodes University,
  Grahamstown, 6139
  South Africa
}
    
\author{E. Schnetter}
\affiliation{Perimeter Institute for Theoretical Physics, 31 Caroline
  St.\ N., Waterloo, ON N2L 2Y5, Canada}
\affiliation{Department of Physics, University of Guelph, 50 Stone
  Road East, Guelph, ON N1G 2W1, Canada}
\affiliation{Center for Computation \& Technology, 216 Johnston Hall,
  Louisiana State University, Baton Rouge, LA 70803, USA}

\date{\today}


\begin{abstract}
We present a new three-dimensional general-relativistic hydrodynamic
evolution scheme coupled to dynamical spacetime evolutions which is
capable of efficiently simulating stellar collapse, isolated neutron
stars, black hole formation, and binary neutron star coalescence. We
make use of a set of adapted curvi-linear grids
(multipatches) coupled with flux-conservative cell-centered adaptive
mesh refinement.  This allows us to significantly enlarge our
computational domains while still maintaining high resolution in the
gravitational-wave extraction zone, the exterior layers of a star, or
the region of mass ejection in merging neutron stars.  The fluid is
evolved with a high-resolution shock capturing finite volume scheme,
while the spacetime geometry is evolved using fourth-order finite
differences.  We employ a multi-rate Runge-Kutta time integration
scheme for efficiency, evolving the fluid with second-order and the
spacetime geometry with fourth-order integration, respectively.  We
validate our code by a number of benchmark problems: a rotating
stellar collapse model, an excited neutron star, neutron star
collapse to a black hole, and binary neutron star coalescence. The
test problems, especially the latter, greatly benefit from higher
resolution in the gravitational-wave extraction zone, causally
disconnected outer boundaries, and application of
Cauchy-characteristic gravitational-wave extraction.  We show that we
are able to extract convergent gravitational-wave modes up to
$(\ell,m)=(6,6)$.  This study paves the way for more realistic and
detailed studies of compact objects and stellar collapse in full three
dimensions and in large computational domains.
The multipatch infrastructure and the improvements to mesh refinement and 
hydrodynamics codes discussed in this paper will be made available as part of the 
open-source Einstein Toolkit.
\end{abstract}


\maketitle



\section{Introduction}
\label{section:introduction}

Some of the most interesting relativistic astrophysical phenomena such
as stellar collapse, black hole formation, or binary neutron star
coalescence, require numerical simulations on large computational
domains, involve many different length scales, and are intrinsically
three-dimensional (3D).  Due to their extreme nature in terms of fluid
densities and velocities, an accurate treatment of
general-relativistic (GR) gravity is required.  Depending on the
problem, magnetic field evolution and neutrino interactions may also
be required. Thus, numerical computations in relativistic astrophysics
are truly multiphysics, and as such, are especially demanding in terms
of computational modeling technology and resources.

Current state of the art 3D GR hydrodynamic simulations in the context
of stellar collapse \cite{ott:12a, ott:11a, kuroda:12, kuroda:10,
  baiotti:07b} or binary neutron star coalescence \cite{bernuzzi:12a,
  baiotti:11, giacomazzo:11, rezzolla:11, etienne:12, kiuchi:12,
  east:12, anderson:08b} (see \cite{faber:12} for a recent review) are
based on Cartesian grids with adaptive mesh refinement (AMR).  As the
domain is enlarged or the resolution increased, such grids pose a
serious bottleneck in terms of the computational power that is
required, even with AMR.  Since Cartesian grids scale as $N^3$ in
terms of the number $N$ of grid points along one spatial direction in
3D, available computational resources are rapidly exhausted when
additional points in each coordinate direction are added.  The
symmetry of the computational problem, however, is essentially
spherical, at least at some distance from the central region of the
simulation.  Thus, Cartesian grids are wasteful with respect to
angular resolution when the problem becomes symmetrically spherical.

For instance, stellar collapse proceeds in approximately spherical or
axisymmetric terms (e.g.
\cite{ott:07prl,ott:07cqg,scheidegger:08,scheidegger:10b}).  At later
times, various hydrodynamic instabilities (e.g.~convection and
instabilities of the shock) break this symmetry. The global features,
however, remain approximately spherical or axisymmetric.

In the case of coalescing binary neutron stars, the central region
containing the two neutron stars is not of spherical symmetry. At
larger distances and in the gravitational-wave (GW) zone, however, the
problem becomes spherical.  The gravitational-wave
extraction zone must generally be located at large radii in order to
limit near-zone effects in the extracted wave.  But even with more
sophisticated techniques such as Cauchy-characteristic extraction
\cite{winicour:09,bishop:97b, reisswig:11ccwave, reisswig:09,
  reisswig:10a, babiuc:11} that allow us to extract gauge-invariant GWs at future null
infinity $\scri^+$, it is necessary to enlarge the domain sufficiently
so that constraint-violating modes generated at the outer boundary are
causally disconnected from the interior evolution and the
wave-extraction zone.  These constraint-violating modes are generated
due to the lack of constraint-preserving outer boundary conditions for
the Einstein equations (see \cite{winicour:12} for a recent review)
for certain types of evolution systems (including the common BSSN
system), and travel at the speed of light \cite{brown:07a,
  ES-Brown2007b} to the interior of the domain.  Without these
systematic errors, the evolution and wave extraction would generally
be more accurate.  Furthermore, in case mass is ejected during and
after merger, enlarging refinement levels to track the evolution of
the ejected material becomes very expensive.

It therefore seems natural to apply spherical grids 
to maintain high resolution also in the outer regions of the domain.
The computational effort when using spherical grids scales linearly with the 
number of radial points $N$, assuming constant angular resolution.
Thus, spherical grids can give a tremendous performance improvement when the
domain is enlarged or the (radial) resolution increased.

Spherical grids have been widely used for many astrophysical problems,
including stellar collapse (e.g.,
\cite{zwerger:97,dimmelmeier:02,dimmelmeier:02a,dimmelmeier:07,abdikamalov:10}),
core-collapse supernovae (e.g.,
\cite{mueller:10,hanke:12,takiwaki:12}), oscillations of neutron stars
(e.g., \cite{dimmelmeier:06,abdikamalov:09b}), neutron star
magnetosphere (e.g., \cite{komissarov:06}), accretion onto black
holes~\cite{font:98}, and simulations of accretion disks (e.g.,
\cite{font:02,zanotti:03,narayan:12}).  Unfortunately, the standard
spherical-polar coordinate system imposes a serious difficulty along
the axis and the poles, where special care must be taken to regularize
the fields and to numerically evolve them \cite{alcubierre:11, montero:11, cordero-carrion:12, baumgarte:12}.  
But even with proper regularization applied, the angular and radial
distribution of grid points is non-optimal in the sense that they
cluster at the poles and at the coordinate origin.  In addition,
spherical grids are less suited in regions where the underlying
symmetry is non-spherical, e.g.,~in the vicinity of a binary neutron
star system, or the highly turbulent and convective region behind the
accretion shock in a core-collapse supernova.

In order to handle multiple regions of different symmetry within the same 
simulation, \textit{multipatch} (sometimes also called \textit{multiblock}) schemes have been developed for
a wide range of physics and engineering applications.
The idea is to cover the simulation domain with multiple curvi-linear coordinate
``patches''. Each patch is locally uniform. Diffeomorphic mappings from
local to the global coordinates enable to represent a wide range of grid
shapes in different regions of the simulation.
One such example is given in Fig.~\ref{fig:7patch-system}. In this setup, a
central Cartesian patch is surrounded by six ``inflated cube'' spherical grid
patches.
This is a natural configuration for our purposes. The aspherical region
of a collapsing star or a merging binary is best modeled by a central
Cartesian patch, capable of AMR. The gravitational-wave zone and/or the outer
layers of a star are best modeled by the more efficient spherical grids.
This allows us to employ large domains at high resolution with modest
computational cost. Notably, the outer boundary can be causally disconnected
from the interior evolution and the gravitational-wave extraction zone.

Within the context of numerical relativity and relativistic
astrophysics, multipatch schemes have already been successfully
applied in a range of different problems ranging from simulations of
accretion disks \cite{korobkin:11,korobkin:12}, horizon finding
\cite{thornburg:04}, wave extraction \cite{pazos:06}, single black
holes \cite{thornburg:04b, dorband:06a}, orbiting black holes
\cite{pazos:09}, relativistic fluid evolutions on fixed backgrounds
\cite{zink:08b}, elliptic and initial data solvers \cite{ansorg:07,
  Pfeiffer:03, gourgoulhon:00, gourgoulhon:02, grandclement:01}, to
characteristic evolutions of Einstein's equations \cite{bishop:97b,
  reisswig:07, gomez:07}.  Multidomain \textit{spectral} methods have
been successfully applied to vacuum binary black hole evolutions
yielding high accuracy and efficiency
\cite{scheel:09,szilagyi:09,chu:09,lovelace:12,buchman:12,macdonald:12}
using a dual-coordinate frame method \cite{scheel:06}.  The same
multidomain spectral code \texttt{SpEC}, coupled to a finite volume
fluid solver, has also been used to simulate neutron star black hole
mergers \cite{foucart:12, foucart:10, duez:08, foucart:08}.  Neither
of the works above, however, make use of AMR for the fluid fields, and
thus are limited in the respective range of astrophysical
applications.  In particular, efficient simulations of stellar
collapse and black hole formation require AMR in the central region of
the collapsing star.  Also, the near-field region in simulations of binary neutron star
coalescence substantially benefit from AMR, in particular when
material is ejected in the post-merger phase.

In the context of vacuum binary black hole merger simulations,
multipatch schemes combined with AMR have been successfully applied
\cite{pollney:11, pollney:09, pollney:10, damour:11, santamaria:10,
  bishop:11, ajith:11}.  
We base our code on the \texttt{Llama} infrastructure developed in \cite{pollney:11}, 
which makes use of the \texttt{Cactus} computational toolkit \cite{goodale:03}
and the \texttt{Carpet} AMR driver \cite{Schnetter-etal-03b,
  ES-Schnetter2006a}.  We extend the original pure vacuum scheme
to include full matter dynamics using the publicly available GR
hydrodynamics code \texttt{GRHydro}, which is part of the
\texttt{EinsteinToolkit} \cite{et:12}.  We thus present the first
successful multipatch scheme capable of AMR that can stably evolve
fluid dynamics coupled to fully GR spacetime dynamics.

In addition, we make a number of improvements:
(i) We extend the AMR driver \texttt{Carpet} to support cell-centered
mesh refinement, which
allows us to apply \textit{refluxing}, a technique to maintain
conservation of mass, energy and momentum fluxes across mesh refinement boundaries
\cite{Berger1984} (see \cite{east:11} for a recent application to GR hydrodynamics).
This greatly improves conservation of mass in our simulations of stellar
collapse, especially in the postbounce evolution.
(ii) We apply enhanced PPM (piecewise parabolic method) reconstruction
\cite{colella:08, mccorquodale:11}, which significantly improves the numerical
accuracy and the behavior of the constraints.
(iii) To improve the execution speed of the simulations, we apply multirate Runge-Kutta (RK) time
integration (e.g.~\cite{schlegel:09, constantinescu:07}) in which the spacetime
is evolved with a standard fourth-order RK method,
whereas the fluid is evolved with a second order RK scheme without significant
loss of accuracy. 
This reduces the number of intermediate steps in the fluid evolution, which
dominates in terms of processor cycles compared to spacetime evolution, in particular when
using a microphysical equation of state.

We apply the new code to a number of benchmark problems,  including the
evolution of a single isolated and perturbed neutron star, the collapse of a
rotating stellar core, the collapse of a neutron star to a black hole,
and the merger of a binary neutron star system.
We investigate the accuracy and convergence of each test problem.
This is an important code verification towards our program
to carry out fully 3D simulations of core-collapse supernovae 
(see \cite{ott:12b} for a recent application of our scheme)
and black hole formation in the context of the collapsar scenario for long gamma-ray bursts. 
The new multipatch scheme allows us to significantly enlarge the computational domain by maintaining
a fixed angular resolution.
This is useful in many ways: (i) we are able to causally disconnect the outer boundary from
the interior evolution and the gravitational-wave extraction zone, thus avoiding systematic errors 
from the approximate and non-constraint preserving artificial outer boundary condition,
(ii) we have a larger wave-extraction zone with higher overall resolution, thus making it possible to extract
higher-order than the dominant GW modes,
(iii) in binary neutron star mergers, ejected material can be tracked out to large radii
with relatively high resolution,
(iv) the number of mesh refinement levels can be decreased, leading to better parallel scaling.
As a result, our multipatch scheme can efficiently evolve models of stellar collapse in full 3D (see also \cite{ott:12b}),
and is capable of more accurate gravitational-wave extraction in models of binary neutron star mergers.
In the latter test problem, we extract convergent gravitational-wave modes up to $(\ell,m)=(6,6)$. 

This paper is organized as follows.
In Sec.~\ref{sec:hydro} and ~\ref{sec:curv}, 
we first review the underlying hydrodynamic and spacetime evolution systems and 
how we solve them numerically.
Subsequently, in Sec.~\ref{sec:multipatch}, we present our approach to multipatches and their
numerical implementation.
We also discuss our implementation of cell-centered AMR (Sec.~\ref{sec:ccamr}), and describe multirate RK
time integration (Sec.~\ref{sec:multirate}).
Finally, in Sec.~\ref{sec:results}, we present detailed tests of isolated perturbed and unperturbed neutron stars, 
collapsing stellar cores, neutron star collapse to a black hole, and
merging binary neutron stars.
We conclude and summarize our findings in Sec.~\ref{sec:summary}.
In an appendix, we presents basic tests with shock tubes (Appendix~\ref{sec:shock}), we review
the enhanced PPM scheme as developed in \cite{colella:08, mccorquodale:11} (Appendix~\ref{sec:ePPM}),
discuss our treatment of the artificial low-density atmosphere (Appendix~\ref{sec:atmo}), 
present an optimized ghost-zone update scheme to improve the parallel scaling (Appendix~\ref{sec:scheduling}),
describe our volume integration scheme for overlapping grids (Appendix~\ref{sec:vol-int}),
and investigate the influence of boundary effects on binary neutron star merger dynamics and wave extraction (Appendix~\ref{sec:BC}).


\section{Methods}
\label{sec:methods}

\subsection{General-Relativistic Hydrodynamics}
\label{sec:hydro}

We base our code on the open-source GR hydrodynamics code \code{GRHydro} that is
part of the \code{EinsteinToolkit} \cite{einsteintoolkitweb} and 
is described in \cite{reisswig:11ccwave,baiotti:05,et:12}.

We introduce \textit{primitive} variables in the form of the fluid density $\rho$,
the fluid's specific internal energy $\epsilon$, and the fluid 3-velocity as seen by
Eulerian observers at rest in the current spatial
3-hypersurface \cite{york:83},
\begin{equation}
v^i = \frac{u^i}{W} + \frac{\beta^i}{\alpha}\,\,,
\label{eq:vel}
\end{equation}
where $u^i$ is the fluid 4-velocity, $W = (1-v^i v_i)^{-1/2}$ is the Lorentz factor, and $\alpha$ and $\beta^i$ are lapse and shift, 
respectively (to be introduced in Sec.~\ref{sec:curv}). 
In terms of
the 3-velocity, the contravariant 4-velocity is then given by
\begin{equation}
u^0  = \frac{W}{\alpha}\,,\qquad
u^i = W \left( v^i - \frac{\beta^i}{\alpha}\right)\,\,,
\end{equation}
and the covariant 4-velocity is
\begin{equation}
u_0  = W(v^i \beta_i - \alpha)\,,\qquad
u_i = W v_i\,\,.
\end{equation}

The evolution equations are written in the Valencia form of GR hydrodynamics~\cite{banyuls:97,font:08} 
as a first-order hyperbolic
flux-conservative evolution system for the \textit{conserved} variables
$D$, $S^i$, and $\tau$ which are defined in terms of the primitive
variables $\rho, \epsilon, v^i$,
\begin{eqnarray}
  D &=& \sqrt{\gamma} \rho W,\nonumber\\
  S^i &=& \sqrt{\gamma} \rho h W^{\,2} v^i,\nonumber\\
  \tau &=& \sqrt{\gamma} \left(\rho h W^{\,2} - P\right) - D\,,
\end{eqnarray}
where $ \gamma $ is the determinant of the 3-metric $\gamma_{ij} $ (see Sec.~\ref{sec:curv}), 
and the quantities $P$, and $h=1+\epsilon+P/\rho$ denote pressure, and specific enthalpy, respectively.
The evolution system then becomes
\begin{equation}
  \frac{\partial \mathbf{U}}{\partial t} +
  \frac{\partial \mathbf{F}^{\,i}}{\partial x^{\,i}} =
  \mathbf{S}\,\,,
  \label{eq:conservation_equations_gr}
\end{equation}
with
\begin{eqnarray}
  \mathbf{U} & = & [D, S_j, \tau], \nonumber\\
  \mathbf{F}^{\,i} & = & \alpha
  \left[ D \tilde{v}^{\,i}, S_j \tilde{v}^{\,i} + \delta^{\,i}_j P,
  \tau \tilde{v}^{\,i} + P v^{\,i} \right]\!, \nonumber \\
  \mathbf{S} & = & \alpha
  \bigg[ 0, T^{\mu \nu} \left( \frac{\partial g_{\nu j}}{\partial x^{\,\mu}} - 
  \Gamma^{\,\lambda}_{\mu \nu} g_{\lambda j} \right), \nonumber\\
  & &\qquad\alpha \left( T^{\mu 0}
  \frac{\partial \ln \alpha}{\partial x^{\,\mu}} -
  T^{\mu \nu} \Gamma^{\,0}_{\mu \nu} \right) \bigg]\,.
\end{eqnarray}%
Here, $ \tilde{v}^{\,i} = v^{\,i} - \beta^i / \alpha $, $
\Gamma^{\,\lambda}_{\mu \nu} $ are the 4-Christoffel symbols, and $T^{\mu\nu}$ is the stress-energy tensor.
The pressure $P=P(\rho,\epsilon,\{X_i\})$ 
is obtained via our equation of state module, which is capable of
handling a set of different equations of state, including microphysical finite-temperature
variants. The $\{X_i\}$ are additional compositional variables of the matter such as the electron fraction $Y_e$,
which are used for microphysical equations of state.
In the present work, however, we resort to simple (piecewise) polytropic and ideal gas ($\Gamma$-law) equations of state.

The above evolution equations are spatially
discretized by means of a high-resolution
shock-capturing (HRSC) scheme using a second-order accurate
finite-volume algorithm.
The equations are kept in semi-discrete form and first-order
(in space) Riemann problems are solved at cell interfaces with the
approximate HLLE solver~\cite{HLLE:88}. 

The states at cell interfaces are reconstructed using a new and improved variant 
of the piecewise parabolic method (PPM) \cite{colella:08, mccorquodale:11, colella:84}.
As noted in \cite{colella:08, mccorquodale:11},
the original PPM scheme \cite{colella:84} has the side-effect of flattening local smooth extrema which are physical,
thus limiting the accuracy. In the present context of simulating compact objects, one naturally has extrema 
at the stellar center(s) where the matter density is largest. We find that the original PPM scheme reduces the accuracy there,
which then strongly affects the overall accuracy of our simulations (see Sec.~\ref{sec:results}, and also Fig.~\ref{fig:tov-constr-1D}).
Ref.~\cite{colella:08}, further refined by Ref.~\cite{mccorquodale:11}, suggests modifications to the original limiter which can distinguish 
between smooth maxima that are part of the solution, and artificial maxima that may be introduced at shocks and other discontinuities. 
While smooth maxima need to be retained as part of the solution, 
artificial maxima must be avoided to suppress Gibbs phenomenon at shocks and other discontinuities.
We summarize the procedure for ``enhanced'' PPM reconstruction in Appendix~\ref{sec:ePPM}.

We note that under certain conditions, the requirement that the 
modulus of the reconstructed primitive velocity must stay below the speed of light $c$ may be violated.
This can happen, since the primitive velocity is a bounded function (bounded by the requirement $v_i v^i \leq c^2$), 
and the enhanced PPM reconstruction scheme does not enforce this constraint
close to any occuring extrema. Thus, the enhanced PPM scheme may reconstruct velocity components 
that result in a velocity modulus equal to or slightly larger than the speed of light near extrema.
To avoid this problem, we reconstruct $W v^i$, i.e.~the Lorentz factor $W$ times the primitive velocity $v^i$.
The quantity $W v^i$ is unbounded and thus does not require special treatment near extrema.

The time integration and coupling with curvature (Sec.~\ref{sec:curv})
are carried out with the Method of
Lines~\cite{Hyman-1976-Courant-MOL-report} (see Sec.~\ref{sec:multirate}).

After each evolution step, we compute the primitive quantities from the evolved conserved quantities.
Since the primitive quantities are implicit functions of the conserved ones, it is necessary to use a numerical root finding algorithm.
As described in, e.g.~\cite{et:12}, this is done via a Newton-Raphson scheme.

In some rare situations, the initial guesses for the root finding procedure 
are not well-posed, and cause the Newton-Raphson scheme
to fail to converge. In particular, we find this behavior at the surface of a 
neutron star, when the latter is threaded by an AMR boundary and refluxing is active.
In this case, we resort to a simple bisection algorithm which converges more slowly, but is more robust.

In regions of the computational domain, where we have physical vacuum, we employ an artificial low density
``atmosphere'' (see Appendix~\ref{sec:atmo}).
In order to reduce the influence of the artificial atmosphere on the curvature evolution, we
exponentially damp the stress-energy tensor $T_{\mu\nu}$ to zero outside a given radius.
More specifically, we introduce the radius dependent stress-energy damping 
$T_{\mu\nu} \rightarrow \lambda(r) T_{\mu\nu}$ with the damping factor
\begin{eqnarray}
  \label{eq:Tmunu-damp}
   \lambda(r) & = & \left\{
    \begin{array}{ll}
      1 & \mathrm{for}\; r \le R_{\rm 0},
      \\
      \frac{1}{2}\left(1 - \tanh\left(\frac{8r-4(R_{\rm 1}+R_{\rm 0})}{R_{\rm 1}-R_{\rm 0}}\right)\right) & \rm{otherwise},
      \\
      0 & \mathrm{for} \; r \ge R_{\rm 1},
    \end{array}
    \right.
\end{eqnarray}
where the damping is applied between the two radii $R_0<R_1$. 

At outer boundaries, we apply a copy-from-neighbor (flat) boundary condition for the evolved fluid quantities.

Finally, in order to be compatible with multipatch discretization, 
we need to introduce additional coordinate transformations
as described in Sec.~\ref{sec:hydro-multiblock} below.

\subsection{Curvature Evolution}
\label{sec:curv}

The spacetime evolution is performed by a variant of the BSSN evolution system
\cite{nakamura:87, shibata:95, baumgarte:99, alcubierre:00} and is implemented
in the \texttt{CTGamma} curvature evolution code \cite{pollney:11}, 
which was developed for arbitrary coordinate systems mapping the spatial domain.

The standard BSSN system is derived from a $3+1$ split of spacetime resulting in a
foliation in terms of spatial hypersurfaces along a timelike vector field.
It introduces the following set of evolved variables
\begin{equation} \label{eq:curv-evolved}
  \phi,\quad \tg_{ab},\quad K,\quad \tA_{ab},\quad \tG^a,
\end{equation}
which are solved according to
\begin{subequations}
\begin{align}
  \p_t \phi      = & -\frac{1}{6} \alpha K 
                  + \frac{1}{6} \p_i\beta^i,
                  \label{eq:evo_phi} \\
  \p_t \tg_{ab} = & -2 \alpha \tA_{ab} 
                  + \beta^i\p_i\tg_{ab} 
                  + 2 \tg_{i(a}\p_{b)}\beta^i
                  \label{eq:evo_tg} \\
                  & - \frac{2}{3}\tg_{ab} \p_i\beta^i,
		  \nonumber \\
  \p_t K        = & -D_iD^i \alpha + \alpha (A_{ij}A^{ij} 
                  + \frac{1}{3} K^2) 
                  + \beta^i\p_iK \label{eq:evo_K}  \\
                  & + 4\pi\alpha\left(\rho_{\rm ADM}+S\right),
                  \nonumber \\
  \p_t \tA_{ab} = & \mathrm{e}^{-4\phi} (-D_aD_b\alpha 
                  + \alpha R_{ab})^\text{TF}
                  + \beta^i\p_i\tA_{ab}
                  \label{eq:evo_tA} \\
                  & + 2\tA_{i(a}\p_{b)}\beta^i
                  - \frac{2}{3}A_{ab}\p_i\beta^i \nonumber \\
                  & - 8\pi \rm{e}^{-4\phi}\alpha \left(S_{ab}\right)^\text{TF}, \nonumber  \displaybreak[3] \\
  \p_t \tG^a    = & \tg^{ij}\p_i\beta_j\beta^a 
                  + \frac{1}{3} \tg^{ai}\p_i\p_j\beta^j
                  - \tG^i\p_i\beta^a
                  \label{eq:evo_tG} \\
                  & + \frac{2}{3}\tG^a \p_i\beta^i
                  - 2\tA^{ai} \p_i\alpha
                  \nonumber \\
                  & + 2\alpha (\tG^a_{ij} \tA^{ij} 
                               - \frac{\kappa}{2} \tA^{ai}\frac{\p_i\hp}{\hp}
                               - \frac{2}{3} \tg^{ai}\p_iK)  \nonumber \\
                  & - 16\pi\alpha\tg^{ai}S_i, \nonumber
\end{align}
\label{eq:bssn}
\end{subequations}
where $D_a$ is the covariant derivative determined by the conformal 3-metric $\tg_{ab}$, and
``TF'' indicates that the trace-free part of the bracketed term is
used.

Above, we show the ``$\phi$''-variant of the BSSN system.
Our curvature evolution code also provides the ``$\chi$''- and ``$W$''-variants of the
evolution system (see \cite{pollney:11} for details).
Here, we employ the $\phi$-variant.

The stress-energy tensor $T_{\mu\nu}$ is incorporated via the projections
\begin{eqnarray}
  \rho_{\rm ADM} & := & \frac{1}{\alpha^2} \left( T_{00} - 2 \beta^i T_{0i} +
  \beta^i \beta^j T^{ij} \right)\,,
  \\
  S & := & \tg^{ij} T_{ij}\,,
  \\
  S_a & := & - \frac{1}{\alpha} \left( T_{0a} - \beta^j T_{aj} \right)\,,
  \\
  (S_{ab})^\text{TF} & := & \left(T_{ab}-\frac{1}{3}\rm{e}^{4\phi}S \tg_{ab}\right)\,.
\end{eqnarray}

After each evolution step, the evolved curvature variables \eqref{eq:curv-evolved} are transformed (via an algebraic relation) to the standard ADM
variables $\left\{g_{ij}, K_{ij}\right\}$ (e.g., \cite{alcubierre:08}), where $g_{ij}$ is the (physical) 3-metric, and $K_{ij}$ the extrinsic curvature.
The ADM variables are used to couple the curvature evolution to the hydrodynamic evolution scheme, i.e., our hydrodynamic
scheme uses the physical 3-metric $g_{ij}$ rather than the evolved conformal 3-metric $\tg_{ab}$ above. 

The \textit{lapse} gauge scalar $\alpha$ is evolved using the
$1+\log$ condition~\cite{bona:95},
\begin{equation}
  \partial_t \alpha - \beta^i\partial_i\alpha 
    = -2 \alpha K,
  \label{eq:one_plus_log}
\end{equation}
while the \textit{shift} gauge vector $\beta^a$ is evolved using the hyperbolic $\tG$-driver
equation~\cite{alcubierre:03a},
\begin{subequations}
\begin{align}
  \partial_t \beta^a - \beta^i \partial_i  \beta^a & 
    = \frac{3}{4} B^a\,, \\
  \partial_t B^a - \beta^j \partial_j B^i &
    = \partial_t \tilde\Gamma^a - \beta^i \partial_i \tilde\Gamma^a
    - q(r)\eta B^a\,,
\end{align}
\end{subequations}
where $\eta$ is a parameter which acts as a (mass dependent) damping
coefficient.
To avoid certain stability issues with the gauge arising in the far-field regime \cite{ES-Schnetter2010a}, 
the damping coefficient is allowed to 
spatially change, either by some dynamic evolution \cite{Muller:2009jx}, 
or by a fixed prescription.
We use the simple prescription for a radial fall-off of $\eta$ given in \cite{ES-Schnetter2010a}
If not stated otherwise, we use a fall-off radius of $R=250\, M_\odot$.

The 3+1 decomposition of the Einstein equations also results in a set of
constraint equations.
The Hamiltonian constraint equation reads
\begin{equation} \label{eq:ham-constr}
H \equiv R^{(3)} + K^2 - K_{ij} K^{ij} -16 \pi \rho_{\rm ADM}= 0\,,
\end{equation}
where $R^{(3)}$ denotes the 3-Ricci scalar,
and the momentum constraint equations read
\begin{equation}
M^a \equiv D_i(K^{ai} - \gamma^{ai}K) - 8\pi S^a = 0\,.
\end{equation}
We do not actively enforce the constraints during evolution, but rather
check how well our numerically obtained metric quantities satisfy the constraints over the course
of the evolution. Thus, this offers a valuable accuracy monitor for the curvature evolution.

The spacetime equations are discretized using fourth-order finite difference operators \cite{diener:05}.
The finite difference stencils are centered. An exception are the advection terms of the form $\beta^i \partial_i$, which use operators
that are upwinded by one stencil point towards the local direction of the shift vector $\beta^i$ \cite{pollney:11}.

Consistent with the order of accuracy of spatial finite difference derivatives, we also apply Kreiss-Oliger dissipation \cite{diener:05} which is 
of one order higher than the spatial discretization order. In the case of fourth-order differencing, we
thus apply fifth-order dissipation operators.
Dissipation is added to the right-hand-sides (RHS) of the curvature evolution quantities at any time integration substep.
The strength of the dissipation can be controlled by a parameter $\epsilon_{\rm diss}\in[0,1]$. Unless otherwise specified, 
we use $\epsilon_{\rm diss}=0.1$ throughout this work.

At outer boundaries, we impose a simple approximate radiative boundary condition as
described in \cite{pollney:11}. Since data from this condition are not strictly constraint satisfying,
constraint violating modes are generated at the boundary, and travel with the speed of light \cite{brown:07a, ES-Brown2007b} 
to the interior of the domain where they introduce a systematic error in the curvature evolution.

\subsection{Multipatches}
\label{sec:multipatch}

We build our code on the \texttt{Llama} infrastructure described in detail in \cite{pollney:11}.
This infrastructure implements multipatches via an arbitrary number of curvi-linear overlapping grid patches using
fourth-order Lagrange and second-order essentially non-oscillatory (ENO) interpolation for exchanging data in inter-patch ghost zones between neighboring patches.
In \cite{pollney:11}, only the pure vacuum problem was considered. Here, we extend the multipatch evolution scheme to include matter.

\subsubsection{Patch Systems}

\begin{figure}[ht!]
  \begin{center}
    \includegraphics[width=\linewidth,trim=20 20 0 20,clip=true]
      {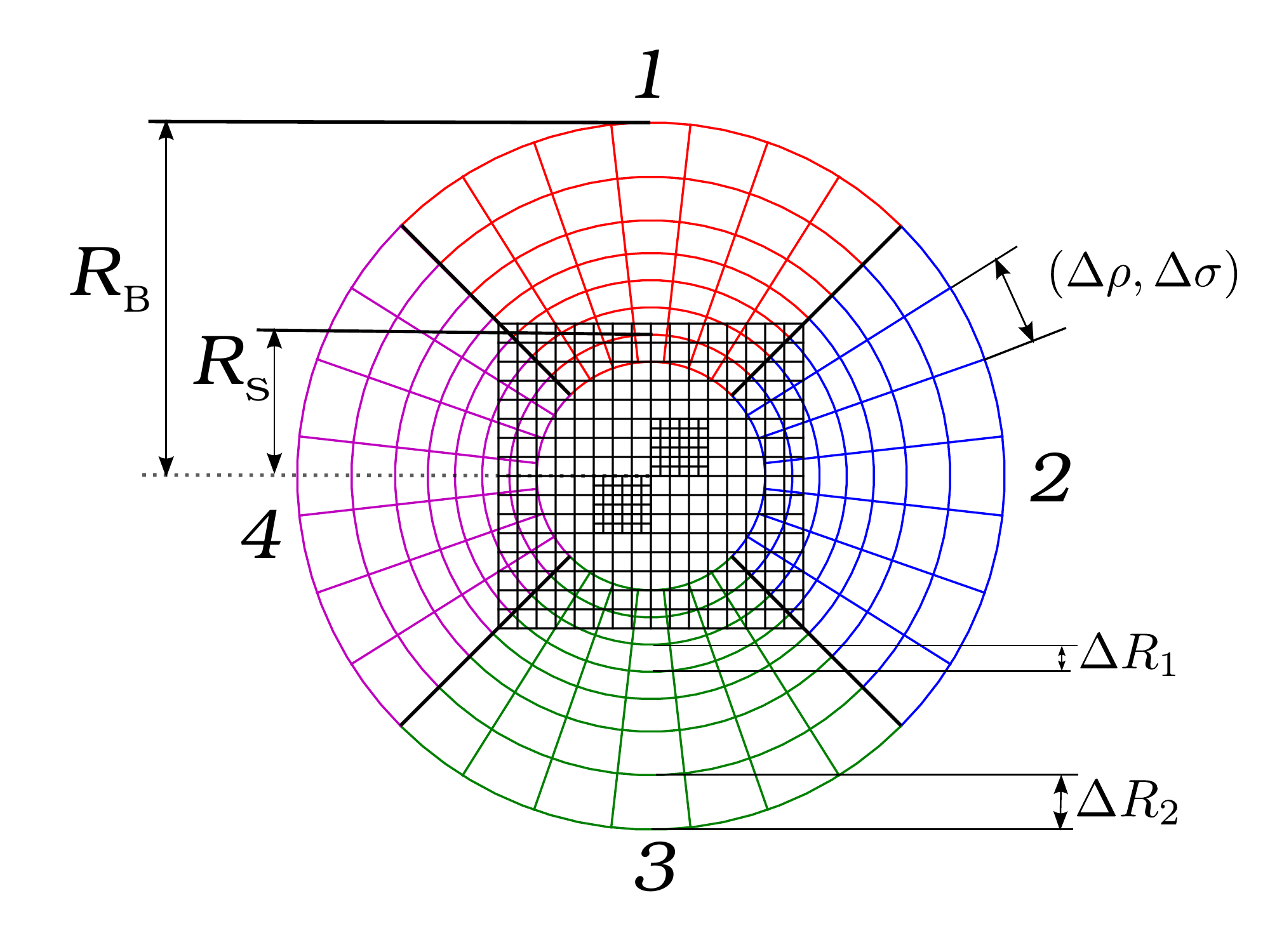}
    \includegraphics[width=0.85\linewidth,trim=100 0 110 0,clip=true]
      {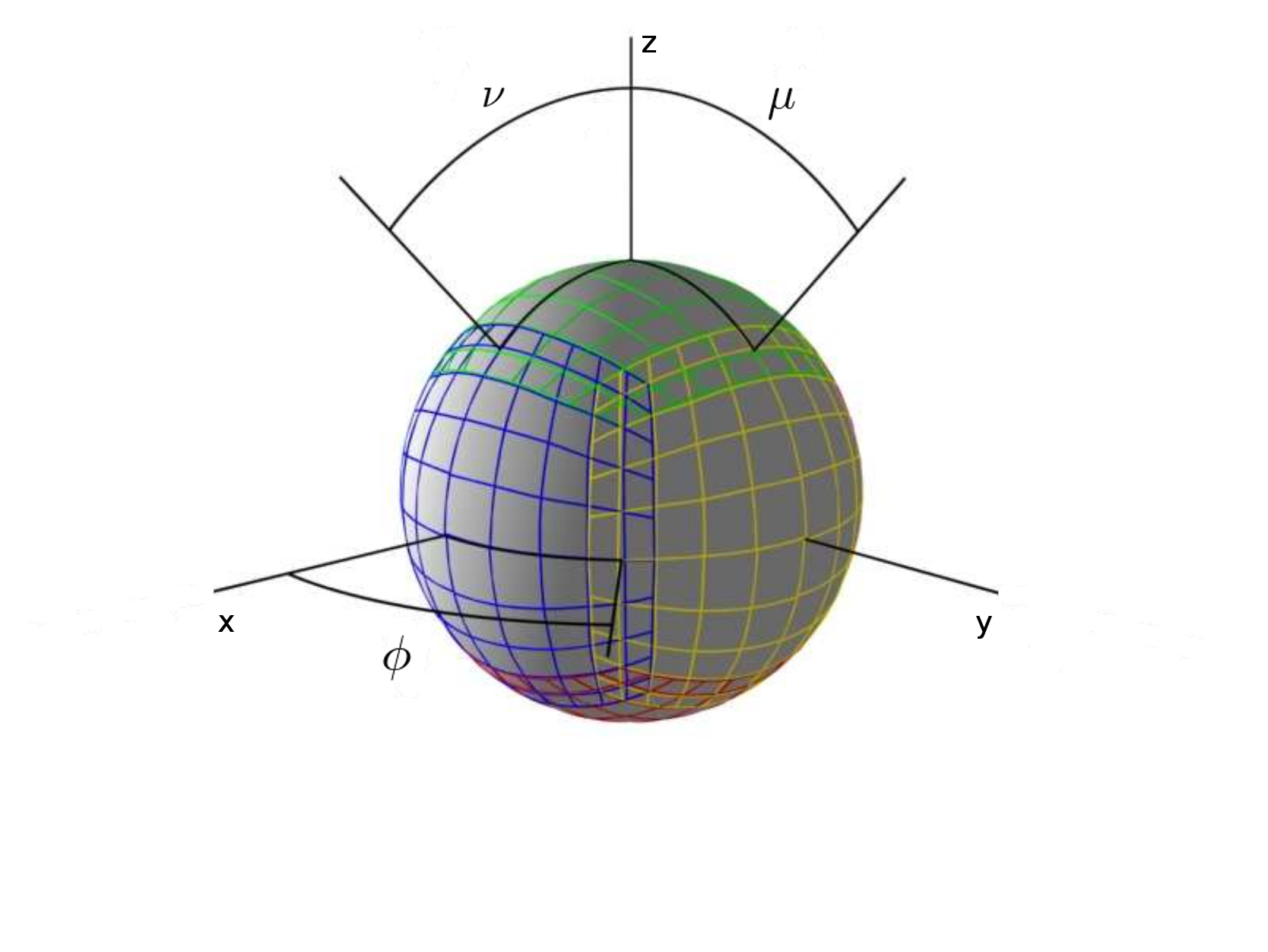}
  \end{center}
  \vspace{-2cm}
  \caption{Depiction of a typical patch system used in our simulations. 
           The upper figure schematically shows a $z=0$ slice of the employed grids: 
           a central Cartesian grid (patch $0$) is surrounded by spherical \emph{inflated-cube} grid patches (patches $1-4$ out of a total of six spherical patches). 
           The central grid is capable of AMR allowing to refine the resolution 
           at the central region of, e.g., a star where the density and curvature gradients become large.
           $R_{\rm B}$ and $R_{\rm S}$ denote the radii of the outer computational boundary and of the boundary between spherical and Cartesian grids, respectively.
           The spherical grid has a fixed angular resolution denoted by $(\Delta\rho,\Delta\sigma)$, while the radial resolution is allowed to stretch from radial resolution
           $\Delta R_1$ to $\Delta R_2$.
           The lower figure shows a radial $R=\mathrm{const}.$ shell of the outer spherical grid, comprised of six inflated-cube grid patches.
           Angular points can be uniquely determined by two out of three angular coordinates $(\mu,\nu,\phi)$ \eqref{eq:ang-coords}.
           Interpolation at patch boundaries reduces to 1D interpolation. Points are almost uniformly distributed across the sphere.}
  \label{fig:7patch-system}
\end{figure}

A useful patch system is shown in Fig.~\ref{fig:7patch-system}: the central Cartesian patch is surrounded by six spherical inflated-cube
patches. 
The \textit{nominal}\footnote{We define the \textit{nominal} grid as the unique set of points covering the entire computational domain, i.e.~the nominal
grid of a single patch excludes ghost points (and additional overlap points; see further below) that are shared with a neighboring patch.} 
grids of the spherical patches have inner radius $R_{\rm S}$, outer radius $R_{\rm B}$, radial
spacing $\Delta R_1$, which is allowed to stretch to $\Delta R_2$ within some finite region,
and angular resolution $(\Delta \rho, \Delta \sigma)$ per angular direction $(\rho,\sigma)$.
Note that the angles $(\rho,\sigma)$ used to define the local coordinates of each inflated-cube patch do not coincide with standard spherical-polar coordinates (see below). 
The central patch contains a hierarchy of refined regions, allowing to place resolution where necessary.
This patch system is particularly useful in problems with spherical symmetry at some radius from the central source. 

Each grid patch defines \textit{local} uniform coordinates $(u,v,w)$ related to the \textit{global} Cartesian $(x,y,z)$ coordinate space
by a diffeomorphic relation. For the central Cartesian patch depicted in Fig.~\ref{fig:7patch-system}, this relation is trivially
given by the identity function. The inflated-cube coordinates, however, are defined by non-trivial coordinate functions.
For each angular patch, we define local angular coordinates $(\rho,\sigma)$ that range over
$(-\pi/4,+\pi/4)\times(-\pi/4,+\pi/4)$ and can be related to global
angular coordinates $(\mu,\nu,\phi)$ (see Fig.~\ref{fig:7patch-system}) which are given by
\begin{subequations} \label{eq:ang-coords}
\begin{align}
  \mu  \equiv \text{rotation angle about the x-axis} &= \arctan (y/z), \\
  \nu  \equiv \text{rotation angle about the y-axis} &= \arctan (x/z), \\
  \phi \equiv \text{rotation angle about the z-axis} &= \arctan (y/x).
\end{align}
\end{subequations}
For each angular patch, we have two unique angles $(\rho,\sigma)$ 
out of the three global angles $(\mu,\nu,\phi)$ that parametrize the local coordinates.
For instance, for the patch normal to the positive $x$-direction, we select
\begin{subequations}
\begin{align}
  \rho \equiv \nu &= \arctan (z/x), \\
  \sigma \equiv \phi &= \arctan (y/x), \\
  R &= f(r),
\end{align}
\end{subequations}
where $r=\sqrt{x^2 + y^2 + z^2}$.
Similarly, the coordinates of the patch along the positive $y$ and $z$ axes are parametrized by $(\rho,\sigma)\equiv(\mu,\phi)$ and $(\rho,\sigma)\equiv(\mu,\nu)$, respectively.
The remaining three patches along the negative axes are related in a similar way. 

In the radial coordinate direction, we apply radial stretching with
an appropriate stretching function $R=f(r)$.
In the stretching region, the physical coordinate radius is stretched, corresponding
to a smooth decrease in radial resolution from spacing $\Delta R_1$ to spacing $\Delta R_2$. 
Outside the stretching region, we keep the radial spacing constant. 
Details can be found in \cite{pollney:11}.

\subsubsection{Spacetime Evolution Scheme}
\label{sec:spacetime-multiblock}

Here, and as described in \cite{pollney:11}, the spacetime evolution is solved in the \textit{global} Cartesian $(x,y,z)$ tensor basis, where
the grid patches are generally distorted, i.e., they are \textit{not} uniform. 
Derivatives are approximated via finite differences in the local coordinate system $(u,v,w)$ of each grid patch, where, as required by our finite difference scheme,
the grid patches are uniform. In order to transform to the global tensor basis, 
Jacobian transformations of the form ${J^i}_j=\partial u^i/\partial x^j$
are applied to the first and second
derivatives at each point, 
\begin{subequations}
\begin{align}
  \frac{\p}{\p x_i} &= \left(\frac{\p u_j}{\p x_j}\right)\frac{\p}{\p u_j}, \\
  \frac{\p^2}{\p x_i \p x_j} &=
    \left(\frac{\p^2 u_k}{\p x_i \p x_j}\right) \frac{\p^2}{\p u_k^2} +
    \left(\frac{\p u_k}{\p x_i} \frac{\p u_l}{\p x_j}\right)
      \frac{\p^2}{\p u_k \p u_l}\,,
\end{align}
\end{subequations}
thus obtaining the derivatives in the global $(x,y,z)$ coordinate space.
The Jacobians are precomputed at each grid point.
The main advantage of solving the equations in the global $(x,y,z)$ basis is simplicity. There is no need for inter-patch coordinate basis transformations.
Perhaps more importantly, the existing code infrastructure, and especially analysis tools, do not need to be changed, since the assumption
of a global Cartesian tensor basis is still maintained.

\subsubsection{Hydrodynamic Evolution Scheme}
\label{sec:hydro-multiblock}

Finite volume schemes work well on general unstructured meshes.
The original implementation of the hydrodynamic evolution code \texttt{GRHydro}, however, assumes uniform coordinates.
Without a major rewrite of the code, we can keep our original scheme by solving the Riemann problem in the \textit{local} frame, where
the coordinates are uniform.
This requires no changes to the core of the scheme. 
Any computation simply carries over to the local coordinate basis. 
Effectively, this means that the primitive and conserved quantities are thus represented 
in the local coordinate basis.

Special attention is required when coupling the hydrodynamics solver to the metric solver (Sec.~\ref{sec:spacetime-multiblock}).
The metric solver explicitly computes the metric components in the \textit{global} frame and is thus generally incompatible with
the hydrodynamic quantities defined in the \textit{local} frame. 
We therefore introduce the additional step of transforming the metric components to the \textit{local} basis
before each hydrodynamic RHS step. 
Correspondingly, after each hydrodynamic step, 
we need to compute the stress-energy tensor $T^{\mu\nu}$ in the \textit{global} basis as required by the metric solver.

Since the various analysis tools explicitly assume a global coordinate frame for the primitive variables,
we introduce a separate set of \textit{global} 
primitive variables.
Effectively, this only requires extra memory for the primitive 3-velocity $\{\tilde{v}^i\}$, 
since the primitive density $\rho$ and $\tau$ are scalars. 
Once the primitive quantities are known in the global frame,
the stress-energy tensor can be directly computed in the global frame.

For clarity, we list the various quantities in their corresponding available coordinate basis in Table~\ref{tab:evolved-quantities}.
\begin{table}[t]
\caption{Required quantities for the hydrodynamic evolution scheme and their coordinate bases. A tilde denotes quantities that need to be obtained 
by applying a Jacobian transformation. The last four quantities are only required for microphysical equations of state.}
\label{tab:evolved-quantities}
\begin{ruledtabular}
\begin{tabular}{llll}
Quantity              & Type   & global   & local \\
\hline
metric                & tensor & $g_{ij}$ & $\tilde{g}_{ij}$ \\
extrinsic curvature   & tensor & $K_{ij}$ & $\tilde{K}_{ij}$ \\
shift                 & vector & $\beta^i$ & $\tilde{\beta}^i$ \\
lapse                 & scalar & $\alpha$  & $\alpha$ \\
prim. density         & scalar & $\rho$    & $\rho$ \\
specific internal energy & scalar & $\epsilon$ & $\epsilon$ \\
prim. velocity        & vector & $\tilde{v}^i$ & $v^i$ \\
cons. density         & densitized scalar & -  &  $D$ \\
cons. internal energy & densitized scalar & -  &  $\tau$ \\
momentum              & densitized vector & -  &  $S_i$ \\
stress-energy tensor  & tensor            & $T^{\mu\nu}$ & - \\
Lorentz factor        & scalar & $W$      & - \\
pressure              & scalar & $P$      & - \\
\hline
prim. electron fraction & scalar & $Y_e$    & $Y_e$ \\
cons. electron fraction & densitized scalar & - & $Y_e^{\rm con}$ \\
temperature           & scalar & $T$      & $T$ \\
entropy               & scalar & $s$      & $s$ \\
\end{tabular}
\end{ruledtabular}
\end{table}

\subsubsection{Inter-Patch Interpolation and Coordinate Transformation}

\begin{figure}
  \begin{center}
    \includegraphics[width=\linewidth,trim=0 0 0 0]{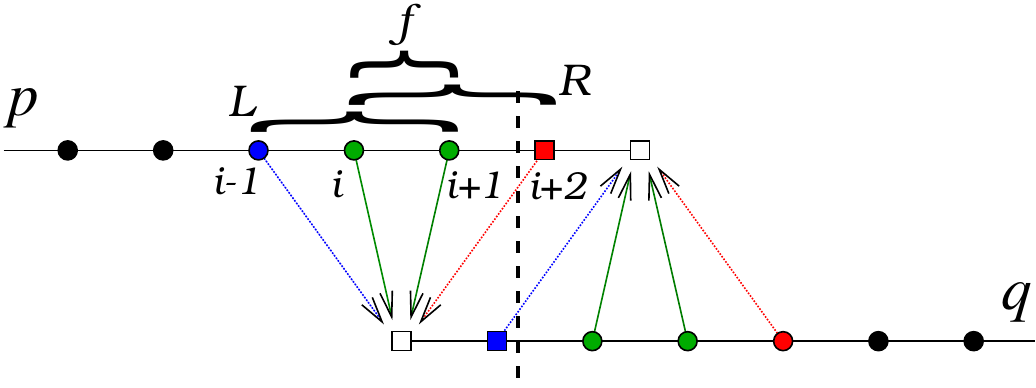}
  \end{center}
  \vspace{-0.5cm}
  \caption{Depiction of the second-order ENO inter-patch interpolation scheme used for the fluid variables
           between two overlapping patches $p$ and $q$. The inter-patch boundary is indicated by the vertical line.
           Each interpolated point in the ghost zones (empty boxes)
           is obtained from an interpolation polynomial whose stencil is selected based on the local smoothness of the interpolated quantity. 
           There are three possible choices: left ($L$) stencil using blue and green points, right ($R$) stencil using green and red points, and first-order ($f$) stencil using
           only green points.
           Since none of the stencil points on $p$ are allowed to be
           inter-patch boundary points of $p$, we need to introduce a certain number of \textit{additional} overlap points (filled boxes)
           to ensure that this is the case.}
  \label{fig:GZ-interp}
\end{figure}

\begin{figure}
  \begin{center}
    \includegraphics[width=0.6\linewidth,trim=0 0 0 0]{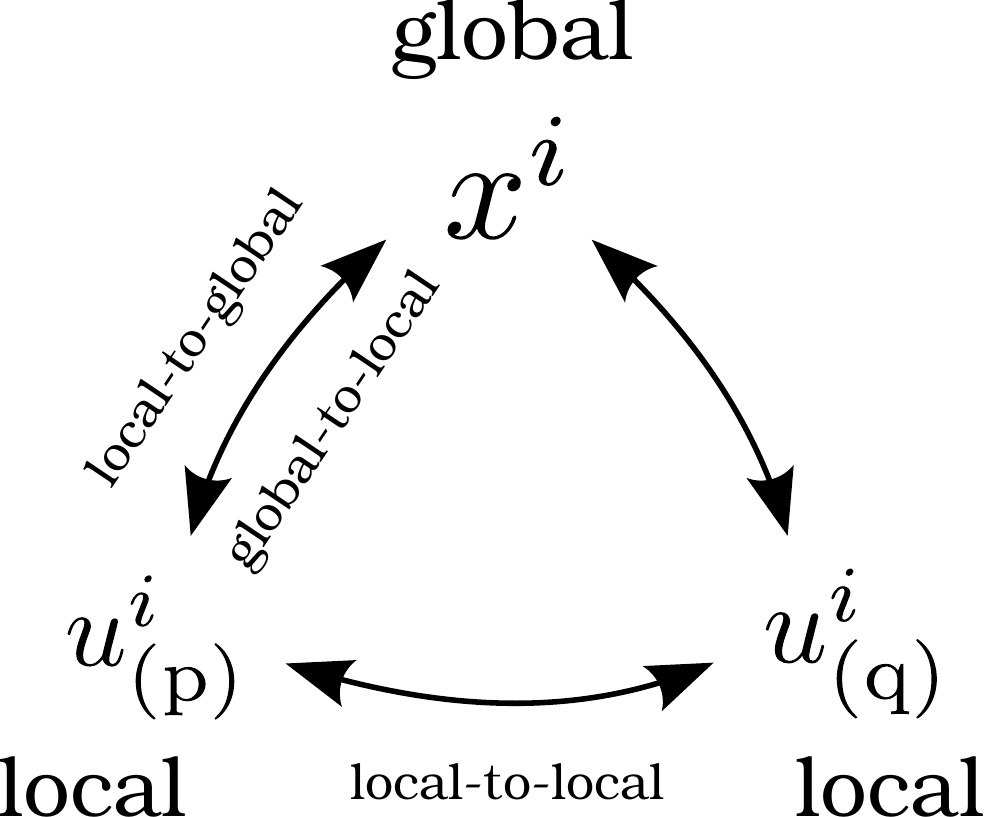}
  \end{center}
  \vspace{-0.5cm}
  \caption{Coordinate systems and their transformations. Local coordinates $u^i_{\rm (p)}$ and $u^i_{\rm (q)}$ of patches
           $p$ and $q$, respectively, are related via ``local-to-local'' transformations.
           ``Local-to-local'' transforms are necessary for fluid variable inter-patch interpolation.
           The global Cartesian coordinates $x^i$ are used to represent the curvature variables and to carry out any analysis on the curvature or fluid variables, 
           such as gravitational-wave extraction, or fluid density oscillation mode analysis.
           Therefore, ``global-to-local'' and ``local-to-global'' transforms are necessary.}
  \label{fig:coord-trans}
\end{figure}

Data in the ghost zones of a given grid patch are exchanged via high-order Lagrange polynomial interpolation for those quantities
that are smooth (such as the curvature evolution variables), and optionally second-order essentially non-oscillatory (ENO) interpolation \cite{shu:98} for those variables that
may contain discontinuities (such as the hydrodynamic evolution variables).
The scheme is depicted in Fig.~\ref{fig:GZ-interp}. Ghost points (indicated by empty boxes) on some patch $p$ must 
be interpolated from points from a neighboring overlapping patch $q$.
The inter-patch boundary is indicated by a vertical line.
For Lagrange interpolation, in order to maintain maximal accuracy, we center the interpolation stencils around the interpolation point.
The ENO operator, on the other hand, is allowed to use second-order off-centered Lagrange interpolation stencils according to the local smoothness of the interpolated fields \cite{shu:98}.
In addition, we check if the interpolant introduces a local maximum and switch to first order in that case.
In order to speed up the computation, we precompute and store all possible stencil configurations
for each inter-patch ghost point.

To yield a consistent boundary treatment, we have to ensure that an interpolation stencil does not 
contain any ghost points from the source patch. For this to be the case, we need to introduce \textit{additional overlap}
points (indicated by colored boxes in Fig.~\ref{fig:GZ-interp}) that lead to an overlap of the evolved region.
Effectively, this means that the equations are solved twice in the additional overlap region, 
which introduces a small computational overhead. 

We note that quantities which are defined in the 
global Cartesian tensor basis such as the curvature evolution variables 
\eqref{eq:curv-evolved} do not need to be transformed between local coordinates
patches.
In our present hydrodynamics scheme, however, the evolved conserved variables are defined in local coordinates. 
Hence, for inter-patch ghost zone interpolation, they must be transformed between local coordinate systems.
Let us denote the local coordinates of source patch $p$ as $u^i_{\rm (p)}$,  
and the local coordinates of target patch $q$ as coordinates $u^i_{\rm (q)}$.
The conserved density, which is a pseudo-scalar of tensor weight $+1$,
transforms as a scalar \textit{tensor density} according to the ``local-to-local'' transformation
\begin{equation} \label{eq:patch-trafo-scalar-dens}
\hat{D}_{\rm (q)} = \left|\det\frac{\p u^i_{\rm (p)}}{\p u^j_{\rm (q)}}\right| \hat{D}_{\rm (p)}
\end{equation}
between local coordinates $u_{\rm (p)}$ of patch $p$ and local coordinates $(u_{\rm (q)})$ of patch $q$.
Hence, after having obtained its interpolated value in the ``old'' basis defined by 
the local coordinates of patch $p$, we need to represent it in the ``new'' basis defined by the local coordinates of patch $q$ according 
to transformation \eqref{eq:patch-trafo-scalar-dens},
before we assign its transformed value to one of the ghost points of $q$.
Similarly, we also need to transform the conserved 3-momentum, which transforms as a densitized contravariant vector according to
\begin{equation} \label{eq:patch-trafo-vect}
 \hat{S}^j_{\rm (q)}= \left|\det\frac{\p u^k_{\rm (p)}}{\p u^l_{\rm (q)}}\right| \frac{\p u_{\rm (q)}^j}{\p u_{\rm (p)}^i} \hat{S}^i_{\rm (p)}\,.
\end{equation}

The various coordinate transformations that are required in our code are depicted in Fig.~\ref{fig:coord-trans}.

\subsection{Cell-centered AMR and Refluxing}
\label{sec:ccamr}

We introduce cell-centered AMR in combination with a \textit{refluxing} scheme at
refinement level boundaries to ensure conservation of rest mass and -- in
the absense of GR effects -- also momentum and
energy of the fluid \cite{Berger1984, berger:89}. 
Because gravity leads to sources and sinks for fluid momentum and energy,
these quantities are generally not conserved in curved spacetimes.
This is reflected in the source terms
of the fluid conservation laws \eqref{eq:conservation_equations_gr}, which are zero only in flat space.
The numerical fluxes in our finite volume scheme between grid cells, however, must be conserved. 
Since we employ subcycling in time where finer
grids take multiple small time steps for each coarse grid time step
\cite{Schnetter-etal-03b}, the conservation properties of
our finite volume approach do not hold at mesh refinement boundaries without refluxing.

\begin{figure}
    \centering
    \includegraphics[width=0.6\linewidth]{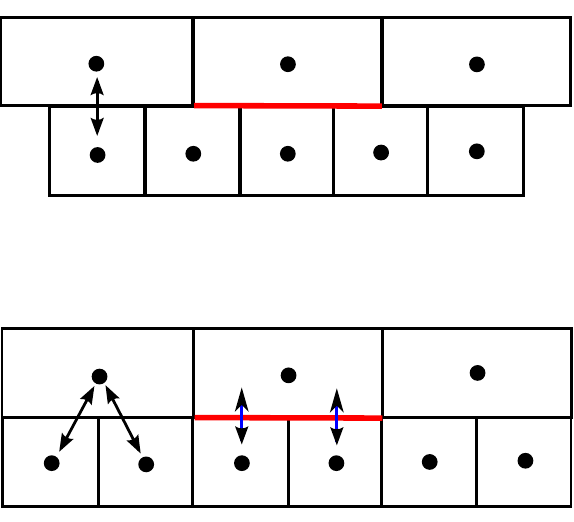}
    \caption{Vertex-centered AMR (upper figure) versus cell-centered AMR (lower figure).
             In cell-centered AMR, two fine grid cell faces always coincide with a coarse grid cell face (red line).
             Thus, it becomes possible to sum up the two fine grid fluxes computed on cell faces to become one coarse grid cell flux.
             Cell-centered quantities, however, always need to be interpolated in the prolongation and restriction operation.
             In the vertex-centered case, every second grid point coincides with one coarse grid point. Thus, interpolation is not necessary for every point, 
             and restriction becomes exact.
    }
    \label{fig:amr-vc-vs-cc}
\end{figure}

In cell-centered AMR schemes, coarse cells are subdivided into
multiple smaller cells, ensuring that coarse grid and fine grid cell
faces align (see red line in the lower part of Fig.~\ref{fig:amr-vc-vs-cc}). In contrast, the cell
centers do not align. This is different from vertex-centered AMR
schemes, where one aligns coarse and fine grid cell centers but not
their faces (red line in the upper part of Fig.~\ref{fig:amr-vc-vs-cc}).

One may argue that vertex-centered schemes are more natural for
wave-type equations such as the Einstein equations, which is why vertex-centered refinement 
was originally implemented in the Carpet AMR driver.
However, refluxing requires cell-centered refinement, and this comes with a
certain added complexity that we describe below.

\paragraph{Prolongation.} Prolongation is the interpolation from
coarse to fine-grid cells. In a vertex-centered scheme (and when
assuming a refinement factor of two), every second
fine-grid point is aligned with a coarse-grid point, and prolongation
there corresponds to a copy. In between coarse-grid points, one needs
to interpolate. Curvature quantities are interpolated via a fifth-order
Lagrange polynomial.
Hydrodynamics quantities are interpolated via a second-order 
ENO interpolator \cite{shu:98} (also see Sec.~\ref{sec:hydro-multiblock})
to avoid oscillations near discontinuities.

In a cell-centered scheme, \textit{every} fine-grid cell requires
interpolation. We interpolate curvature quantities via a fourth-order
Lagrange polynomial, and interpolate
hydrodynamics quantities via a second-order ENO interpolator.

\paragraph{Restriction.} Restriction transfers fine-grid information
to the next coarser grid, after both have been evolved in time, and
are aligned in time again.
Different discretization errors will have led to slightly different
results, and one overwrites the coarse-grid results by respective fine-grid 
results. For a vertex-centered scheme, this is straightforward,
since each coarse-grid point is aligned with a fine-grid point, and
hence the variable on the fine-grid point can simply be copied.

For cell-centered schemes, things are more complex, since restriction
also requires interpolation. We interpolate curvature quantities via
a third-order Lagrange polynomial. 
Hydrodynamics quantities are averaged, corresponding to
linear interpolation. This is a conservative operation, so that
e.g.\ the mass in a coarse-grid cell is the sum of the masses in all
contained fine-grid cells.

The distinction between curvature and hydrodynamics quantities is
crucial to achieving high accuracy. If one does not use higher-order
operations for the curvature quantities, then the accuracy of the
overall simulation is significantly reduced. On the other hand, one needs to employ
a conservative interpolation scheme for the hydrodynamics quantities,
but can accept a lower order of accuracy there.
For restricting curvature quantities, we therefore use third-order
polynomial interpolation.

\begin{figure}
  \includegraphics[width=0.8\linewidth]{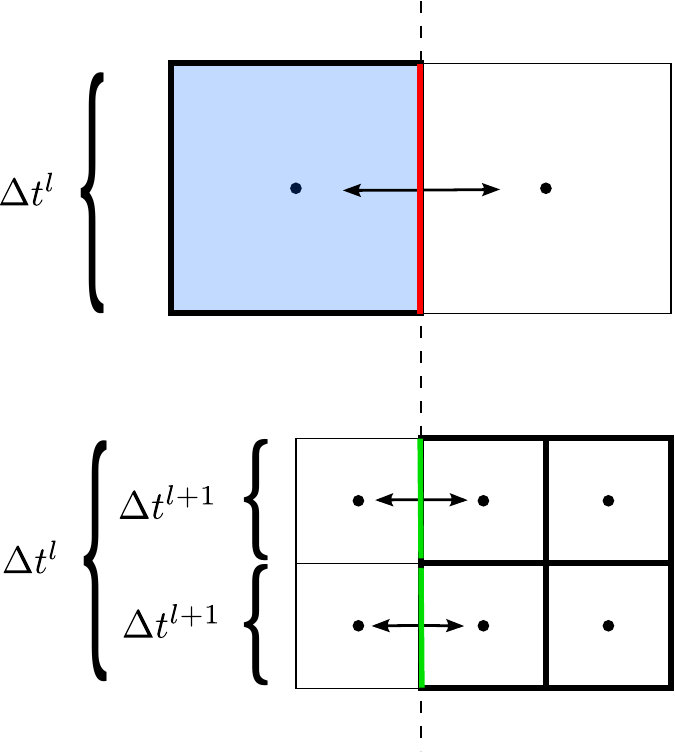}
  \caption{AMR time evolution, showing fluxes across cell faces, for
    both coarse (upper row) and fine cells (lower row). Time moves
    upwards; the fine grid (lower panel) takes multiple steps for each coarse grid
    step (upper panel). In Berger-Oliger AMR, the coarse and fine levels are evolved
    independently, and the sum of the fine grid fluxes crossing the green faces are not
    guaranteed to be equal to the coarse grid flux crossing the red face. At the end of a
    time step, the neighboring bold-faced coarse and fine cells may
    be in an inconsistent state, requiring \emph{refluxing} to add a correction
    to the light blue coarse grid cell.}
  \label{fig:amr-refluxing}
\end{figure}

\paragraph{Refluxing.} Refluxing is an algorithm to ensure
conservation across mesh refinement boundaries \cite{berger:89, east:11}.
Since coarse and fine
grids are evolved in time independently, it is not guaranteed that the
fluxes leaving the fine grid are identical to those entering an abutting
coarser grid (see Fig.~\ref{fig:amr-refluxing}).
Refluxing integrates the coarse grid and
fine grid fluxes across these faces, and then adjusts the coarse grid
cell just outside the refined region according to the flux difference.

We outline the generic refluxing algorithm for a conserved quantity $f$ in the steps below.
\begin{enumerate}
\item
We start with a fine grid level $l+1$ and a coarse grid level $l$ which are
momentarily aligned in time, i.e.~$t^l_i=t^{l+1}_{2j}$, where $i$ denotes the $i$-th step
on the coarse level, and $j$ denotes the $j$-th step on the fine grid. Due to subcycling in time,
for any coarse-grid time step, there are twice as many fine-grid time steps, i.e.~$i=2j$.
\item
At the refinement boundary (red line of Fig.~\ref{fig:amr-vc-vs-cc}, or red and green lines in Fig.~\ref{fig:amr-refluxing}),
we store integrated coarse and fine grid \textit{flux registers} $I^{l}$ and $I^{l+1}$ for some conserved quantity $f$. 
Due to the 2:1 mesh refinement, there are four integrated fine grid
flux registers for every integrated coarse grid flux register. (Only
two are visible in Fig.~\ref{fig:amr-vc-vs-cc}).
At $t^l_i=t^{l+1}_{2j}$, all registers are zero.
\item
Each refinement level is independently integrated forward in time until 
the two refinement levels are aligned in time again, i.e.,~until we have $t^l_{i+1}=t^{l+1}_{2j+2}$.
During each integration step, the hydrodynamic evolution scheme computes fluxes $F$ for a quantity $f$ located at all cell interfaces.
At the refinement boundary, we use the computed fine grid fluxes $F^{l+1}$ on the fine grid cell interfaces, and coarse grid fluxes $F^l$ on the
coarse grid cell interfaces to integrate coarse and fine grid flux registers
forward in time, i.e., we independently integrate
\begin{equation}
\partial_t I^{l+1} = F^{l+1},\quad \partial_t I^{l} = F^{l}\,,
\end{equation}
at the refinement boundary.
\item
After restriction, when $t^l_{i+1}=t^{l+1}_{2j+2}$, we use $I^{l+1}$ and $I^{l}$
to compute a correction for conserved quantity $f$.
The correction is obtained as follows.
\begin{enumerate}
 \item 
 The integrated fine grid flux register $I^{l+1}$ is restricted to the coarse grid via
 \begin{equation}
  I_{\rm fine}^{l}= \mathcal{R} I^{l+1}\,,
 \end{equation}
 where $\mathcal{R}$ denotes the cell interface restriction operator. Note that 
 since the flux registers are stored on cell \textit{faces}, this operator
 is different from the operator used for the fluid state vector. 
 \item
 A correction $C_f^l$ for conserved quantity $f$ on coarse grid level $l$ is now obtained via
 \begin{equation}
 C_f^l = (I^{l}_{\rm fine} - I^l) / \Delta^l x
 \end{equation}
 where $\Delta^l x$ denotes the grid spacing of refinement level $l$. 
\end{enumerate}
\item
The correction $C_f^l$ is added to the coarse grid cell on level $l$ next to the refinement
boundary (blue cell in Fig.~\ref{fig:amr-refluxing}), i.e.
\begin{equation}
f_{\rm corrected}^l = f^l + C_f^l\,.
\end{equation}
This completes the refluxing operation. We repeat the steps $1$-$5$ until the evolution is complete.
\end{enumerate}
The steps above are performed for any of the evolved conserved quantities $D$, $S^i$, $\tau$, and $Y_e^{\rm con}$.

We note that the state thus obtained in the corrected coarse
grid cells may be thermodynamically inconsistent, e.g., near the
surface of a star, and may need to be projected onto a self-consistent
state. This is to be expected with our atmosphere treatment, as we
discuss in Appendix~\ref{sec:atmo}.

\subsection{Time Integration and Multirate Runge-Kutta Schemes}
\label{sec:multirate}

We carry out time integration using the Method of Lines (MoL) \cite{Hyman-1976-Courant-MOL-report}.
MoL is based on a separate treatment of the spatial derivatives (the right-hand sides), and the time derivatives.
This allows one to employ integration methods for ordinary differential equations (ODE) such as Runge-Kutta (RK) schemes for the time integration.

We evolve the spacetime and hydrodynamic sector of our evolution system simultaneously using full matter-spacetime coupling.
The coupling between the two sectors is achieved via source terms.
The spacetime evolution is sourced by the stress-energy tensor computed by the hydrodynamic sector. Vice versa, the hydrodynamic part contains additional source terms which
are a result of the coupling to a curved spacetime metric.
Written in simplified form, our system is given by
\begin{eqnarray}
\p_t \mathbf{g} &=& \mathbf{F}(\mathbf{g},\mathbf{q}) \label{eq:intmetric}\,, \\
\p_t \mathbf{q} &=& \mathbf{G}(\mathbf{g},\mathbf{q}) \label{eq:intmatter}\,,
\end{eqnarray}
where $\mathbf{g}$ denotes curvature evolution quantities, $\mathbf{q}$ denotes fluid evolution quantities, and
$\mathbf{F}$ and $\mathbf{G}$ denote the RHS functions.

Traditionally, spacetime metric and hydrodynamic variables are evolved simultaneously using the same time integration scheme.
A standard choice in our case is the classical fourth-order Runge-Kutta (RK4) method.
The timestep is chosen such that the Courant-Friedrich-Lewy (CFL) factor, defined as $C=\Delta t /
\Delta x$, becomes $C=0.4$. The CFL factor is limited by the stability region of the numerical scheme,
which in turn is limited by the speed of light.

We observe two important points in our simulations. First, the error in our numerical evolution is in most cases
not dominated by the time integration (see Sec.~\ref{sec:results}). 
The choice of $\Delta t$ is not guided by accuracy requirements, but rather by the restrictions imposed by the CFL condition.
This is unfortunate since a larger timestep would speed up our simulation with only small negative impact on the accuracy.
Second, we find that the CFL factor is largely determined by the spacetime evolution.
In the Cowling approximation, i.e.~when the spacetime sector is not evolved and held fixed at its initial setup, 
we typically can use more than twice as large CFL factors (up to $C\approx 1$) without encountering any numerical instabilities. 

Since our timestep is fixed, rather than enlarging the timestep $\Delta t$ (and hence $C$), we switch to the classical second-order Runge-Kutta 
(RK2) method instead. This scheme has a smaller stability region by roughly a factor of two compared to RK4.
Due to the less restrictive CFL factor for the fluid evolution compared to the curvature evolution, however, we can still use the same timestep as for the curvature evolution with RK4.
The advantage of the RK2 schemes is that it require half as many RHS evaluations compared to RK4. 
The accuracy of RK2, however, is typically much lower than that of an RK4 integration. 
In practice, we find that the reduction in accuracy is not a severe limitation for most cases (see Sec.~\ref{sec:results}).

We therefore apply the RK2 integrator for the hydrodynamic sector,
while maintaining the RK4 integrator for the spacetime part. 

A scheme for coupling different parts of a system of equations with different RK integrators is 
given by \textit{multirate} RK schemes (e.g.~\cite{schlegel:09, constantinescu:07}).
Here, we make the simple Ansatz of performing one RK2 intermediate RHS evaluation for two RK4 intermediate RHS evaluations.
That is, the additional RK4 intermediate RHS evaluations simply use the results from the last intermediate RK2 step.

To be more explicit, given the equation
\begin{equation}
\p_t y = f(t,y)\,,
\end{equation}
where $f$ corresponds to the RHS,
we write a generic RK scheme according to
\begin{eqnarray}
y_{n+1} &=& y_n + \Delta t \sum_{i=1}^s b_i\, k_i\,, \\
k_i &=& f(t_n + c_i \Delta t\,, y_n + \Delta t \sum_{j=1}^s a_{ij} k_j)\,.
\end{eqnarray}
The coefficients $b_i$, $c_i$, and $a_{ij}$ can be written in the standard Butcher
notation (see, e.g.~\cite{numrep}).

In our multirate scheme, we use two different sets of coefficients. 
The coefficients for the RK2 scheme are arranged such that RHS evaluations coincide with
RK4 RHS evaluations.
We list the corresponding multirate Butcher tableau in Table~\ref{tab:butcher}.

\begin{table}[t]
\caption{Butcher tableau for an explicit multirate RK4/RK2 scheme. The right table (separated by the double vertical line) shows
the coefficients $b_i$ (bottom line), $c_i$ (first vertical column), and $a_{ij}$ for the classical RK4 scheme.
The left table shows the corresponding RK2 coefficients evaluated at timesteps that coincide with RK4 timesteps.}
\label{tab:butcher}
\begin{ruledtabular}
\begin{tabular}{l|llll||l|llll}
0 &     &   &   &          &   0   & & \\ 
0 & 0   &   &   &          &   1/2 & 1/2\\
0 & 0   & 0 &   &          &   1/2 & 0   & 1/2 &     & \\
1 & 1   & 0 & 0 &          &   1   & 0   & 0   & 1/2 & \\
\hline
 & 1/2 & 0 & 0 & 1/2 &     & 1/3 & 1/6 & 1/6 & 1/3\\
\end{tabular}
\end{ruledtabular}
\end{table}

\subsection{Gravitational Wave Extraction}
\label{sec:waveextract}

GWs are extracted in the \textit{wave-extraction zone} of our simulation.
We define the wave-extraction zone as the region on the computational grid which is at sufficient distance
from the gravitating source to avoid near-zone effects, and at the same time offers sufficient resolution 
to resolve the waves. Beyond the wave-extraction zone, we typically use radial stretching to gradually decrease the radial resolution 
up to a certain radius (e.g., Fig.~\ref{fig:7patch-system}).

We use the techniques described in detail in \cite{reisswig:11ccwave}.
Among those are (i) the standard slow-motion weak-field quadrupole formalism (see e.g., 
\cite{thorne:80,dimmelmeier:02,shibata:04,dimmelmeier:08,ott:07cqg,ott:07prl}) which is purely based on
the quadrupolar matter distribution and does not take into account any curvature effects, (ii)
Regge-Wheeler-Zerilli-Moncrief (RWZM) extraction based on gauge-invariant
spherical perturbations 
about a fixed Schwarzschild background (see \cite{nagar:05} for a review),
(iii) Newman-Penrose extraction based on complex spin-weighted components of the Weyl tensor \cite{penrose:63, newman:62, pollney:11}, 
and (iv) Cauchy-characteristic extraction (CCE) \cite{winicour:09,bishop:97b,
reisswig:11ccwave, reisswig:09, reisswig:10a, babiuc:11}
making use of nonlinear nullcone evolutions of the Einstein equations out to future null infinity $\scri^+$
(see \cite{reisswig:12null} for a new high-order algorithm).
The latter extraction technique is the only one capable of determining the gravitational radiation content unambiguously
and without finite-radius and gauge errors \cite{reisswig:11ccwave,
reisswig:09, reisswig:10a, babiuc:11}.

The curvature-based techniques (ii)-(iv) require one or two integrations in time in order to compute the strain, which may lead to strong non-linear and unphysical 
artificial drifts. This can be overcome
by the \emph{fixed frequency integration} (FFI) technique presented in \cite{reisswig:11}.
FFI requires the choice of a cut-off frequency $f_0$, which ideally must be below the physical frequency components
contained in the signal. For instance, for a typical binary neutron star inspiral signal, $f_0^{m} < m \Omega_{\rm orbital} / 2\pi$, where $\Omega_{\rm orbital}$
is the initial orbital frequency, and $m$ is the associated harmonic $m$-mode number.

The energy and angular momentum that is lost due to the emission of GWs can be computed in terms of spin-weighted spherical harmonic coefficients
of $\Psi_4$ as derived in \cite{lousto:07a, ruiz:08}.
We use the expressions for the radiated energy flux $dE_{\rm rad}/dt$ and angular momentum flux $dJ_{\rm rad}/dt$ in terms of the Weyl scalar $\Psi_4$ from \cite{ruiz:08}.
In the expressions for $dE_{\rm rad}/dt$ and $dJ_{\rm rad}/dt$, 
we evaluate the appearing time integrals of the harmonic modes using FFI with $f_0^m = m f_0$ for each given $m$-mode.
In order to obtain the total radiated energy $E_{\rm rad}$ and angular momentum $J_{\rm rad}$ from their fluxes, respectively, we time integrate
in the time domain\footnote{FFI cannot be applied since the radiated fluxes are non-oscillatory.}.

\subsubsection{Numerical Setup}

We report the numerical settings employed for the various wave extraction
techniques that are used in this work. 
Since we are not interested in the numerical convergence properties of the wave
extraction methods themselves (this has been analyzed elsewhere,
e.g.~\cite{reisswig:11ccwave, reisswig:09, reisswig:10a, pazos:06,
pollney:09, pollney:11}), 
we stick to fixed settings for all test cases
and numerical resolutions considered in Sec.~\ref{sec:results}.
Guided by previous work \cite{reisswig:11ccwave,reisswig:10a}, we find that
the numerical error in the wave extraction is negligible provided appropriate
settings. 

The most involved GW extraction technique is CCE.
In that method, we solve the Einstein equations along null hypersurfaces
between a worldtube $\Gamma$ and future null infinity $\scri^+$. The worldtube $\Gamma$ is
typically located at some radius $R_\Gamma$ in the wave-extraction zone, and is simulation dependent \cite{reisswig:10a} (and references therein).
Specific to the present work, the CCE grid consists of $N_r=301$ points
along the radial direction.
Each radial shell is discretized by two stereographic patches comprised of
$N_{\rm ang}=81$ points per direction per patch.
At the inner-boundary worldtube $\Gamma$, we use up to $\ell_{\rm max}=8$ harmonic modes
for the decomposed Cauchy metric data. The metric data is decomposed on spheres
with $N_\theta=120$ and $N_\phi=240$ points in $\theta$ and $\phi$ direction,
respectively. 
The compactification parameter\footnote{See \cite{reisswig:10a} for a description of CCE relevant parameters.} 
$r_{\rm wt}$ is set to the particular extraction
radius for a given simulation, e.g. $r_{\rm wt}=100M_\odot$.
In all cases, the innermost radial compactified coordinate point is given by
$x_{\rm in}=0.49$. Together with an appropriate setting of $r_{\rm wt}$, this
ensures that the worldtube $\Gamma$ is located close to the first few radial
points on the characteristic grid.
The timestep and extraction radius must be picked on a case by case basis.
The wave-extraction zone is alwaus located on the spherical ``inflated-cube'' grids.
For the stellar collapse model A3B3G3 (Sec.~\ref{sec:A3B3G3}), the wave-extraction zone is located
between radii $1000\,M_\odot < R_\Gamma < 2500 M_\odot$. For all remaining tests, the wave-extraction zone
is located at $100\,M_\odot < R_\Gamma < 250 M_\odot$. 
The wave-extraction output frequency is dictated by the timestep of the spherical ``inflated-cube'' grids.

The remaining wave-extraction techniques are much simpler and only require
single spheres at some finite radius $R$. 

To project metric data from the 3D grid onto spheres, we use fourth-order Lagrange interpolation.

\subsection{Horizon Finding and Hydrodynamic Excision at the Puncture}
\label{sec:hydro-exc}

To track the appearance and shape of an apparent horizon, we use \code{AHFinderDirect} \cite{thornburg:04} 
which is part of the \code{EinsteinToolkit} \cite{et:12}.
As soon as an apparent horizon is found during an evolution, we excise the 
fluid variables within a fraction of the radius of the apparent horizon and
set them to their corresponding atmosphere values.
We get stable evolutions when excising  
about $85\%$ of the interior of the apparent horizon volume.

In order to compute angular momentum $J_{\rm AH}$ and mass $M_{\rm AH}$ of a black hole, 
we use the isolated / dynamical horizon framework provided by \texttt{QuasiLocalMeasures} \cite{dreyer:02}, 
which is part of the \code{EinsteinToolkit}. This framework defines mass and angular momentum in terms of
particular closed 2-surfaces, such as the apparent horizon.

The spherical surface defining the apparent horizon shape uses $N_\theta=41$ points along the $\theta$-direction
and $N_{\phi}=80$ points along the $\phi$-direction.


\section{Results}
\label{sec:results}

We revisit a number of ``benchmark`` problems commonly found in the
literature: an isolated perturbed and unperturbed neutron star, a
rotating core collapse model, a collapsing neutron star to a black
hole, and a binary neutron star coalescence.  Basic code tests such as
shock tubes can be found in the Appendix.  We describe our analysis in
more detail in corresponding sections below.

\subsection{Isolated Neutron Star}
\label{sec:tov}

\begin{table}[t]
\caption{Initial parameters and properties of the (perturbed) TOV star used to construct the initial data.
The density perturbation is only applied in the perturbed TOV test case. Units are in $c=G=M_\odot=1$.}
\label{tab:NS-parameters}
\begin{ruledtabular}
\begin{tabular}{lll}
\hline
Polytropic scale & $K$ & $100$ \\     
Polytropic index & $\Gamma$ & $2$ \\
Central rest-mass density & $\rho_c$ & $1.28\times10^{-3}$ \\
ADM mass $[M_\odot]$      & $M_{\rm ADM}$      & $1.4002$ \\
Baryonic mass $[M_\odot]$ & $M_B$             & $1.5062$ \\
Equatorial radius $[M_\odot]$ ([km]) & $R_e$      & $9.586$ $(14.16)$\\
\hline
Density pert. mode        & $\ell$     & $2$  \\ 
Density pert. amplitude   & $\lambda$  & $0.01$ \\
Monopole fundamental mode [kHz]  & $F$ & $1.458$ \\
First overtone [kHz]  & $H_1$ & $3.971$ \\
Quadrupole fundamental mode [kHz]  & ${}^2f$ & $1.586$ \\
First overtone [kHz]  & ${}^2p_1$ & $3.726$
\end{tabular}
\end{ruledtabular}
\end{table}

\begin{figure}
  \includegraphics[width=1.\linewidth]{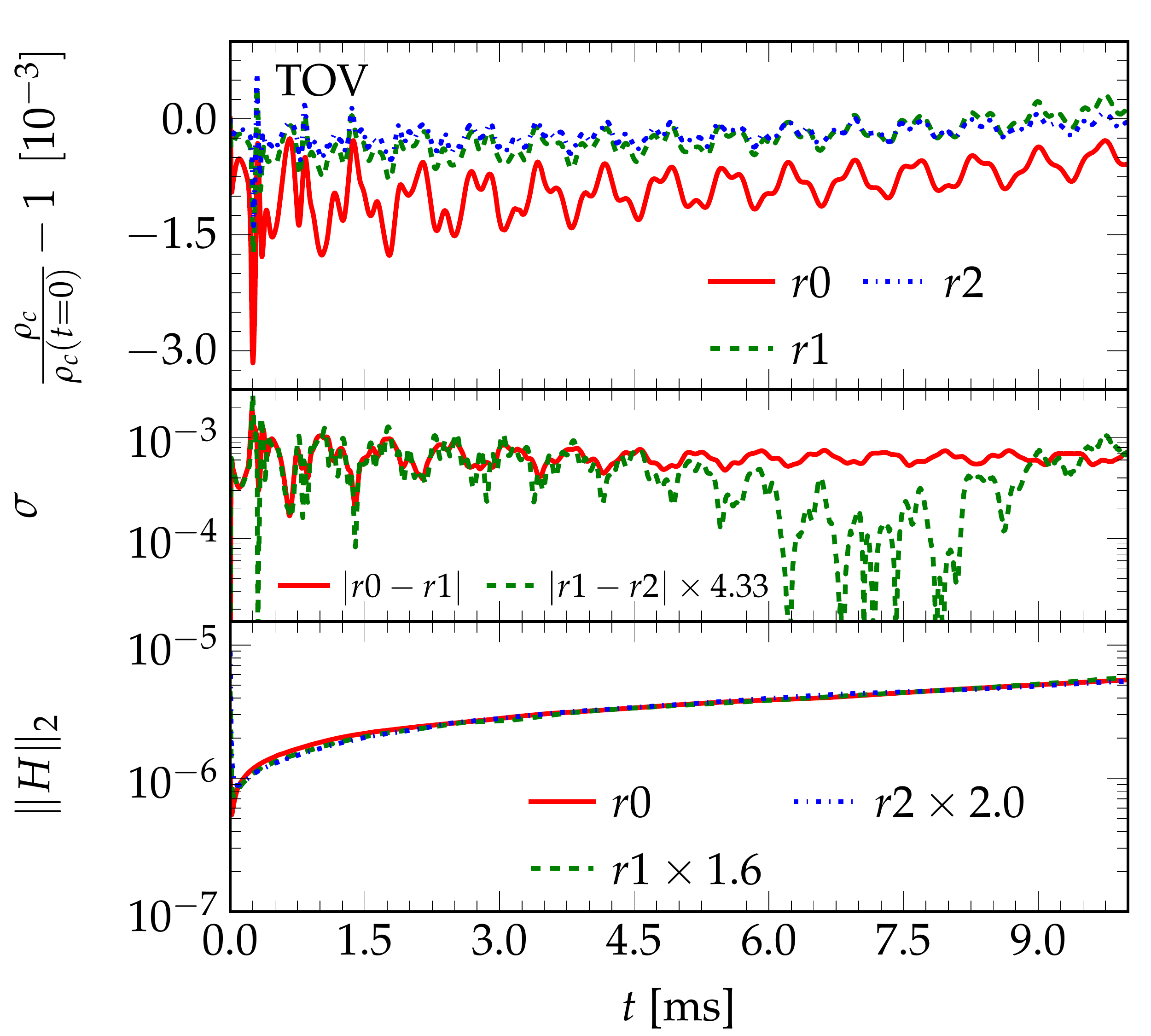}
  \vspace{-0.7cm}
  \caption{Unperturbed TOV star: normalized central density $\rho_c(t)/\rho_c(t=0)-1$ on the three resolutions $r0$, $r1$, and $r2$ (top panel), 
           difference in normalized central density between low and medium resolutions, and medium and high resolution (center panel),
           and the $L_2$-norm of the Hamiltonian constraint $\norm{H}_2$ on all three resolutions (bottom panel). 
           As the resolution is increased, the amplitude of the central density oscillations, the offset, and the slope decrease as expected.
           The differences in resolutions of the central density are scaled for second-order convergence.
           The $L_2$-norms of the Hamiltonian constraint $\norm{H}_2$ are scaled for first-order convergence. 
           The resolution study is performed using cell-centered AMR and ePPM.} 
  \label{fig:TOV-conv}
\end{figure}

\begin{figure}
  \includegraphics[width=1.\linewidth]{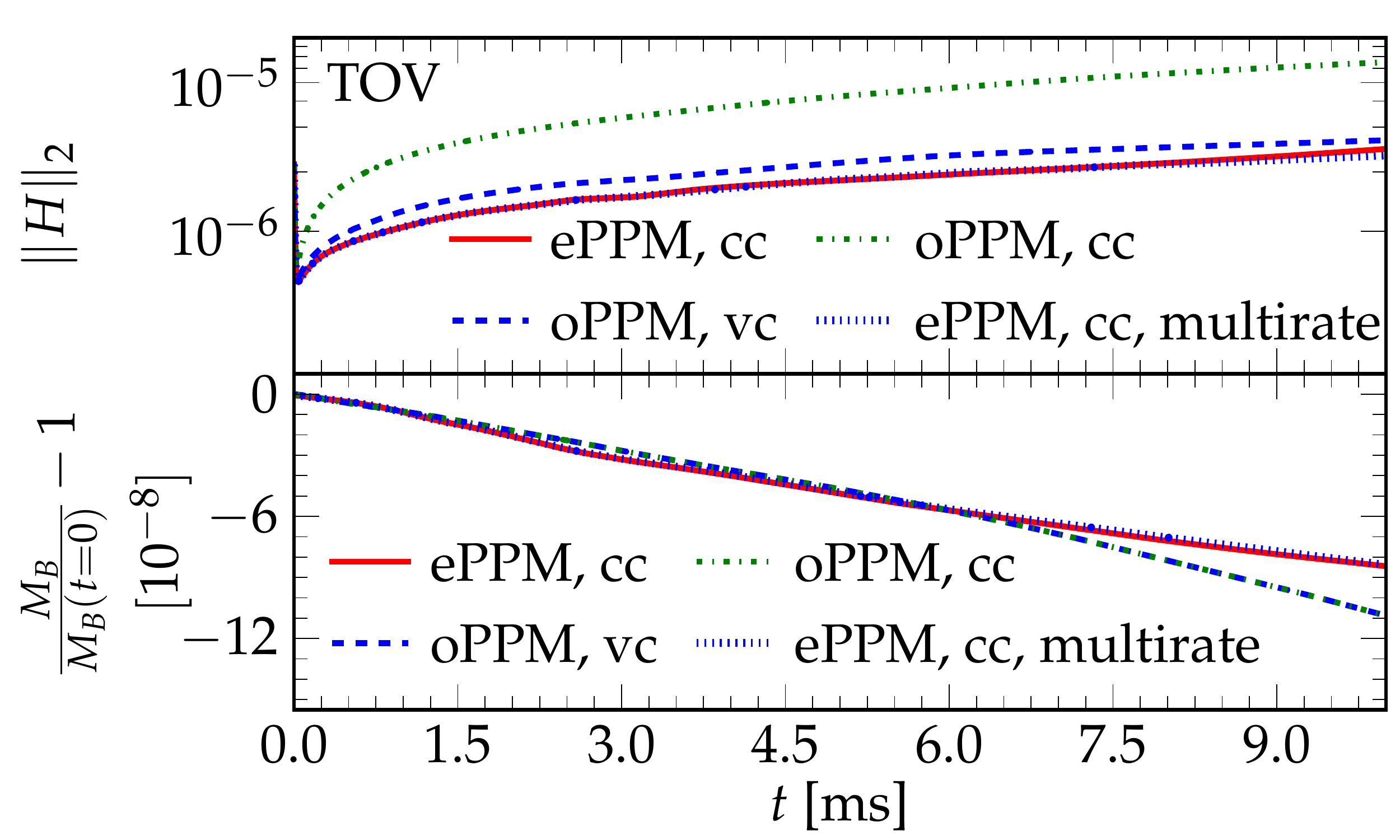}
  \vspace{-0.4cm}
  \caption{Unperturbed TOV star: the $L_2$-norm of the Hamiltonian constraint $\norm{H}_2$ (upper panel),
          and conservation of total baryonic mass $M_B$ (lower panel) for different numerical setups.
          We compare vertex-centered (vc) with cell-centered (cc) AMR using
          oPPM and/or ePPM. In addition, we also show a simulation with ``ePPM, cc'' using multirate time integration.
          $\norm{H}_2$ is strongly effected by the choice of numerical scheme, while $M_B$ is essentially uneffected. 
          The setup ``ePPM, cc'' performs best, while ``oPPM, cc'' performs worst.
          The standard scheme ``vc, oPPM'' used in other codes (e.g.~\cite{et:12, baiotti:05}) is slightly worse than the new scheme ``ePPM, cc''.
          Multirate time integration leads to nearly identical results.} 
  \label{fig:TOV-comp}
\end{figure}

\begin{figure}
  \includegraphics[width=1.\linewidth]{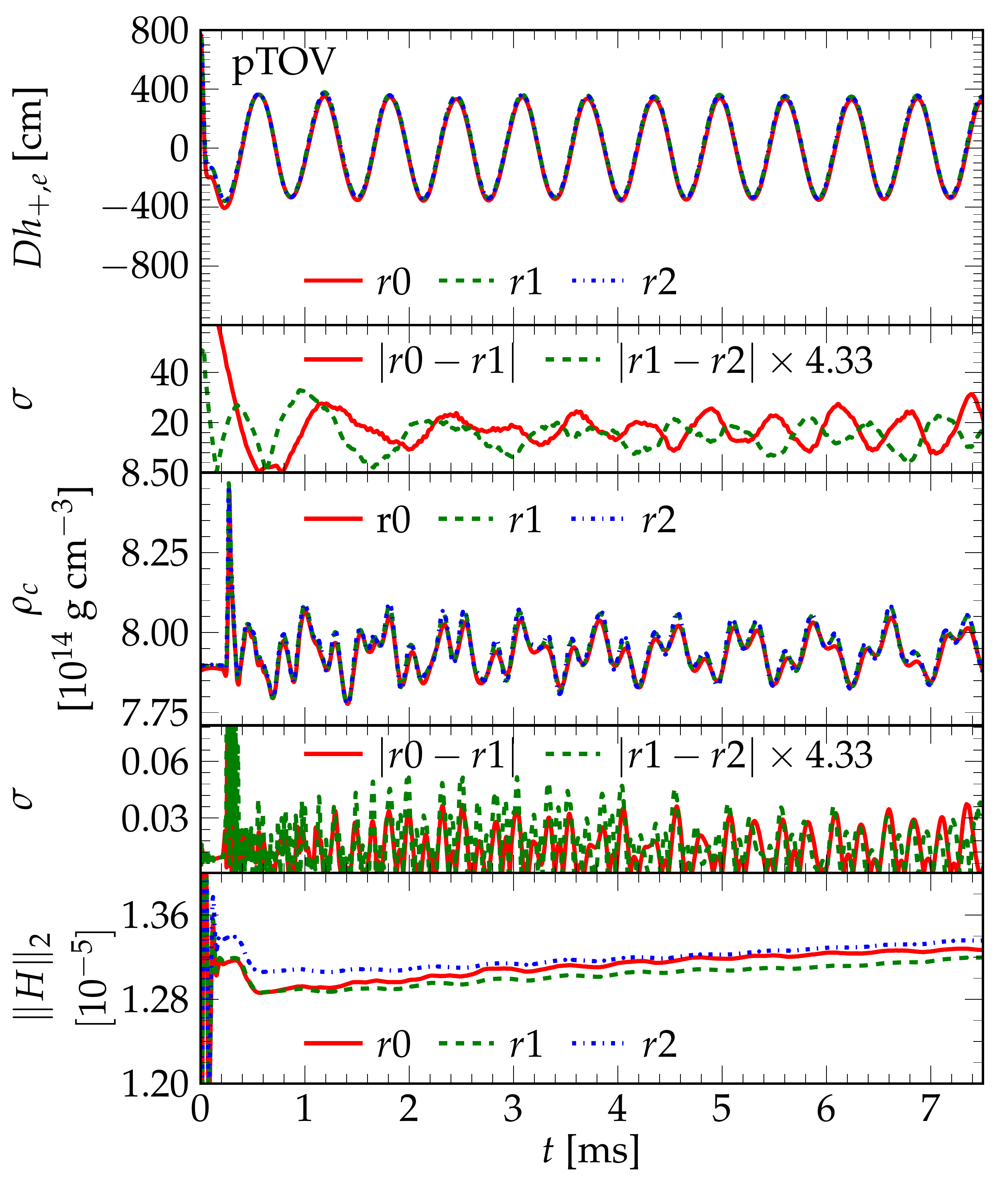}
  \vspace{-0.7cm}
  \caption{Perturbed TOV star: the top panel shows the ``+'' polarization of the GW strain $D h_{+,e}$ as emitted in the equatorial plane 
           and rescaled by distance $D$ for the three resolutions $r0$, $r1$, and $r2$. The waveforms are computed with CCE.
           In the panel below, we show the differences in GW strain between $r0$ and $r1$, and $r1$ and $r2$, where the latter is rescaled for
           second-order convergence. In the third panel from the top, we show the absolute central density evolution $\rho_c(t)$ for the three resolutions.
           Below, we show the differences in central density scaled for second-order convergence.
           In the bottom panel, we show the $L_2$-norms of the Hamiltonian constraint $\norm{H}_2$. Since the initial data for the perturbed case 
           are not constraint satisfying, the constraints do not exhibit clean convergence. The convergence study is performed using cell-centered AMR and ePPM.} 
  \label{fig:eTOV-conv}
\end{figure}

\begin{figure}
  \includegraphics[width=1.\linewidth]{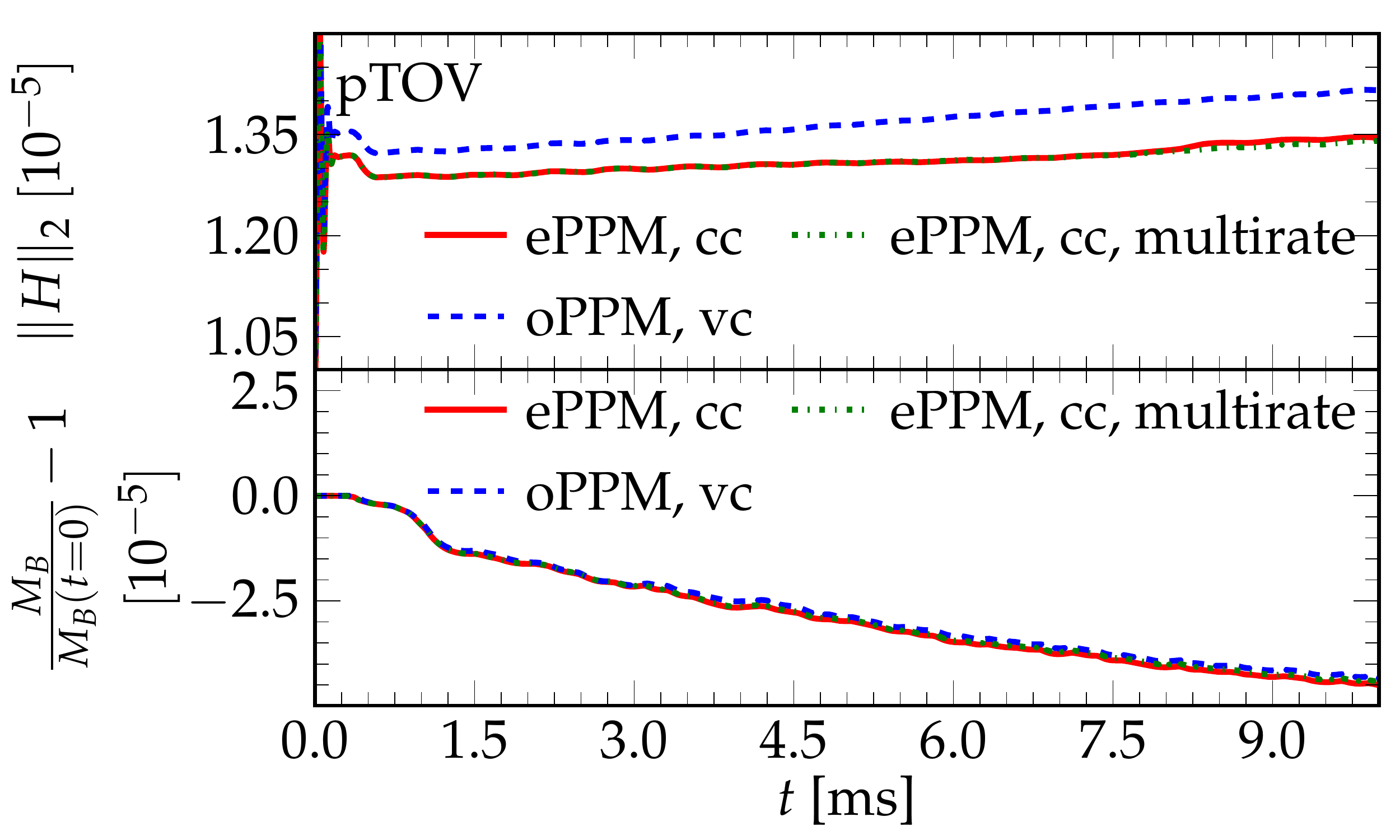}
  \vspace{-0.4cm}
  \caption{Perturbed TOV star: the impact of different numerical settings on the $L_2$-norm of the Hamiltonian constraints $\norm{H}_2$ (upper panel), and on
           the conservation of baryonic mass $M_B$ (lower panel). The setup using vertex-centered (vc) AMR and oPPM (blue dashed curve) leads to larger constraint violations
           than the setup using cell-centered (cc) AMR and ePPM. Multirate time integration does not change the accuracy of the results.
           In all cases, $M_B$ is nearly equally well conserved.} 
  \label{fig:eTOV-comp}
\end{figure}

\begin{figure}
  \includegraphics[width=1.\linewidth]{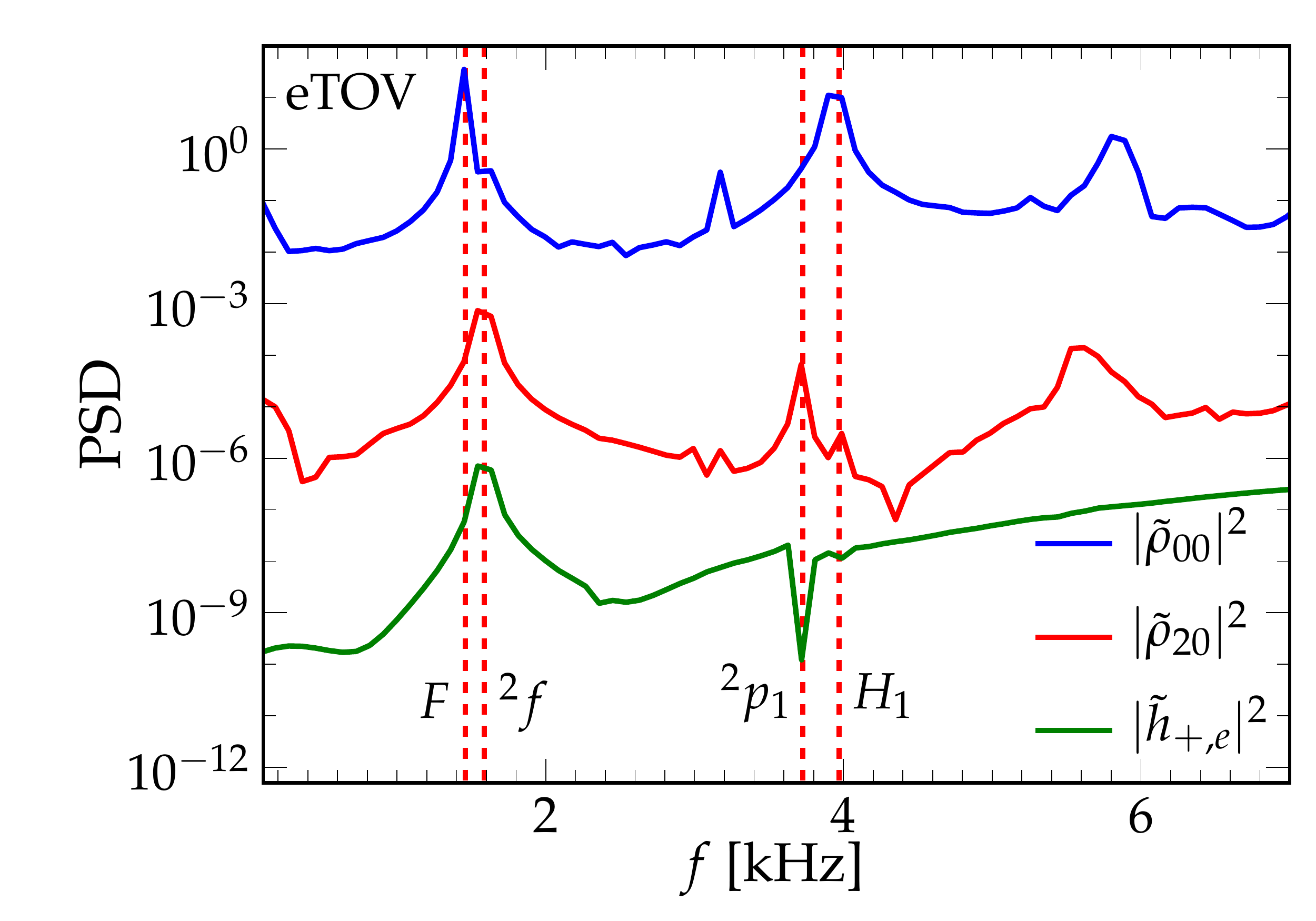}
  \vspace{-0.7cm}
  \caption{Perturbed TOV star: Power spectrum of $\rho_{00}$, $\rho_{20}$ and $h_{+,e}$ (individually scaled for better visibility), and the first few 
           fundamental neutron star oscillation modes (vertical lines) computed in \cite{dimmelmeier:06}.
    }
  \label{fig:eTOV-freq}
\end{figure}

We investigate convergence and accuracy of an isolated unperturbed neutron star and
an isolated perturbed neutron star using full GR matter-spacetime coupling in three spatial
dimensions. 
The neutron stars are given by the solution of the Tolman-Oppenheimer-Volkoff (TOV) equations \cite{tolman:39, oppenheimer:39}.

This test aims at showing the correctness of our cell-centered AMR scheme, and enhanced
PPM reconstruction.

\subsubsection{Initial Conditions and Equation of State}

We use a polytropic equation of state $P=K\rho^{\Gamma}$ with scale $K=100$ and
index $\Gamma=2$ in the initial data construction.
Although this choice does not represent a realistic choice for real neutron
stars, these parameters have been used in previous work (e.g.~\cite{dimmelmeier:06, baiotti:09}), 
and can be used as code verification. 
During evolution, we use an ideal fluid $\Gamma$-law equation of state with $\Gamma=2$.
The key parameters
are given in Table~\ref{tab:NS-parameters}. 
The initial data are generated via Hachisu's self-consistent field method
\cite{komatsu:89a,komatsu:89b} which requires as input the central density $\rho_c$ of
the star, and a polar-to-equatorial axes ratio between $0$ and $1$ to define
rotation.
In the present case, we set $\rho_c=1.28\times10^{-3}$ and use an axes ratio of
$1$ (no rotation). In the case of the perturbed TOV star, we perturb the star by a
spherical harmonic $(\ell,m)=(2,0)$ density perturbation of amplitude $\lambda=0.01$.

\subsubsection{Numerical Setup}

The grid is similar to the one depicted in Fig.~\ref{fig:7patch-system},
except that here, we have just one refinement region.
The fine grid spacing is $\Delta x=0.2\,M_\odot$ for the low resolution ($r0$), $\Delta
x=0.125\,M_\odot$ for the medium resolution ($r1$), and
$\Delta x=0.1\,M_\odot$ for the high resolution simulation ($r2$).
The fine grid extends to $R=11\,M_\odot$ and encompasses the entire star.
The inter-patch boundary between central Cartesian patch and outer spherical grid
is located at $R_{\rm S}=65\,M_\odot$.
We use $15$, $24$ and $30$ cells per angular direction per spherical patch for
the low, medium and high resolutions, respectively. 
The radial resolution is chosen based on the Cartesian coarse grid resolution
$\Delta r=1.6\,M_\odot$, $\Delta r=1.0\,M_\odot$, and $\Delta r=0.8\,M_\odot$,
for low, medium, and high resolutions, respectively. 
We use radial stretching outside the wave extraction zone to efficiently extend the computational domain 
so that the outer boundary is causally disconnected from wave-extraction zone and interior evolution. 
Accordingly, we stretch the radial resolution to $\Delta r=6.4\,M_\odot$, $\Delta
r=4.0\,M_\odot$, and $\Delta r=3.2\,M_\odot$ for
low, medium, and high resolution simulations, respectively, in the region between radii
$R_1=100 M_\odot$ and $R_2= 800 M_\odot$.
The outer boundary is located
at $R_{\rm B}=3500\,M_\odot$.

\subsubsection{Discussion}

\paragraph{Unperturbed TOV star.}
We first consider a single isolated non-rotating, unperturbed TOV star with parameters reported in Table~\ref{tab:NS-parameters}.
In the top panel of Fig.~\ref{fig:TOV-conv}, we show the normalized central density evolution $\rho_c(t)/\rho_c(t=0)$ as a function of time 
on the three resolutions $r0$, $r1$, and $r2$, using our new cell-centered AMR and enhanced PPM scheme.
In an ideal setting, the central density evolution should be constant as a function of time since the TOV solution represents a static fluid configuration.
Numerical errors induced by interpolation from the initial data solver grid onto the evolution grid, 
however, lead to an artificial excitation of the star, and, hence, to non-trivial central density oscillations,
which must converge to zero as the resolution is increased.
Due to the interpolation of the fluid initial data onto the evolution grid, we observe a large initial spike
and an overall offset in the density oscillations.
We additionally see an overall non-zero slope
in the central density evolution caused by numerical errors during evolution.
As the resolution is increased, we consistently observe that the amplitudes of the oscillations decrease, the offset becomes smaller, and the overall slope is reduced.
In the center panel, we show 
the difference in normalized central density $\rho_c(t)/\rho_c(t=0)$ between resolutions $r0$ and $r1$, and $r1$ and $r2$.
We perform a three-level 
convergence test by computing the ratio of the 
differences in a given quantity $F$ between the three resolutions,
\begin{equation} \label{eq:conv-rate}
 C=\frac{|F^{\rm medium}-F^{\rm low}|}{|F^{\rm
high}-F^{\rm medium}|}\,.
\end{equation}
The ratio $C$ defines the \textit{measured} convergence rate of the solution (e.g.~\cite{alcubierre:08}). Given three resolutions
with spacing $\Delta x_{\rm low}$, $\Delta x_{\rm medium}$, and $\Delta x_{\rm high}$, 
the \textit{theoretical} convergence rate for a particular order of convergence $p$ can be computed via
\begin{equation} \label{eq:conv-rate-from-order}
 C=\frac{|\Delta x_{\rm medium}^p-\Delta x_{\rm low}^p|}{|\Delta x_{\rm
high}^p-\Delta x_{\rm medium}^p|}\,.
\end{equation}
Given our numerical resolutions, according to \eqref{eq:conv-rate-from-order}, we expect that the difference between medium and high resolution, $r1$ and $r2$,
decreases by a factor of $C=4.33$ for second-order convergence compared with the difference between medium and low resolution, $r1$ and $r0$.

In the bottom panel of Fig.~\ref{fig:TOV-conv}, we show the time evolutions of the $L_2$-norm of the Hamiltonian 
constraint $\norm{H(t)}_2$ \eqref{eq:ham-constr} for the three resolutions $r0$, $r1$, and $r2$.
As the resolution is increased, the error drops consistent with first-order convergence, since the rescaled medium and high resolution curves are on top of each other.
We note that while the fluid body itself is smooth, the surface of the star is non-smooth, hence inducing a dominant first-order error
(compare Fig.~\ref{fig:tov-constr-1D}).

In the top panel of Fig.~\ref{fig:TOV-comp}, we show the $L_2$-norm of the Hamiltonian constraint $\norm{H}_2$ of a static TOV star using vertex-centered (vc) AMR, and cell-centered (cc) AMR.
Both AMR setups are run with the oPPM and ePPM reconstruction method.
In addition, we also perform a simulation using cell-centered AMR and ePPM reconstruction with multirate time integration.
We observe that the setup ''ePPM, cc'' exhibits the lowest constraint violations. The setup ``ePPM, cc, multirate''
is right on top of the red curve, hence indicating comparable accuracy.
The setup ``vc, oPPM'', which is the setup used in previous work (e.g.~\cite{et:12, baiotti:05, reisswig:11ccwave, ott:12a, ott:11a}) yields slightly 
less accurate evolution.
Finally, the setup ``cc, oPPM'' yields significantly reduced accuracy compared to all other setups. 
This is mainly due to the oPPM scheme, which is known to reduce the order of accuracy at smooth maxima to first order (see Appendix~\ref{sec:ePPM}, Fig.~\ref{fig:tov-constr-1D}).
This effect is not seen in the vertex-centered setup ``vc, oPPM'', since the central density is exactly located on a grid point.

In the bottom panel of Fig.~\ref{fig:TOV-comp}, we show conservation of mass for the considered numerical setups.
In all cases, the total mass loss is on the order of $10^{-7}$ over the course of the evolution. 
Since the AMR boundaries are all located in the vacuum region outside the star,
refluxing at AMR boundaries is not relevant.
The mass loss is entirely due to interaction with the artificial low-density atmosphere in the vacuum region (see also Appendix~\ref{sec:atmo}).

\paragraph{Perturbed TOV star.}
As a second test, we apply an initial $(\ell=2,m)=(2,0)$ density perturbation with amplitude $\lambda=0.01$ onto the same TOV star considered above.
A more complete study of this configuration including variations on perturbation parameters has been performed in \cite{dimmelmeier:06, baiotti:09}.
Numerical grids and setups are identical to those of the static TOV star, and we perform
the same analysis as above. In addition, we also analyze the non-trivial $(\ell,m)=(2,0)$ mode of the GW signal that is induced by fundamental mode oscillations.
In the upper panel of Fig.~\ref{fig:eTOV-conv}, we plot the ``+'' polarization of the GW signal $D h_{+,e}$ as emitted in the equatorial plane from the three resolutions
$r0$, $r1$, and $r2$. Since only the $(\ell,m)=(2,0)$ mode is excited, the entire wave signal can be written as
\begin{equation} \label{eq:l2m0-strain}
D h_{+,e} = D h_+^{20}\, {}_{-2}Y_{20}(\theta=\frac{\pi}{2},\phi=0).
\end{equation}
Here, $D$ is the distance from the source. 
We compute $h_+^{20}$ with CCE and use an FFI cut-off frequency of $f_0=812$ Hz (see Sec.~\ref{sec:waveextract}).
We also show the differences of the GW strain between low and medium, and medium and high resolutions, where the
latter is scaled for second-order convergence. 
In addition, we show the central density evolution $\rho_c(t)$ for the three resolutions which converge. 
Similar to the above, we plot the differences between low and medium, and medium and high resolutions
scaled for second-order convergence.
We also show the $L_2$-norm of the Hamiltonian constraints $\norm{H}_2$ of the three resolutions.
Since the initial data solver does not take into account the effects of the perturbation onto the initial spacetime metric, 
the constraints do not converge initially,
and only slowly converge at later times. 
In the present plot, we have not used any rescaling. We note, however, that the slopes of
the medium and high resolutions are slightly smaller than for the low resolution case.

When comparing the strain $D h_{+,e}^{\rm CCE}$ as computed with CCE to the strain $D h_{+,e}^Q$ as computed from the RWZM formalism,
we generally find that the strain computed via the RWZM formalism is
prone to numerical noise. In addition, we find that the finite-radius
error and gauge error inherent in the waveform obtained from RWZM master 
functions at radii $R=100\,M_\odot$ and $R=250\,M_\odot$ is on the order of $10\%$.
A similar behavior applies to the strain $D h_{+,e}^{\rm NP}$ as extracted via the NP formalism at a finite radius.

Finally, we also check that the correct fundamental oscillation modes are excited.
In Fig.~\ref{fig:eTOV-freq}, we compare the frequency spectrum of the density $\rho$ and 
the strain $D h_{+,e}$ to the eigenmodes found in \cite{dimmelmeier:06}.
In order to compute the spectrum of $\rho$, we first project $\rho$ from the 3D grid onto spherical shells
inside the star, and then decompose in terms of spherical harmonics.
The vertical lines in Fig.~\ref{fig:eTOV-freq} correspond to the fundamental monopole mode $F$ and its first overtone
$H_1$, and the fundamental quadrupole mode ${}^2f$ and its first overtone ${}^2p_1$.
As expected, the spectrum of the strain $D h_{+,e}$ and the $(\ell,m)=(2,0)$ mode of the density $\rho_{20}$ both peak
at the correct quadrupole eigenmode frequencies.
Likewise, the spectrum of the $(\ell,m)=(0,0)$ density mode correctly peaks at the monopole eigenmode frequencies.

\subsection{Rotating Stellar Collapse}
\label{sec:A3B3G3}

\begin{figure}
  \includegraphics[width=1.\linewidth]{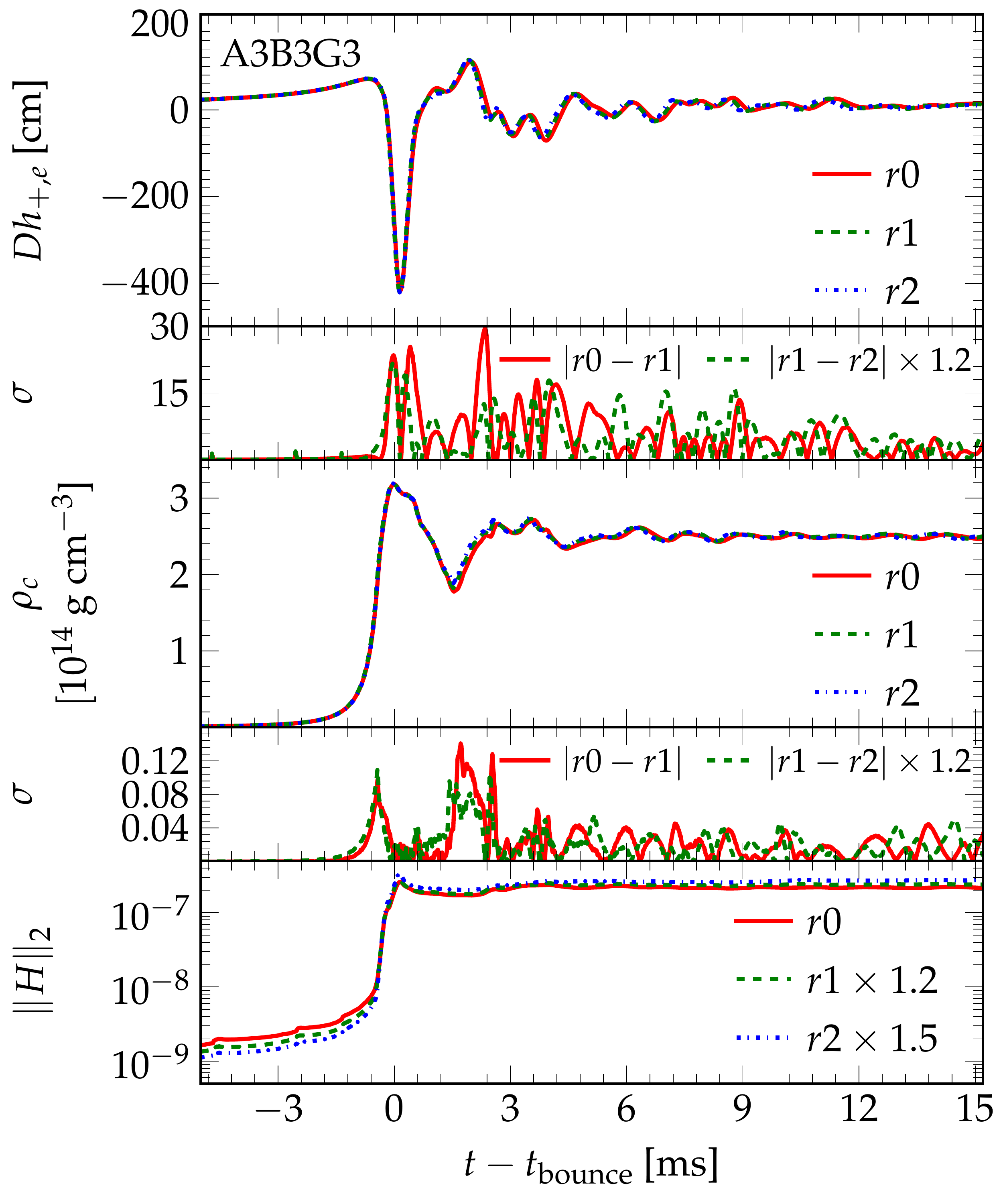}
  \caption{Stellar collapse: The GW strain $Dh_{+,e}$ extracted via the quadrupole formula (upper panel), the central density
$\rho_c$ (third panel from the top), and the $L_2$-norm of Hamiltonian constraint $\norm{H}_2$, all on the three resolutions $r0$, $r1$, and $r2$.
The panels directly below the top panel and the third panel from the top show the difference in strain and central density between low and medium
resolutions, and medium and high resolutions. The differences are
scaled for second-order convergence.
The $L_2$-norm of Hamiltonian constraint is scaled for first-order
convergence. Before core bounce, the constraint exhibits second order
convergence. After shock formation, the convergence rate is reduced to first
order. The convergence study is performed using cell-centered AMR and ePPM.
    }
  \label{fig:A3B3G3-convergence}
\end{figure}

\begin{figure}
  \includegraphics[width=1.\linewidth]{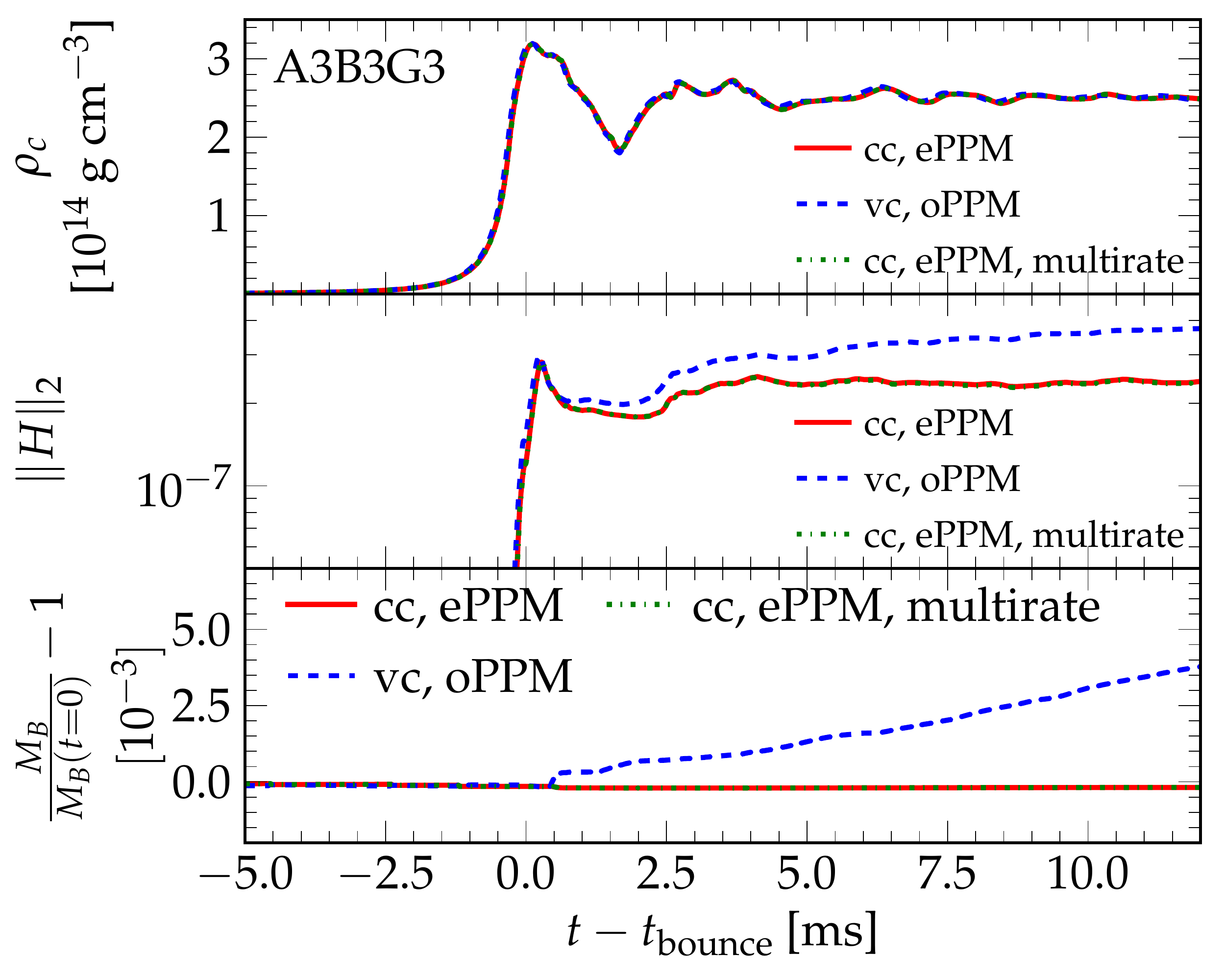}
  \caption{Comparison of vertex-centered (vc) AMR with oPPM versus
    cell-centered (cc) AMR with ePPM for stellar collapse model
    A3B3G3. We show the central density $\rho_c$ (upper panel), tthe
    $L_2$-norm of the Hamiltonian constraint $\norm{H}_2$ (middle
    panel), and conservation of total baryonic mass $M_B$ (bottom
    panel). Due to refluxing in the cell-centered case, the mass is
    almost perfectly conserved, while in the vertex-centered case, the
    mass is rapidly growing (bottom panel). Due to ePPM, the
    constraints in the cell-centered case exhibit almost no growth
    after core bounce, while in the vertex-centered case with oPPM the
    constraints are clearly growing (lower panel). The results are not
    changed when multirate time integration is used.  The comparison
    is done using baseline resolution $r1$.  }
  \label{fig:A3B3G3-comparison}
\end{figure}

\begin{figure}
  \includegraphics[width=1.\linewidth]{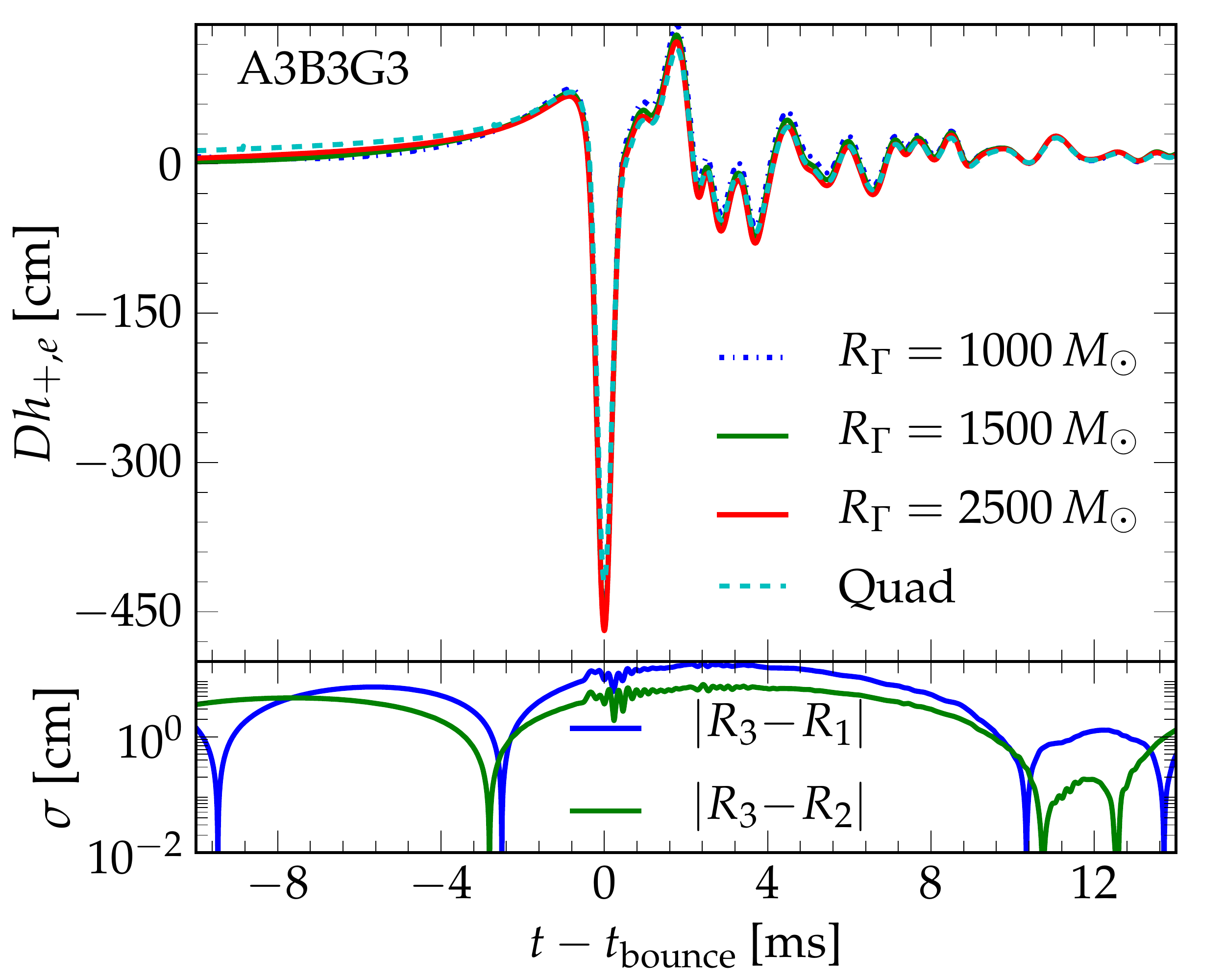}
  \caption{The GW strain extracted $D h_{+,e}$ from the rotating
    stellar collapse model A3B3G3 (upper panel). We show the strain
    extracted via CCE from different worldtube locations
    $R_\Gamma=1000\,M_\odot$, $R_\Gamma=1500\,M_\odot$, and $R_\Gamma=2500\,M_\odot$, as
    well as the strain computed via the quadrupole formula.  Larger
    CCE worldtube radii permit lower FFI cut-off frequencies without
    introducing unphysical drifts in the GW strain. All waveforms
    extracted via CCE are in good agreement to within a few percent
    with the waveform computed via the quadrupole formula.  The lower
    panel shows the differences in strain amplitude of the inner
    extraction radii to the outermost extraction radius. The
    differences converge as the extraction radius is increased. The
    comparison is done using baseline resolution $r1$.  }
  \label{fig:A3B3G3-GW}
\end{figure}

\begin{table}[t]
\caption{Initial parameters and properties of the rotating stellar
  collapse model A3B3G3. Units are in $c=G=M_\odot=1$.}
\label{tab:A3B3G3-parameters}
\begin{ruledtabular}
\begin{tabular}{lll}
\hline
Polytropic scale & $K$ & $0.4640517$ \\     
Initial polytropic index & $\Gamma_{1,\rm ini}$ & $1.\bar{3}$ \\
Evolved polytropic index 1 & $\Gamma_1$           & $1.31$ \\
Evolved polytropic index 2 & $\Gamma_2$           & $2.5$ \\
Thermal polytropic index & $\Gamma_{\rm th}$    & $1.5$ \\
Central rest-mass density & $\rho_c$ & $1.6193\times10^{-8}$ \\
Axes ratio                &          & $0.93$ \\
Degree of differential rotation [km] & $A$ & $500$ \\
Rotational / binding energy $[\%]$ & $T/|W|$          & $0.9$ \\
Equatorial radius $[M_\odot]$ & $R_e$      & $1.0661\times10^3$ \\
Baryonic mass $[M_\odot]$          & $M_{B}$        & $1.4596$ \\
ADM mass $[M_\odot]$               & $M_{\rm ADM}$  & $1.4596$ \\
ADM ang.~mom.~$[M_\odot^2]$        & $J_{\rm ADM}$  & $2.4316$ \\
Spin                               & $a$     & $1.1413$
\end{tabular}
\end{ruledtabular}
\end{table}

We investigate convergence and accuracy of the benchmark rotating
stellar collapse model A3B3G3, which has been previously
considered in the literature \cite{zwerger:97,dimmelmeier:02}.  This
tests the ability of the code to simulate the collapse of a rapidly
differentially spinning iron core in full 3D with causally
disconnected outer boundaries, albeit with simplified microphysics.
We show that due to larger wave extraction radii, the waveforms
extracted via curvature-based methods such as CCE are more accurate
than what has been computed before \cite{reisswig:11ccwave}.

\subsubsection{Initial Data and Equation of State}

For the purpose of this test, we employ a \emph{hybrid}
equation of state \cite{janka:93,dimmelmeier:02,dimmelmeier:02a}
that combines a 2-piece piecewise polytropic pressure $P_{\rm P}$ with
a thermal component $P_{\rm th}$, i.e., $P=P_{\rm P}+P_{\rm th}$.  To
model the stiffening of the equation of state at nuclear density $\rho_{\rm
nuc}\cong2\times10^{14}\,\rm{g}\,\rm{cm}^{-3}$, we assume that the
polytropic index $\Gamma$ jumps from $\Gamma_1$ below nuclear density
to $\Gamma_2$ above. 
The equation of state parameters are given in Table~\ref{tab:A3B3G3-parameters}.

The initial data are constructed from $n=3$ ($\Gamma_{1,\rm ini}=\Gamma_1
= 4/3$) polytropes in rotational equilibrium generated via Hachisu's
self-consistent field method \cite{komatsu:89a, komatsu:89b} which not only
provides fluid, but also spacetime curvature initial data. While being set up as
marginally stable polytropes with $\Gamma_{1,\rm ini} = 4/3$, during
evolution, the initial sub-nuclear polytropic index $\Gamma_{1}$ is
reduced to $\Gamma_1 < \Gamma_{1,\rm ini}$ to accelerate
collapse. Following previous
studies~\cite{zwerger:97,dimmelmeier:02,ott:07cqg}, we use $\Gamma_2 =
2.5$ in the super-nuclear regime.

In the present test, we revisit model A3B3G3 from
\cite{zwerger:97,dimmelmeier:02}.
This configuration uses $\Gamma_1 = 1.31$. It is
strongly   differentially rotating, with its initial central angular velocity
dropping by a factor of two over $A=500\,\mathrm{km}$. This, in
combination with $T/|W| = 0.9\%$, leads to rapid rotation in the
inner core, resulting in a very strong GW signal at core bounce and
dynamics that are significantly affected by centrifugal effects. It
produces a ``Type-I'' GW signal with a centrifugally-widened broad
peak at core bounce \cite{zwerger:97, dimmelmeier:02}.

\subsubsection{Numerical Setup}
\label{sec:CC-numerics}

We use five refinement levels located at the center of the domain.
The refinement boxes of each level have a half-width of $R_{\rm rl}=\left[192M_\odot, 144M_\odot, 98M_\odot, 40M_\odot, 12M_\odot\right]$, respectively.
The coarsest level is comprised of cubed-sphere multipatch grids
(Fig.~\ref{fig:7patch-system}). The
inner radius of the spherical grids is $R_{\rm S}=384M_\odot$, and the outer boundary
is $R_{\rm B}=16000M_\odot$.
Initially, only the coarsest level is active. Additional levels are
progressively added as the central density increases during collapse.
The initial stellar radius of model A3B3G3 is $R_e=1066.1M_\odot=1574.84\,\rm{km}$ in the equatorial
plane. Thus, the inter-patch boundaries thread the star in this
particular setup.
The finest refinement level is picked such that the protoneutron star is fully
contained on that level.
The GW extraction zone extends
to a radius of $R=2500M_\odot$. 
Beyond that radius, we apply radial stretching up to a radius $R=6000M_\odot$. 
In this stretching region, the
radial grid spacing is increased by a factor of $16$, and the resolution becomes too coarse for
reliable wave extraction.

For our baseline resolution (denoted by $r1$), we pick a
radial grid spacing of $\Delta r=8.0M_\odot$ on the non-stretched spherical inflated-cube grids,
and a Cartesian resolution of $\Delta x = 8.0 M_\odot$
on the central Cartesian patch. 
Given our five refinement levels above, this results in a resolution of $0.25 M_\odot=369.3\,\rm{m}$
for the protoneutron star.
The angular resolution of the cubed-sphere grids is set to $N_{\rm ang}=30$
cells per patch and direction.
This makes a total of $N_{\rm ang, total}=120$ points across the equatorial
plane. 

In addition to our baseline resolution $r1$, we also consider a low resolution
run $r0$, and a high resolution run $r2$ to check for convergence. 
Resolution $r0$ uses $\Delta r = \Delta x = 9.6 M_\odot$ and $N_{\rm ang}=24$ ($20\%$ lower), 
and resolution $r2$ uses $\Delta r
= \Delta x = 6.4 M_\odot$ and $N_{\rm ang}=36$ ($20\%$ higher).

In all considered cases, we set the damping coefficient of the $\Gamma$-driver gauge
condition to $\eta=1/2$.
Dissipation is set to $\epsilon_{\rm diss}=0.1$ on the fine levels, and $\epsilon_{\rm diss}=0.01$ on
the multipatch grid.
The atmosphere level is set to be $10^{-10}$ of the central density, and we damp 
the stress-energy tensor in the atmosphere starting using \eqref{eq:Tmunu-damp} with $R_0=1300 M_\odot$ and $R_1=1400 M_\odot$.

\subsubsection{Discussion}
 
In Fig.~\ref{fig:A3B3G3-convergence}, we show convergence of the plus polarization
of the GW strain $Dh_{+,e}$ measured in the equatorial plane, the
central density $\rho_c$, and the $L_2$-norm of the
Hamiltonian constraint $\norm{H}_2$. The GW strain is
computed using the quadrupole formula, though a similar analysis and result
applies to all extraction methods.
All three quantities are shown for the three resolutions $r0$, $r1$, and $r2$,
using multipatches, cell-centered AMR, refluxing, and enhanced PPM (see
Sec.~\ref{sec:CC-numerics}).
We align the results from all three resolutions at the time when the central density
$\rho_c$ reaches its maximum at core bounce.
We observe first order convergence in $\norm{H}_2$ after core bounce.
In the prebounce phase, $\norm{H}_2$ exhibits second-order convergence. This behavior is
expected since the numerical scheme reduces to first order at the shock front after bounce where the
error are greatest.

In Fig.~\ref{fig:A3B3G3-convergence},
we also show the absolute difference of the GW strain $Dh_{+,e}$ and the central density
$\rho_c$ between low ($r0$) and medium ($r1$) resolutions, and
medium and high ($r2$) resolutions. 
The convergence behavior of the two quantities 
is less clean than what can be
observed for the Hamiltonian constraint due to their oscillatory nature.
The convergence is between the expected first
and second-order accuracy.

In Fig.~\ref{fig:A3B3G3-comparison}, we compare
vertex-centered AMR with original PPM reconstruction versus cell-centered AMR with
refluxing and enhanced PPM.
In addition, we show the behavior of the latter case when multirate RK time
evolution is applied.
As is clear from the bottom two panels, the cell-centered scheme with
refluxing and enhanced PPM (``cc, ePPM``) outperforms the
vertex-centered scheme with original PPM (''vc, oPPM``).
While in the cell-centered case, $\norm{H}_2$ essentially remains
constant after core bounce, it clearly grows in the vertex-centered case. 
Even worse, the vertex-centered case exhibits a rapid growth in total baryonic mass after
core bounce.
The evolution with multirate RK performs equally well
as the ''cc, ePPM`` setup, which uses standard RK4 time integration. The multirate setup
offers a speed up of $\sim20\%$ for the current test problem. The speed-up can be
significantly larger when full microphysics and neutrino transport is employed
(e.g.~\cite{ott:12b}).

In Fig.~\ref{fig:A3B3G3-GW}, we revisit our study of extracting
gravitational radiation using curvature-based methods
\cite{reisswig:11ccwave}. In \cite{reisswig:11ccwave}, we found a radial
dependence of the accuracy of the curvature-based extraction methods.
This study made use of purely Cartesian simulation domains, and 
was thus limited in terms of possible domain sizes and extraction radii. The
maximum extraction radius was limited to
$R=1000M_\odot$. This is still fairly close and means that the waveforms are
extracted well inside the star. Our curvature-based extraction methods,
however, assume vacuum, i.e.~a vanishing stress-energy tensor at the extraction
location.
In \cite{reisswig:11ccwave}, we thus conjectured that increased
extraction radii that are located outside the star would further improve the
accuracy of the extracted waveforms.
Given our new multipatch setup, we can confirm this conjecture.
We have placed three extraction radii at $R=\left[1000M_\odot, 1500M_\odot, 2500M_\odot\right]$
in a region with constant radial spacing $\Delta r=8.0M_\odot$ where the radial
direction is not yet stretched.
The upper panel of Fig.~\ref{fig:A3B3G3-GW} shows the ``+''
polarization of the GW strain $Dh_{+,e}$ measured in the equatorial plain
extracted via CCE.
As a comparison, in the same panel, we also show $Dh_{+,e}$ computed via
the quadrupole formula.
We apply FFI to compute the strain $D h$ from $\Psi_4$ extracted with CCE (see Sec.~\ref{sec:waveextract}).
In \cite{reisswig:11ccwave}, we conjectured that the low cut-off frequency that
must be picked for FFI can be reduced as the extraction radius is increased.
Here, we confirm that this is indeed the case. While
extraction radius $R=1000M_\odot$ requires a low cut-off frequency $f_0=100\,\rm{Hz}$
which is well inside the LIGO sensitivity band, we find that at radius $R=1500M_\odot$
we can get away with $f_0=60\,\rm{Hz}$. At radius $R=2500M_\odot$, we can further
reduce this to $f_0=30\,\rm{Hz}$ without introducing artificial non-linear
drifts in the strain.
In the bottom panel of Fig.~\ref{fig:A3B3G3-GW}, we show the difference in GW
amplitude of the waveforms computed from the inner extraction radii to the
waveform computed from the outer most extraction radius.
We confirm that as the extraction radius is increased, the differences
further decrease similar to what has been found in
\cite{reisswig:11ccwave}.

The waveform
computed via the quadrupole formula does not suffer from
amplification of low frequency errors \cite{reisswig:11ccwave}.
We observe that the waveforms extracted via CCE at larger radius and
decreased $f_0$ more closely resemble the monotonically rising signal in the
prebounce phase that the waveform computed via the quadrupole formula exhibits.
Overall, in accordance with \cite{reisswig:11ccwave},
we still measure the same deviations between GW amplitudes computed from CCE and the quadrupole formula
to within a few percent at core bounce.
This is not surprising, since the error in CCE due to different worldtube
extraction locations is much smaller than the observed deviation from the
waveform extracted via the quadrupole formula.

Finally, we note that we have also computed the GW strain via the
RWZM formalism (not shown). In our previous
more detailed study on GW extraction in the context of rotating stellar
collapse \cite{reisswig:11ccwave}, we found that the RWZM formalism leads
to waveforms which are contaminated by high frequency noise. Unfortunately, in
the current study, which allows us to use larger extraction radii than
$R=1000M_\odot$, we find that the systematic high-frequency noise inherent in the RWZM
waveforms is not reduced, but instead even increases with increased
extraction radius. As already conjectured in
\cite{reisswig:11ccwave}, this is most likely due to the perturbative manner the
waves are extracted from the spacetime in the RWZM formalism. In this
formalism, the spherical background geometry is projected out, which can
result in very small values for the aspherical perturbation
coefficients that are prone to numerical noise and cancellation
effects. At larger radii, the aspherical perturbations are even smaller since
they fall of as $1/r$, and thus are harder to capture accurately.
The RWZM approach may therefore be less suited for the
extraction of the generally weak GW signals emitted in core collapse.

\subsection{Neutron Star Collapse}

\begin{figure}
  \includegraphics[width=1.\linewidth]{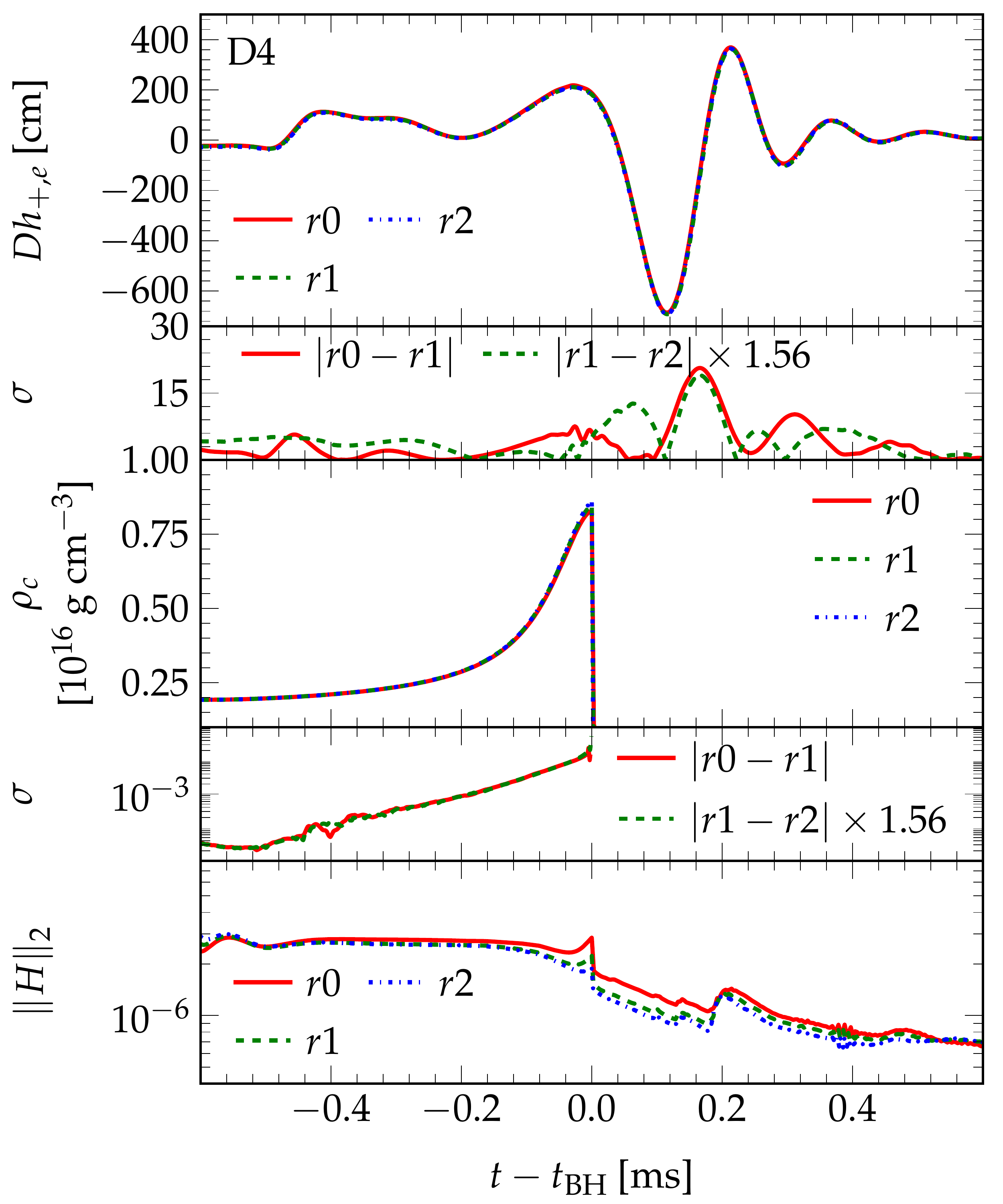}
  \caption{Rotating neutron star collapse: convergence analysis of the ``+'' polarization of the GW strain $D h_{+,e}$
as emitted in the equatorial plane and extracted via CCE (top two panels), central density $\rho_c$ evolution (next two panels),
and $L_2$-norm of the Hamiltonian constraint $\norm{H}_2$ (bottom panel).
The differences in $D h_{+,e}$ and $\rho_c$ between medium and high resolution are scaled for second-order convergence.
At $t-t_{\rm BH}=0$, the density drops to zero due to hydrodynamic excision within the horizon.
The $L_2$-norm of the Hamiltonian constraint (bottom panel) does not converge initially due to numerical artifacts from the initial data solver,
however, later converges at second-order during black hole formation $t-t_{\rm BH}\sim 0$ and black hole ring-down $t-t_{\rm BH}>0$.
The convergence study is performed using cell-centered AMR with ePPM.
    }
  \label{fig:rnsD4-convergence}
\end{figure}

\begin{figure}
  \includegraphics[width=1.\linewidth]{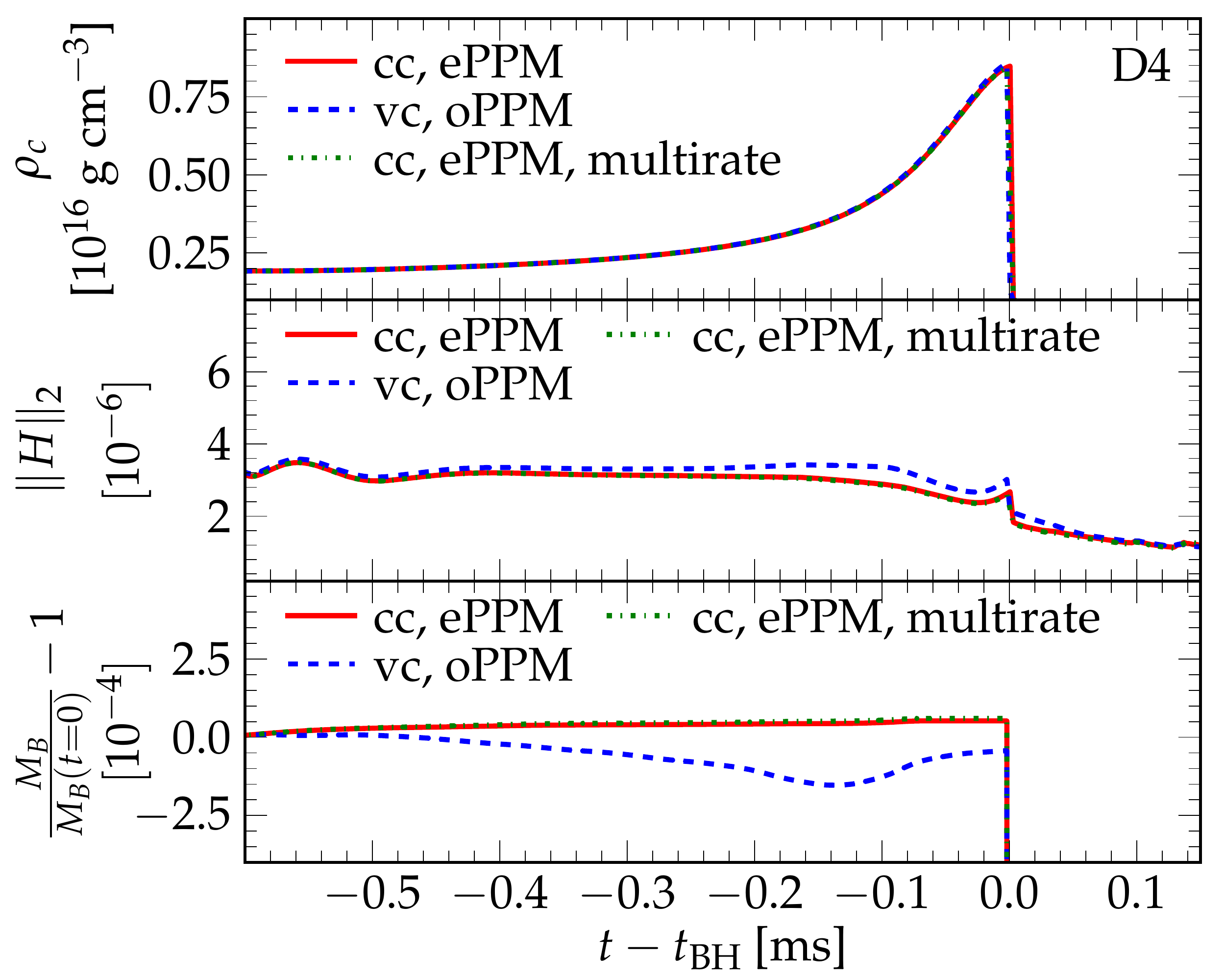}
  \caption{Rotating neutron star collapse: we compare vertex-centered (vc) AMR and oPPM reconstruction
with cell-centered (cc) AMR and ePPM reconstruction. The latter setup is also shown using multirate
RK time integration.
The top panel compares the central density evolution profile $\rho_c(t)$.
The center panel compares the evolution of the $L_2$-norm of the Hamiltonian constraint $\norm{H}_2$.
The bottom panel compares the conservation of baryonic mass $M_B$.
The setup ``vc, oPPM'' produces slightly larger violations in the Hamiltonian constraints, especially in the late collapse phase 
shortly before the black hole forms.
Due to refluxing, the cell-centered case exhibits much better conservation of baryonic mass.
Multirate RK time integration does not lead to different results.
The comparison is done using baseline resolution $r1$.
    }
  \label{fig:rnsD4-comparison}
\end{figure}

\begin{figure}
  \includegraphics[width=1.\linewidth]{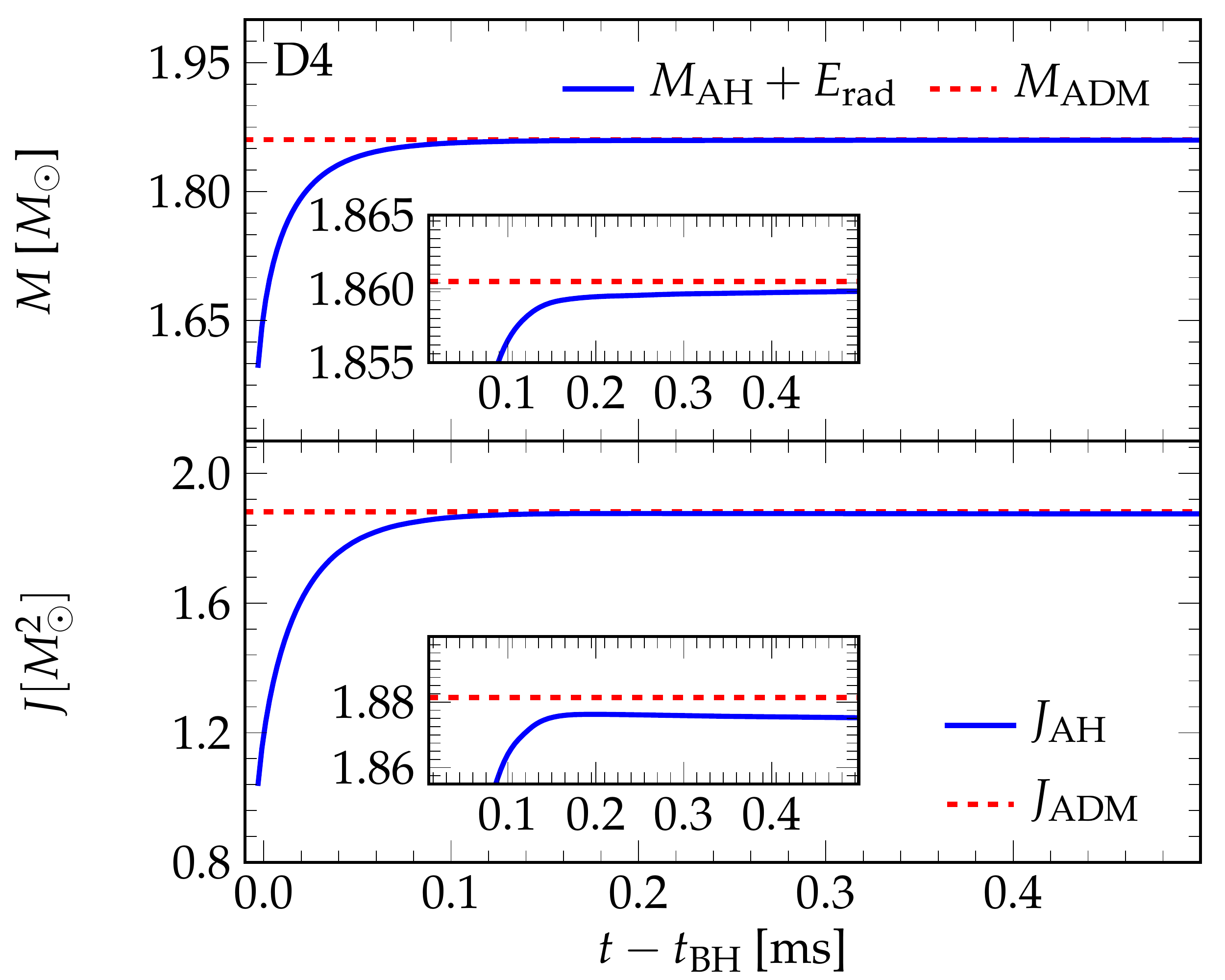}
  \caption{Rotating neutron star collapse: we show the total ADM mass $M_{\rm ADM}$ (top panel; red dashed line)
and the mass of the apparent horizon $M_{\rm AH}$ plus energy radiated in GWs $E_{\rm rad}$ as a function of time (blue straight line).
The total ADM angular momentum $J_{\rm ADM}$ of the spacetime (red dashed line),
and the angular momentum $J_{\rm AH}$ as measured on the apparent horizon (blue straight line) 
is shown in the bottom panel. The inset plots show a close-up of the time evolution of $M_{\rm AH}+E_{\rm rad}$ and
$J_{\rm AH}$.
As all matter becomes trapped in the event horizon, both, $M_{AH}+E_{\rm rad}$ and $J_{AH}$, quickly asymptote to the conserved ADM values of the spacetime.
Due to systematic (atmosphere) and numerical errors, the asymptoted values do not agree with the initial ADM values.
Note that the mass radiated in GWs is negligible compared to the total mass of the black hole and thus barely contributes to $M_{AH}+E_{\rm rad}$.
No angular momentum is radiated in GWs.
The results are shown for resolution $r2$ using cell-centered AMR with ePPM.
    }
  \label{fig:rnsD4-AH}
\end{figure}

\begin{figure}
  \includegraphics[width=1.\linewidth]{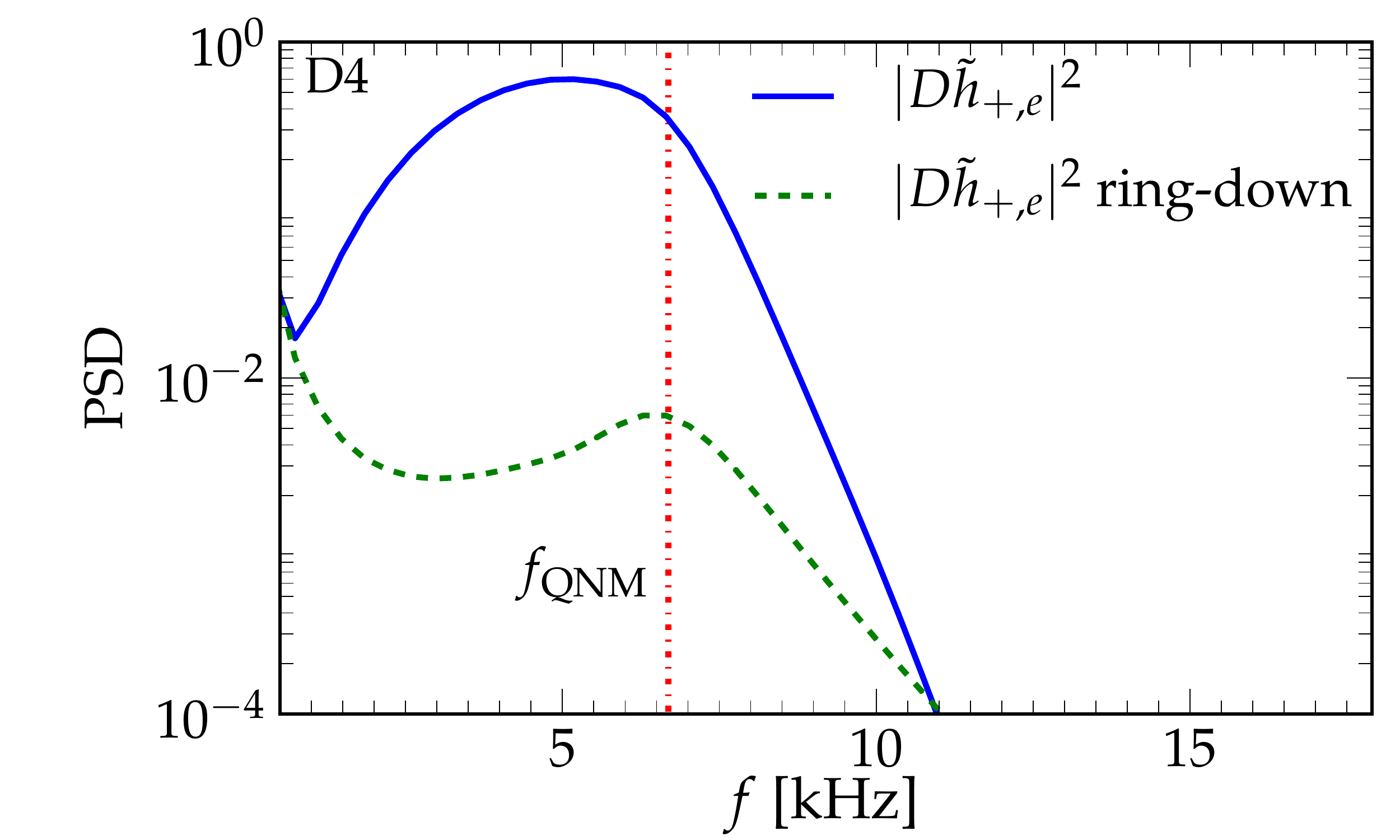}
  \caption{Rotating neutron star collapse: power spectral density of ``+`` polarization of GW strain $D h_{+,e}$ as emitted in the
equatorial plane and extracted via CCE. The blue straight line is the spectrum of the entire waveform, while the green dashed line is the spectrum of the ring-down
signal. The red vertical line denotes the $(\ell,m)=(2,0)$ prograde fundamental ($N=0$) quasi-normal mode frequency $f_{\rm QNM}=6.68\,\rm{kHz}$ 
of a spinning black hole of mass $M=1.8602\, M_\odot$ and dimensionless spin $a=0.5435$
as computed in \cite{berti:09}. Mass and spin of the nascent black hole are determined on its apparent horizon using the isolated horizon framework.
The analysis is done using baseline resolution $r1$ with cell-centered AMR with ePPM.
    }
  \label{fig:rnsD4-QNM}
\end{figure}

\begin{table}[t]
\caption{Initial parameters and properties of the collapsing neutron star. ADM mass $M_{\rm ADM}$ and angular momentum $J_{\rm ADM}$ are computed
         from the initial data solver at spatial infinity $i^0$. The radiated energy $E_{\rm rad}$ and angular momentum $J_{\rm rad}$
         are computed from waves extracted via the method of CCE including modes up to $\ell=6$.
         The apparent horizon mass $M_{\rm AH}$ and angular momentum $J_{\rm ADM}$ are computed on the apparent horizon surface 
         after the black hole has settled to an approximate Kerr state.
         The data are reported for high resolution simulation $r2$. The value in brackets denotes the numerical error in the last reported digit.
         Units are in $c=G=M_\odot=1$.}
\label{tab:RNSD4-parameters}
\begin{ruledtabular}
\begin{tabular}{lll}
\hline
Initial polytropic scale & $K_{\rm ini}$ & $100$ \\     
Evolved polytropic scale & $K$ & $98$ \\     
Polytropic index & $\Gamma$ & $2$ \\
Central rest-mass density & $\rho_c$ & $3.116\times10^{-3}$ \\
Axes ratio                & & $0.65$ \\
Rotational / binding energy $[\%]$ & $T/|W|$          & $7.68$ \\
Equatorial radius $[M_\odot]$ & $R_e$      & $9.6522$ \\
Baryonic mass $[M_\odot]$      & $M_{B}$           & $2.0443$ \\
ADM mass $[M_\odot]$      & $M_{\rm ADM}$           & $1.8605$ \\
ADM ang.~mom.~$[M_\odot^2]$ & $J_{\rm ADM}$         & $1.8814$ \\
Spin                      & $a$  & $0.5435$ \\
Rad.~energy $[M_\odot]$  & $E_{\rm rad}$     & $8.14(3)\times10^{-7}$  \\
Rad.~ang.~mom.~$[M_\odot^2]$   & $J_{\rm rad}$ & $0(1)\times10^{-10}$ \\
AH mass $[M_\odot]$          & $M_{\rm AH}$          & $1.8602(3)$ \\
AH ang.~mom.~$[M_\odot^2]$     & $J_{\rm AH}$        & $1.874(7)$ \\
\end{tabular}
\end{ruledtabular}
\end{table}

Three-dimensional collapse of an isolated neutron star to a black hole is a
valuable test of accuracy and convergence of our code for
black hole formation in massive stars.
We consider the uniformly rapidly rotating model $D4$ previously studied in
\cite{baiotti:05,baiotti:07b} as a benchmark problem. 
Apart from showing convergence and consistency with previous results, we improve
the simulations by causally disconnecting the outer boundary from the
interior evolution and the wave-extraction zone. We show that cell-centered AMR
with refluxing leads to better conservation of mass than vertex-centered AMR.
We also employ CCE for GW
extraction.

\subsubsection{Initial Data and Equation of State}

The initial condition is
given by a stable relativistic polytrope.
Specifically, we use a polytrope $P=K\rho^\Gamma$ with $\Gamma=2$ and
$K_{\rm ini}=100$ in the initial data construction.
The initial data are
generated via Hachisu's self-consistent field method \cite{komatsu:89a,
komatsu:89b}.
The central density is set to $\rho_c=3.116\times10^{-3}=1.924\times10^{15}\, \rm{g}\,\rm{cm}^{-3}$.
We use an axes ratio of $0.65$, which results in
$\beta=T/|W|=7.6796\times10^{-2}$ corresponding to a dimensionless spin of
$a=J/M^2=0.54354$.
In order to induce the gravitational collapse, 
we introduce an artificial pressure depletion of $2\%$ by setting
$K=98$ at the onset of the evolution.
During evolution, we use an ideal fluid $\Gamma$-law equation of state with $\Gamma=2$.
The initial parameters and properties of the test case are summarized in Table~\ref{tab:RNSD4-parameters}.

\subsubsection{Numerical Setup}

The GW extraction is carried out on the cubed-sphere
grid setup shown in Fig.~\ref{fig:7patch-system}.
We pick the radius of the outer boundary such that the wave-extraction zone and the
interior evolution are causally disconnected from the outer boundary, which we set to $R_{\rm B}=800M_\odot$.

For our baseline grid setup $r1$, we make use of a radial and Cartesian
resolution of $\Delta r = \Delta x = 1.28M_\odot$ and $N_{\rm
ang}=25$ cells per patch and per angular direction.
The boundary between central Cartesian and cubed-sphere grids is located at
$R_{\rm S}=65M_\odot$. The radial coordinate spacing is increased from $\Delta r$
to $2\Delta r$ in the region between $R=250M_\odot$ and $R=600M_\odot$.

We employ five additional levels of AMR with half-widths $R_{\rm
rl}=\left[30M_\odot,18M_\odot,11M_\odot,5M_\odot,3M_\odot\right]$ located at the center of the Cartesian domain.
With an initial radius of $R_{\rm NS}\approx 10M_\odot$ along the equatorial plane,
this means that the finest two levels thread through the neutron star. These two
levels are required to resolve the black hole formed in
the collapse. For our baseline resolution $r1$, we therefore have a grid spacing of
$\Delta x = 0.16 M_\odot = 0.24\,\rm{km}$ on the third finest level encompassing the entire neutron star, and a
resolution of $\Delta x = 0.04M_\odot = 0.06\,\rm{km}$ on the finest level containing the black hole.

In addition to $r1$, we also use a low resolution $r0$ with
a coarse grid spacing of $\Delta r = \Delta x = 1.6M_\odot$
and $N_{\rm ang}=20$ cells per patch and per angular direction,
and a high resolution setup $r2$ with
a coarse grid spacing of $\Delta r = \Delta x = 1.024M_\odot$
and $N_{\rm ang}=31$ cells per patch and per angular direction.

We set the damping coefficient of the $\Gamma$-driver gauge
condition to $\eta=1/2$, 
and exponentially
damp $\eta$ to zero starting from radius $R_\eta=65M_\odot$ .

The artificial low-density atmosphere is $10^{-8}$ of initial central density.
We also perform a simulation with an atmosphere density $10^{-10}$ of the central density, 
however, we find only negligible differences in the accuracy of our results.

\subsubsection{Discussion}

Following initial pressure depletion, the uniformly rotating polytrope collapses. 
During collapse, the central density $\rho_c$ increases until time $t-t_{\rm BH}=0$, 
the time when an apparent horizon, and thus a black hole forms.
After formation of the horizon, the matter inside the horizon is excised from the grid, and 
the remaining exterior matter is rapidly dragged into the nascent black hole,
leaving behind the artificial low-density atmosphere.
Upon formation, the black hole is highly excited and radiates GWs until it settles to a Kerr state.
This produces a characteristic ring-down GW signal with a particular quasi-normal mode frequency
which depends only on mass and spin of the black hole.

In Fig.~\ref{fig:rnsD4-convergence}, we show the emitted GW signal $D h_{+,e}$, and the evolution of the central density $\rho_c$ for the three
resolutions $r0$, $r1$ and $r2$, respectively. The simulations are performed using cell-centered AMR, refluxing, and ePPM reconstruction.
The GW signal is extracted using CCE and we use
FFI with a cut-off frequency of $f_0=1\,\rm{kHz}$ to obtain $D h_{+,e}$. We note that the only significant non-zero signal is contained in the $(\ell,m)=(2,0)$ 
wave mode\footnote{Earlier studies \cite{baiotti:05,baiotti:07b} also found an $(\ell,m)=(4,0)$ wave mode. In our case, 
this mode is three orders of magnitudes smaller than the $(\ell,m)=(2,0)$ mode amplitude and comparable to the level of numerical noise.
Since the earlier study did not use causally disconnected outer boundaries, did not compute the waveform at future null infinity $\scri^+$, 
and had less resolution in the wave-extraction zone, we argue that a $(\ell,m)=(4,0)$ could have been excited because of numerical artifacts and systematic errors.}
and we use \eqref{eq:l2m0-strain} to get $D h_{+,e}$.
When comparing the waveform obtained from CCE to the one obtained from RWZM (not shown), we notice that the waveforms from RWZM are more susceptible to numerical noise and
contain spurious high-frequency oscillations. This is consistent with our findings in \cite{reisswig:11ccwave} (see also Sec.~\ref{sec:A3B3G3}). The waveforms extracted via RWZM
are similar to those obtained in \cite{baiotti:05,baiotti:07b}, which also use RWZM extraction. 
We thus believe that the results of \cite{baiotti:05,baiotti:07b} also
suffer from the same spurious high-frequency noise.

We align all quantities at the coordinate time when an apparent horizon appears ($t-t_{\rm BH}=0$).
By computing the differences in low and medium, and medium and high resolutions, 
we get an estimate for the convergence of our simulations. 
In panels below the emitted GW signal $D h_{+,e}$, and central density evolution $\rho_c$ 
of Fig.~\ref{fig:rnsD4-convergence}, respectively, we show the differences in GW signal and central density using the three different resolutions.
The differences between medium and high resolutions are scaled for second-order convergence.
At black hole formation, the GW signal and central density exhibit clear second-order convergence.
During collapse, while the central density shows second-order convergence, the convergence of the GW signal
is somewhat obscured due to the oscillatory nature of the latter, especially when the signal is not perfectly in phase. 
In the lower panel of Fig.~\ref{fig:rnsD4-convergence}, we show the $L_2$-norms of the Hamiltonian constraint $\norm{H}_2$ for the 
three resolutions.
Since the artificial initial pressure depletion is not constraint satisfying, the constraints do not converge initially.
For this reason, we do not introduce any rescaling for convergence.
However, the slopes for higher resolutions are smaller, resulting in somewhat smaller constraint violations at later times. 
At the time when an apparent horizon appears, and during ring-down, the constraints exhibit second-order convergence.

In Fig.~\ref{fig:rnsD4-comparison}, we compare performance of cell-centered AMR with ePPM, 
vertex-centered AMR with oPPM, and cell-centered AMR with ePPM and multirate RK time integration using baseline resolution $r1$.
The vertex-centered case with oPPM exhibits slightly larger constraint violations than the cell-centered setup using ePPM.
Before the horizon forms, baryonic mass should be exactly conserved.
In practice, this is not the case, even in the cell-centered case with refluxing. One reason for non-conservation is the artificial low-density atmosphere
(see Appendix~\ref{sec:atmo}). Another reason is the bufferzone prolongation in regions that thread the surface of the star.
Here, prolongation involving cells in the atmosphere can amplify mass non-conservation.
We note, however, that the cell-centered case with refluxing performs better than the vertex-centered case.
The simulation using multirate time integration performs equally well compared to the same simulation using standard RK4 time integration. 

In Fig.~\ref{fig:rnsD4-AH}, we show the mass and spin evolution of the apparent horizon. After $t-t_{\rm BH}=0$,
horizon mass and spin are quickly growing until they asymptote towards the ADM mass and angular momentum of the spacetime, respectively.
For a given spacetime, ADM mass and angular momentum are always constant.
Both quantities are calculated in the initial data solver and evaluated at spatial infinity.  
Since all matter falls into the horizon, the black hole mass plus the 
radiated energy must be equal to the ADM mass. The same applies to the angular momentum.
In the present case, we have $M_{\rm ADM}=1.8605\,M_\odot$. The black hole settles to a horizon mass of $M_{\rm AH}=1.8602\,M_\odot$.
Thus, the difference is $0.016 \%$.
Similarly, the angular momentum initially is $J_{\rm ADM}=1.8814 \,M_\odot^{2}$, and the black hole settles to $J_{AH}=1.874 \,M_\odot^{2}$.
This makes a difference of $0.39 \%$.
The radiated energy is $E_{\rm rad}=8.14\times 10^{-7} M_\odot$ and hence is tiny compared to the rest mass of the system.
This value agrees to the estimate given in \cite{baiotti:05,baiotti:07b}.
Since the only significant non-zero GW mode is the $(\ell,m)=(2,0)$ mode, no angular momentum is radiated.
We find that by decreasing the atmosphere level and increasing the resolution, the differences in horizon mass and angular momentum compared to the initial ADM values
are decreased. Hence, the error in mass and angular momentum conservation is due to systematic (atmosphere) and numerical error.

In Fig.~\ref{fig:rnsD4-QNM}, we investigate the power spectrum of the emitted GW signal $D \tilde{h}_{+,e}$.
The blue straight curve is the power spectrum of the entire signal which peaks at $f_{\rm peak}=5.06\,\rm{kHz}$.
The green dashed curve is produced by first applying a time-domain window function around the black hole ring-down part of the 
waveform before taking the Fourier transform. Thus, the green dashed curve is the power spectrum of the black hole ring-down part of the waveform.
This curve peaks at $f_{\text{peak, ring-down}}=6.47\,\rm{kHz}$. We can compare this frequency with the theoretically obtained quasi-normal (QNM) ring-down
frequency for a perturbed black hole in vacuum. For the black hole mass $M_{\rm AH}=1.8602\,M_\odot$ and dimensionless spin $a=J_{\rm AH}/M_{\rm AH}^2=
0.5414$, the $(\ell,m)=(2,0)$ prograde fundamental ($N=0$) quasi-normal frequency is $f_{\rm QMN}=6.68\,\rm{kHz}$ \cite{berti:09}.
Thus, the relative difference is $\sim 3.3 \%$. This is consistent with \cite{baiotti:07b} who find ''good agreement`` (unfortunately they do not provide numbers).
Note that we do not expect the two values to exactly coincide. The theoretical QNM frequency
is strictly only valid for perturbed Kerr black holes in vacuum. Since matter is crossing the horizon initially, the ring-down
signal will naturally be affected by black hole growth and spin-up.

\subsection{Binary Neutron Stars}
\label{sec:bns}

\begin{figure}
  \includegraphics[width=1.\linewidth]{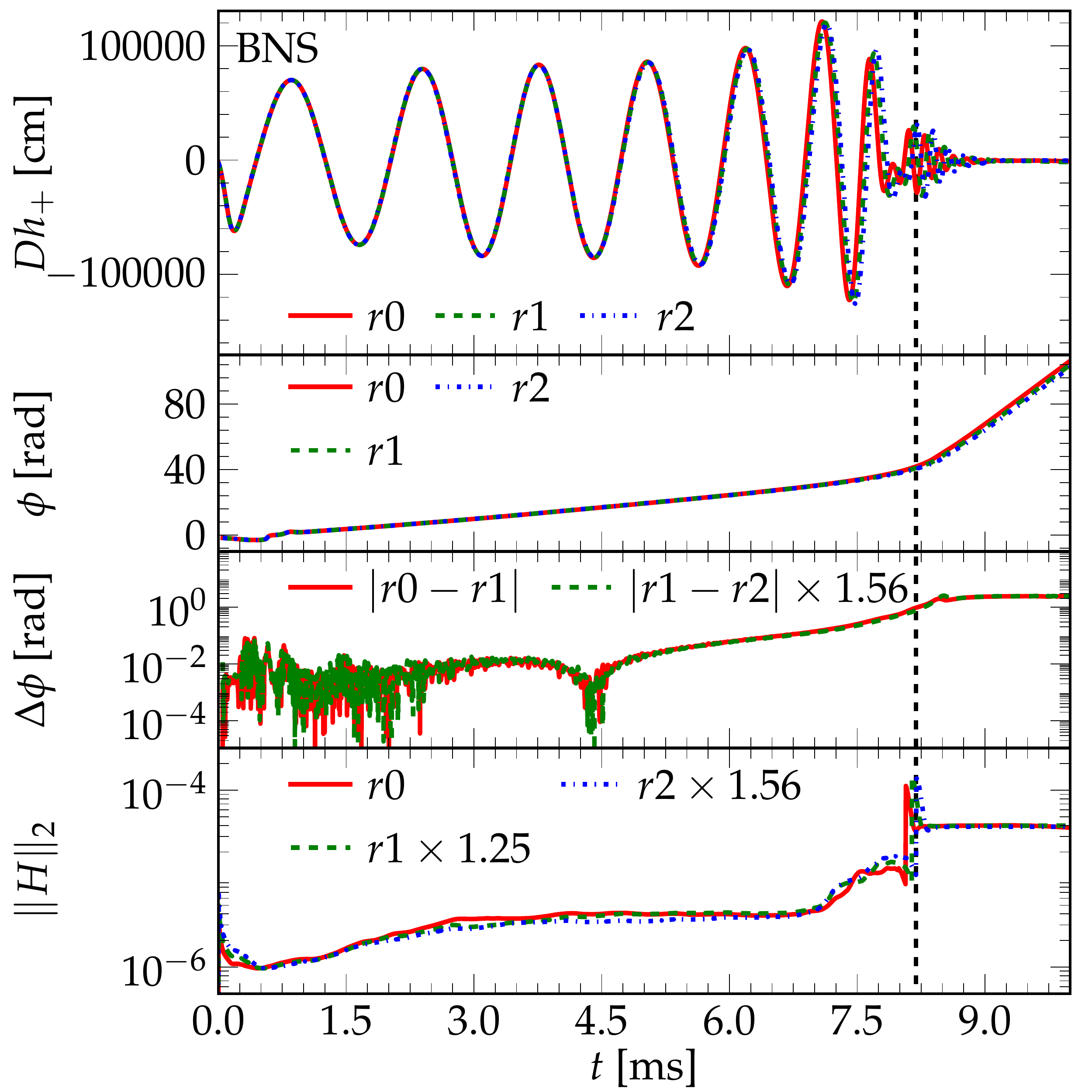}
  \caption{Binary neutron stars: Convergence study of the $(\ell,m)=(2,2)$ mode of the GW strain $D h$, 
  and the $L_2$-norm of the Hamiltonian constraint $\norm{H}_2$. The top panel shows the ``+'' polarization of the $(\ell,m)=(2,2)$ mode for
  all three resolutions. The panel below shows the GW phase $\phi$ of the $(\ell,m)=(2,2)$ mode.
  The third panel from the top shows the difference in phase $\phi$, scaled for
  second-order convergence. The vertical dashed line indicates appearance of an apparent horizon in the high-resolution simulation.
  The bottom panel shows the $L_2$-norm of the Hamiltonian constraint scaled for first-order convergence.
  The simulations were performed using cell-centered AMR, refluxing, and ePPM reconstruction.}
           \label{fig:bns-conv}
\end{figure}

\begin{figure}
  \includegraphics[width=1.\linewidth]{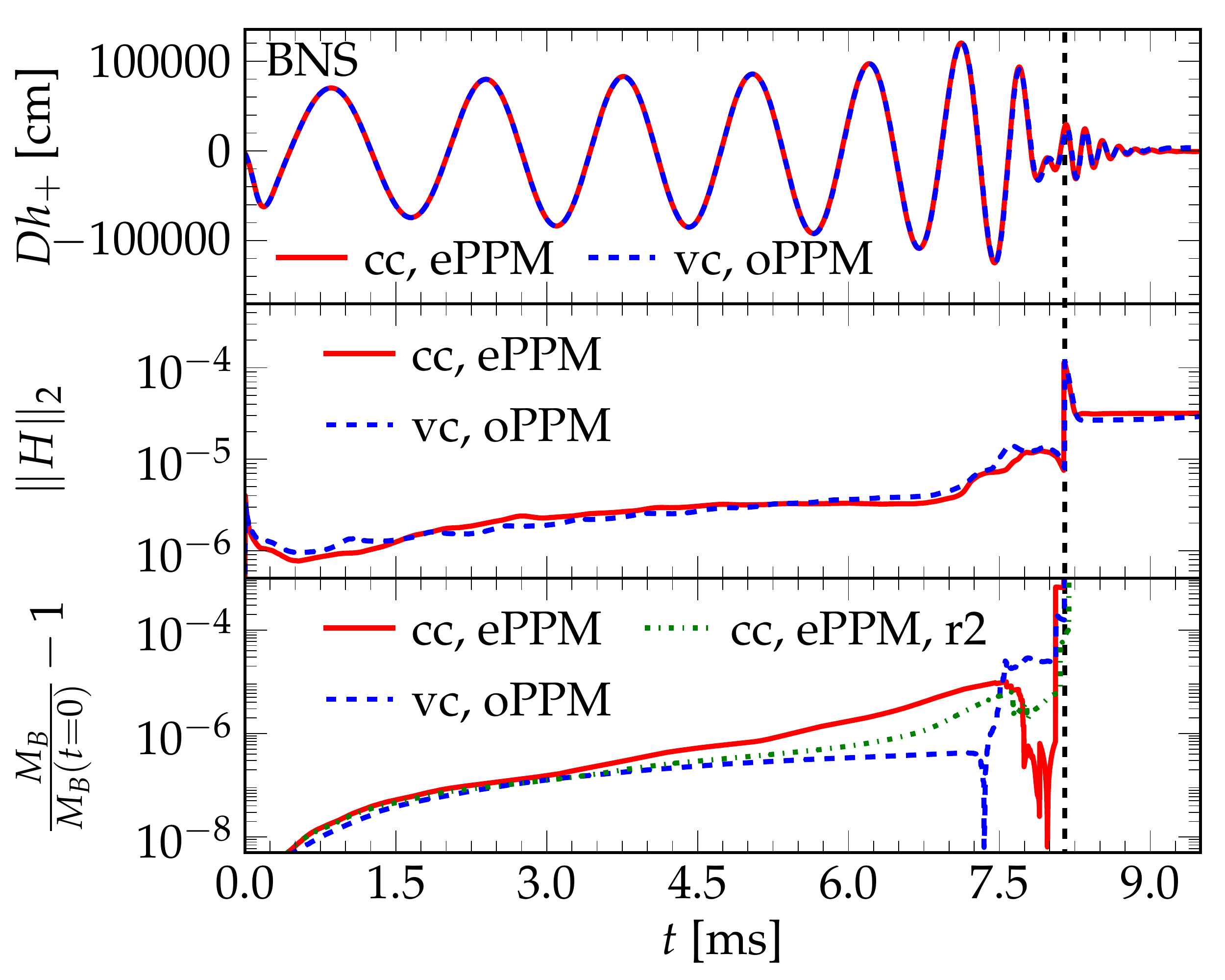}
  \caption{Binary neutron stars: comparison between cell-centered (cc) AMR with ePPM and vertex-centered (vc) AMR with oPPM.
           The top panel shows the $(\ell,m)=(2,2)$ mode of the ``+'' polarization of the GW strain $D h$. 
           The center panel shows the $L_2$-norm of the Hamiltonian constraint $\norm{H}_2$. The bottom panel shows conservation
           of baryonic mass $M_B$. The vertical dashed line indicates the appearance of an apparent horizon in the baseline resolution simulation.
           The simulations were performed using resolution $r1$, though for the conservation of mass, we also show the high resolution ($r2$) result.
           The error in mass conservation converges with better than second-order as the resolution is increased up to the point when a new refinement level is switched on at $t\sim7.5$ ms.}
           \label{fig:bns-compare}
\end{figure}

\begin{figure}
  \includegraphics[width=1.\linewidth]{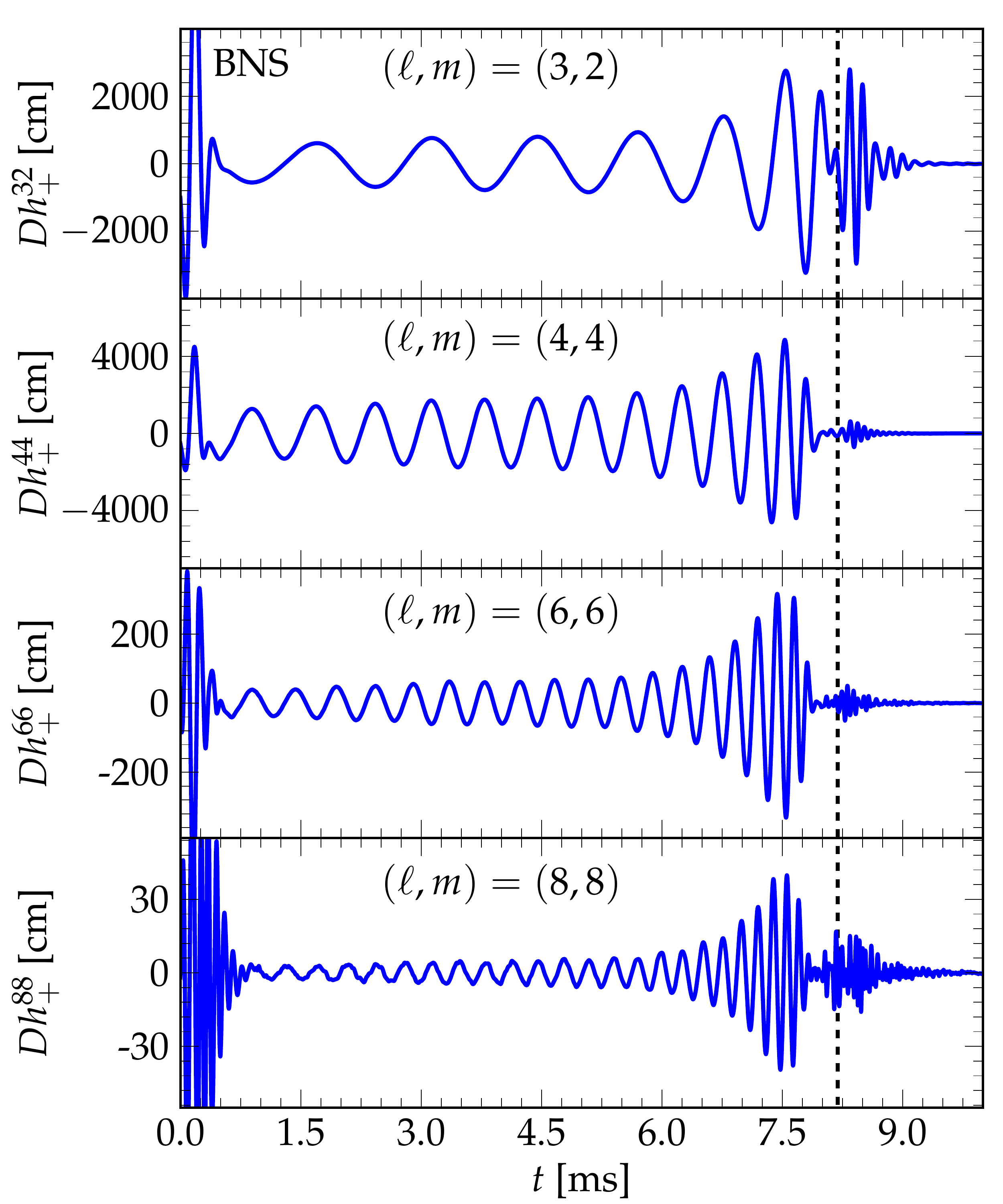}
  \caption{Binary neutron stars: GW modes $(\ell,m)=(3,2),\,(4,4),\,(6,6),\,(8,8)$ of ``+'' polarization of the strain $D h$ 
           unambiguously extracted via CCE.
           The waveforms are shown for high resolution simulation $r2$.
           The vertical line indicates the time of appearance of an apparent horizon. 
           Following appearance of an apparent horizon, a black hole ring-down signal is visible.}
           \label{fig:bns-modes}
\end{figure}

\begin{figure}
  \includegraphics[width=1.\linewidth]{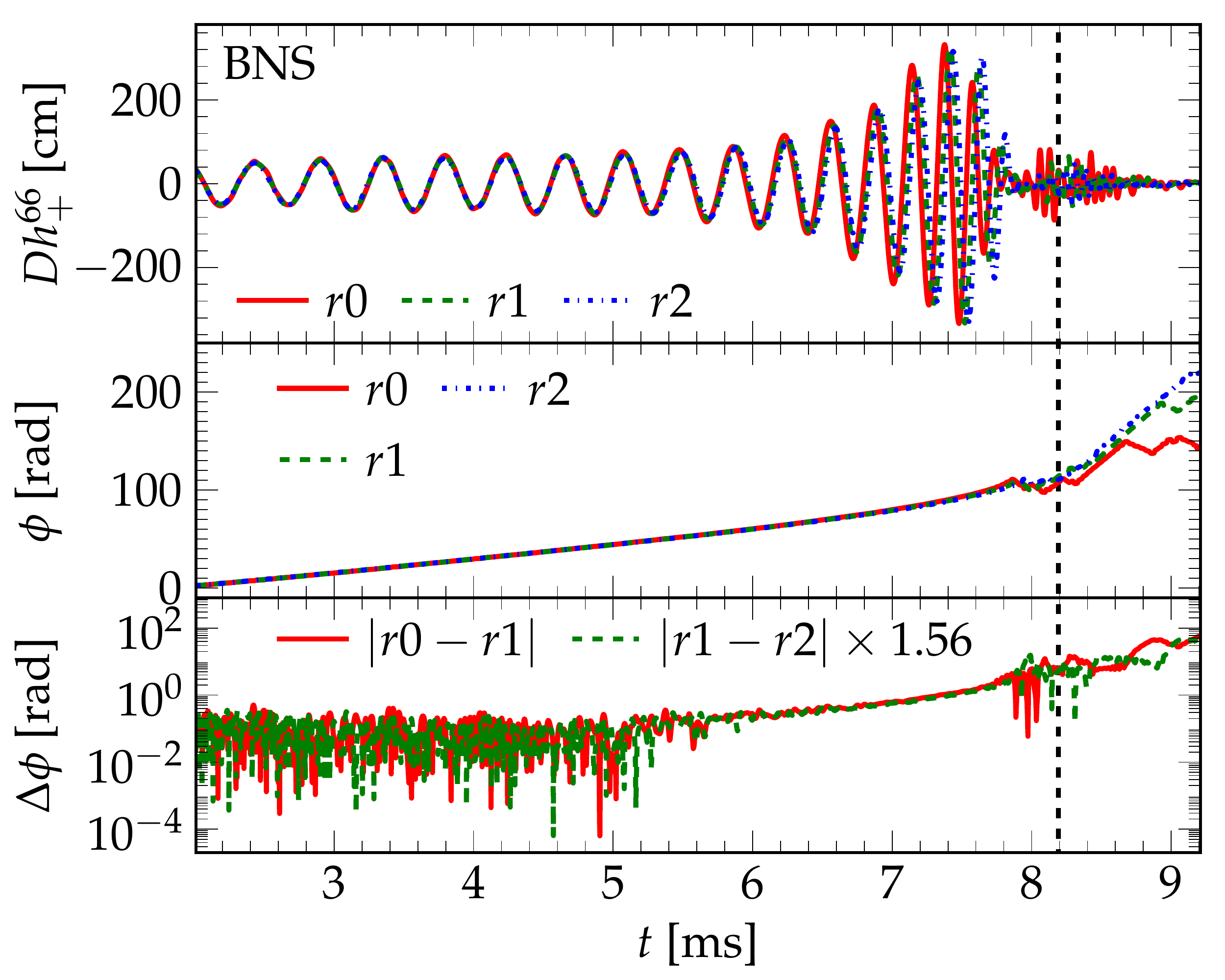}
  \caption{Binary neutron stars: Phase convergence of the $(\ell,m)=(6,6)$ mode of the GW strain $D h$. The top panel shows the ``+'' polarization component $D h_{+}^{66}$,
           and the panel below shows the phase $\phi$, for low $r0$, medium $r1$, and high $r2$ resolutions.
           The bottom panel shows the phase differences between low and medium, and medium and high resolutions, scaled for second-order convergence.
           Convergence is maintained throughout inspiral and merger. In the ring-down phase, the coarsest resolution $r0$ is insufficient 
           to accurately resolve this mode, and the results cease to converge properly.
           The vertical line indicates appearance of an apparent horizon in the high resolution simulation.}
           \label{fig:bns-amp-66-conv}
\end{figure}

\begin{figure}
  \includegraphics[width=1.\linewidth]{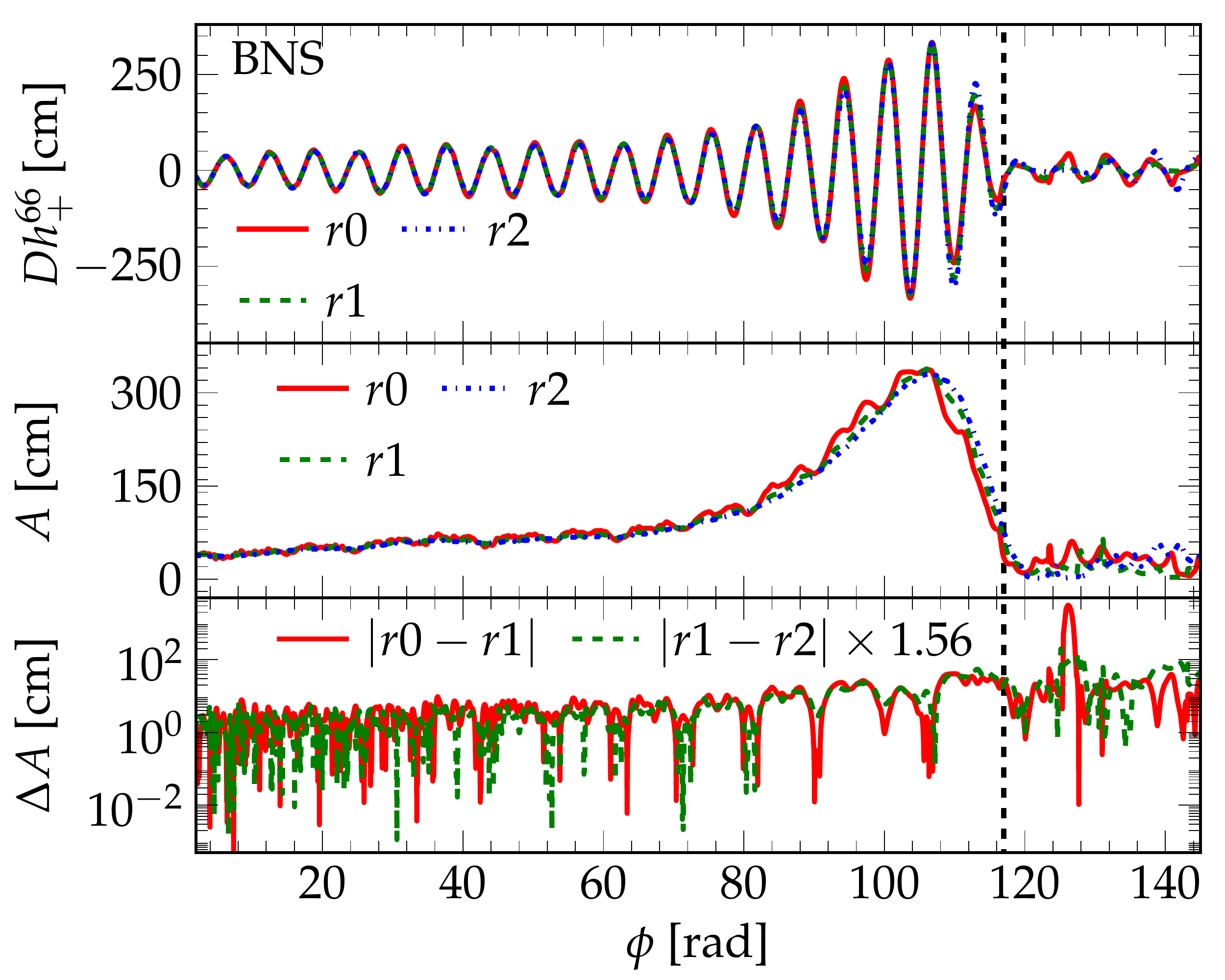}
  \caption{Binary neutron stars: Amplitude convergence of the $(\ell,m)=(6,6)$ mode of the GW strain $D h=D h(\phi)$ as a function of phase $\phi$. 
           The top panel shows the ``+'' polarization component $D h_{+}^{66}(\phi)$,
           and the panel below shows the amplitude $A(\phi)$, for low ($r0$), medium ($r1$), and high ($r2$) resolutions.
           The bottom panel shows the amplitude differences between low and medium, and medium and high resolutions, scaled for second-order convergence.
           Convergence is maintained throughout inspiral and merger. In the ring-down phase, however, the coarsest resolution $r0$ is insufficient 
           to accurately resolve this mode, and the results cease to properly converge.
           The vertical line indicates appearance of an apparent horizon in the high resolution simulation.}
           \label{fig:bns-phase-66-conv}
\end{figure}

\begin{table}[t]
\caption{Parameters of the binary neutron star system. ADM mass $M_{\rm ADM}$ and angular momentum $J_{\rm ADM}$ are computed
         by the initial data solver at spatial infinity $i^0$. The radiated energy $E_{\rm rad}$ and angular momentum $J_{\rm rad}$
         are computed from waves extracted via CCE including modes up to $\ell=6$.
         The apparent horizon mass $M_{\rm AH}$ and angular momentum $J_{\rm ADM}$ are computed after the black hole has settled to an approximate Kerr state.
         Gravitational disk mass $M_{\rm disk}$ and angular momentum $J_{\rm disk}$ are calculated from energy and angular momentum conservation.
         The data are reported for simulation $r2$. The value in brackets denotes the numerical error in the last reported digit.
         Units are in $c=G=M_\odot=1$.}
\label{tab:BNS-parameters}
\begin{ruledtabular}
\begin{tabular}{lll}
\hline
Lorene initial data set  & \multicolumn{2}{l}{\code{G2\_I12vs12\_D5R33\_60km}} \\
Initial separation [km] & $d$ & $45$ \\ 
Polytropic scale & $K$ & $123.6$ \\     
Polytropic index & $\Gamma$ & $2$ \\
Initial orbital frequency $[Hz]$ & $\Omega_{\rm ini}$ & $302$ \\
ADM mass $[M_\odot]$      & $M_{\rm ADM}$ & $3.2515$ \\
ADM ang.~mom.~$[M_\odot^2]$ & $J_{\rm ADM}$ & $10.1315$ \\
Rad.~energy $[M_\odot]$ ($\%$)  & $E_{\rm rad}$ & $2.51(5)\times10^{-2}$ $(0.77\%)$ \\
Rad.~ang.~mom.~$[M_\odot^2]$ ($\%$)  & $J_{\rm rad}$ & $1.206(9)$ $(11.9\%)$ \\
AH mass $[M_\odot]$          & $M_{\rm AH}$  & $3.2249(3)$ \\
AH ang.~mom.~$[M_\odot^2]$     & $J_{\rm AH}$  & $8.75(2)$ \\
AH spin                        & $a$  & $0.841(2)$ \\
Grav.~mass disk          $[M_\odot]$          & $M_{\rm disk}$  & $1.4(4)\times10^{-3}$ \\
Bary.~mass disk         $[M_\odot]$          & $M_{B,\rm disk}$  & $1.3(2)\times10^{-3}$ \\ 
Ang.~mom.~disk          $[M_\odot^2]$          & $J_{\rm disk}$  & $0.16(4)$ \\
\end{tabular}
\end{ruledtabular}
\end{table}

We investigate accuracy and convergence of 
the inspiral and coalescence of a binary neutron star (BNS) system.
Previous studies in full general relativity were restricted by the employed
purely Cartesian grids (e.g. \cite{thierfelder:11, bernuzzi:12a, bernuzzi:12b, gold:11, baiotti:11,
giacomazzo:11, rezzolla:11, etienne:12, kiuchi:12}, also see \cite{faber:12} for a recent review), and thus 
the accuracy of the GW extraction was limited.

For the first time in the context of binary neutron star mergers, we
use CCE for GW extraction at future null infinity $\scri^+$ (see Sec.~\ref{sec:waveextract}).
This removes finite radius and gauge errors and, combined with our multipatch grid,
allows us to extract the higher than leading order modes.

Finally, we also compare vertex centered AMR with oPPM with
cell-centered AMR with refluxing and ePPM.

\subsubsection{Initial Conditions and Equation of State}

The particular system we evolve is the initial data set
\code{G2\_I12vs12\_D5R33\_60km} produced by the \code{LORENE}
code~\cite{LORENE:web,gourgoulhon:00}. 
This system, with the same parameters as described below, has also been
considered in \cite{baiotti:10,baiotti:09a}.

The system consists of two neutron stars initially
described by a polytropic equation of state $P = K \rho^\Gamma$ with $K =
123.6$ and $\Gamma = 2$ with an initial coordinate separation of
$45\,$km. We evolve the system using a $\Gamma$-law equation of state of the
form
\begin{equation}
    P = (\Gamma-1) \rho \epsilon\text{.}
    \label{eqn:NSNSgammalaw}
\end{equation}
These parameters yield neutron stars of individual baryonic mass $M_B =
1.78\,\mmsun$ and ADM-mass in isolation of $M_\text{NS} =  1.57\,\mmsun$. 
The total ADM
mass of the system is $M_{\rm ADM} =  3.2515\,\mmsun$, and the total ADM angular momentum
is $J_{\rm ADM} = 10.1315\,M_\odot^{2}$.
The initial orbital angular frequency of the binary is $\Omega_{\rm ini}=302\,\rm{Hz}$.
The initial parameters and properties are listed in Table~\ref{tab:BNS-parameters}.

\subsubsection{Numerical Setup}

The numerical setup consists of the six spherical \emph{inflated-cube} grids
that surround the central Cartesian cube. The inner spherical radius of the inflated
cube grids is located at a coordinate radius of $R_{\rm S}=75.84\,\mmsun$ and the
outer (spherical) boundary is located at a radius of $R_{\rm B}=2800\,\mmsun$. The
radial resolution at the inner spherical inter-patch boundary matches the coarse-grid Cartesian
resolution of the central cube and is $\Delta x = 1.5\,\mmsun = 2.22\,\rm{km}$, $\Delta x = 1.2\,\mmsun = 1.77\,\rm{km}$ and
$\Delta x 0.96\,\mmsun = 1.42\,\rm{km}$ for the low, medium and high resolution runs, respectively.  In
the region $250\,\mmsun < r < 800\,\mmsun$ we smoothly transition to a coarser
resolution of $6.0\,\mmsun$, $4.8\,\mmsun$ and $3.84\,\mmsun$ for low ($r0$), medium ($r1$)
and high resolution ($r2$), respectively. The angular resolution is constant along
radial distances and we use $21$, $25$ and $31$ angular grid points per
angular direction and spherical patch for the three resolutions. We use 4
initial levels of mesh refinement in the inner Cartesian cube to resolve the
neutron stars. We surround each neutron star with a set of nested, refined cubes
of
half-width $13\,\mmsun$, $17.875\,\mmsun$ and $26.125\,\mmsun$, where the
finest level completely covers the neutron star. All refined cubes surrounding
the stars are contained in the common, coarse cube of half-width $R_{\rm S}$.
In each
refined level the resolution is twice that of the previous level. 
On the finest level, the neutron stars are covered 
with a resolution of $\Delta x = 0.1875\,M_\odot = 0.278\,$km, $\Delta x = 0.15\,M_\odot = 0.222\,$km and
$\Delta x = 0.12\,M_\odot = 0.176\,$km for the three resolutions $r0$, $r1$ and $r2$, respectively.

When the two neutron stars are about to come into contact, we remove the
nested set of cubes surrounding each individual star and surround
the binary with a common set of nested cubes of half-width $R_{\rm S}$, $30\,\mmsun$,
$15\,\mmsun$ and $7.5\,\mmsun$ ensuring uniform resolution in the central
region. Once the lapse function drops to values that indicate 
that an apparent horizon is about\footnote{This is a consequence of the 
$1+\log$ slicing condition \eqref{eq:one_plus_log} which locally slows down 
time evolution (i.e.~$\alpha<1$) in regions of strong curvature. A closed surface of lapse of $\alpha\lesssim0.3$ 
has been found to approximately resemble the apparent horizon shape.}
to form, we switch on a final level of radius $3.5\,\mmsun$ and resolution
$9.38\times10^{-2}\,\mmsun$, $7.5\times10^{-2}\,\mmsun$ and
$6.00\times10^{-2}\,\mmsun$ for the low, medium and high resolution runs
respectively. This level allows us to handle the steep metric gradients
developing inside of the newly formed apparent horizon.

During inspiral, we track the center of mass of each neutron star to keep
the two fluid bodies close to the center of their refined regions. We compute the
center of mass of an individual neutron star by integrating over the conserved density 
within a radius $R=4.0\,M_\odot$ of the densest point on the grid. This
method produces smoother tracks than directly using the location
of the densest point, and helps reducing the jitter in the mesh refinement boxes
observed otherwise.

We set the damping coefficient of the $\Gamma$-driver gauge
condition to $\eta=1$.

We set the dissipation strength to $\epsilon_{\rm diss}=0.1$ everywhere on the grid.
The artificial low-density atmosphere is $10^8$
times lower than the initial central density.

\subsubsection{Discussion}

While the two neutron stars orbit each other, they lose energy due to gravitational radiation, inspiral, and finally merge.
The nascent hypermassive neutron star remnant has a mass which is well above the maximum mass of neutron stars.
It forms a black hole on a dynamical timescale.
The black hole is initially highly excited, and relaxes to a Kerr state by emitting gravitational ring-down radiation.

In Fig.~\ref{fig:bns-conv}, we show convergence of the dominant $(\ell,m)=(2,2)$ mode of
the GW strain $D h$, the $L_2$-norm of the Hamiltonian constraint $\norm{H}_2$.
The upper panel shows the ``+'' polarization of the $(\ell,m)=(2,2)$ mode of the GW strain for the resolutions $r0$, $r1$, and $r2$.
The waveform is extracted via CCE. To obtain $D h$, we use a cut-off parameter $f_0=507$ Hz, which is below the initial instantaneous $(\ell,m)=(2,2)$ 
mode frequency $f_{\rm ini}^{22}$ determined from the initial orbital frequency by $f_{\rm ini}^{22}=2\Omega_{\rm ini}$.
To assess the phase convergence, we plot the differences in phase between low $r0$ and medium $r1$ resolution, and medium and high $r2$ resolution, 
scaled for second-order convergence.
We also plot the $L_2$-norm of the Hamiltonian constraint $\norm{H}_2$ scaled for first-order convergence.
Similar to the isolated neutron star tests in Sec.~\ref{sec:tov}, the dominant constraint error is generated at the contact discontinuity at the neutron star surface, where
our scheme locally reduces to first-order accuracy.

In Fig.~\ref{fig:bns-compare}, we compare cell-centered (cc) AMR and ePPM reconstruction with vertex-centered (vc) AMR and oPPM.
The $(\ell,m)=(2,2)$ mode of the GW strain $D h$ and the $L_2$-norm of the Hamiltonian constraint $\norm{H}_2$
do not show any significant differences between the two
numerical setups at this point.
After black hole and disk formation, the vertex-centered scheme exhibits a slightly larger slope in constraint growth.
In the bottom panel, we show conservation of total baryonic mass $M_B$. During early inspiral, both setups conserve mass to a high degree, only affected by
small errors due to our artificial atmosphere (see Appendix~\ref{sec:atmo}). Note that both neutron stars are completely contained on their finest grids.
Thus, there are no refinement boundaries directly influencing the evolution of the two fluid bodies.
As the inspiral progresses, we find that mass conservation is violated in the cell-centered case to a higher degree than in the vertex-centered case (though the error 
converges as the resolution is increased).
This appears to be an artifact of buffer zone prolongation close to the neutron star surface in combination with low density matter slightly above and at atmosphere values. 
Due to numerical errors, small amounts of mass are leaking out of the neutron star during inspiral and interact with the atmosphere. As this low density matter 
reaches the buffer zones, numerical
errors due to prolongation, which are by construction larger in the cell-centered case, tend to amplify the negative effects of the atmosphere treatment.
In experiments with isolated neutrons stars, however, we find that when the refinement boundaries are sufficiently far removed, 
and/or the atmosphere level is further decreased, mass can be conserved to a higher degree.

We also compare the simulations to a setup using multirate RK time integration and cell-centered AMR with ePPM. 
Unfortunately, due to the large fluid bulk velocities in the inspiral phase, the orbital phase accuracy is significantly affected by 
the lower order fluid time integration.
Thus, we do not recommend application of
multirate RK schemes in the context of binary neutron star mergers, especially when orbital phase accuracy is paramount.
The problem may be ameliorated by the use of co-rotating coordinates (see, e.g., \cite{duez:08}).

In order to demonstrate the potential of the multipatch scheme for more accurate wave extraction, we show in Fig.~\ref{fig:bns-modes} 
some of the higher harmonic GW modes that are emitted during inspiral, merger, and ring-down.
We show (from top to bottom) the $(\ell,m)=(3,2)$, $(\ell,m)=(4,4)$, $(\ell,m)=(6,6)$, and $(\ell,m)=(8,8)$
modes of ``+'' polarization of the strain $D h$. 
The modes are extracted from a simulation using resolution $r2$, cell-centered AMR, 
and ePPM.
All modes up to $(\ell,m)=(4,4)$ show a clean inspiral, merger and ring-down signal, and converge with resolution (see below). 
For higher modes, our lowest resolution run $r0$ is insufficient to also allow for clean convergence of the corresponding ring-down signals. Accordingly,
those should be taken with a grain of salt.
As an example, in Figs.~\ref{fig:bns-amp-66-conv} and~\ref{fig:bns-phase-66-conv},
we show convergence of phase and amplitude of the $(\ell,m)=(6,6)$ mode of the GW strain, respectively.
Fig.~\ref{fig:bns-amp-66-conv} shows the GW amplitude $A$ reparametrized in terms of the gravitational phase $\phi$ to disentangle
phase from amplitude.
Both figures indicate that second-order convergence is maintained during inspiral up to merger. The ring-down part, however, does not exhibit
clean second-order convergence. In that case, the coarse resolution becomes insufficient, and the result ceases to converges properly.
We note that for the highest extracted mode, $(\ell,m)=(8,8)$, the coarsest resolution is insufficient to allow for clean convergence also in the inspiral phase.

We compute the radiated energy $E_{\rm rad}$, radiated angular momentum $J_{\rm rad}$,
the horizon mass $M_{\rm AH}$, and horizon angular momentum $J_{\rm AH}$.
For the computation of the radiated quantities, we include modes
$\ell \leqslant 6$ as extracted via CCE.
After the black hole has formed and settled to an approximate Kerr state,
some amount of material is located in an accretion disk surrounding the black hole. 
Hence, we do not expect that horizon mass and radiated energy balance with the total ADM mass at this time.
Rather, the difference denotes the gravitational mass of the accretion disk that has formed.
Likewise, the same is true for the balance of angular momentum.
Given the horizon mass, the spacetime's total ADM mass, and the radiated energy, we estimate the gravitational mass of the accretion disk to be
$M_{\rm disk}=M_{\rm ADM}-M_{\rm AH}-E_{\rm rad}=(1.4\pm0.4)\times10^{-3}\, M_\odot$. 
The disk's baryonic mass is $M_{B,\rm disk}=(1.3\pm0.2)\times10^{-3}\,M_\odot$, which we compute by
integrating over all material outside of the apparent horizon and within a radius $R<40\,M_\odot$.
Both, baryonic and gravitational mass agree within their error bars.
We note that the mass of the disk, though clearly visible in density contour plots of our simulation (not shown), 
is tiny and thus not much above the numerical error.
Given the horizon angular momentum, the spacetime's total ADM angular momentum, and the radiated angular momentum, we estimate the disk's angular momentum to be
$J_{\rm disk}=J_{\rm ADM}-J_{\rm AH}-J_{\rm rad}=0.16\pm0.04\, M_\odot^2$.
For convenience, we list spacetime, black hole, disk, and radiated mass (and angular momentum) in Table~\ref{tab:BNS-parameters}.
All error bars are estimated using medium and high resolution results. The results for mass and spin of the black hole
agree to the values that were found in \cite{baiotti:10}.

In our binary neutron star merger problem, we also investigate the error inherent to finite-radius GW extraction.
We compare $\Psi_4$ as extracted via the NP formalism at a finite radius with $\Psi_4$ as extracted via CCE at future null infinity $\scri^+$.
We align two given waveforms in the early inspiral phase by minimizing their phase difference over an interval $t\in[2.5\,\rm{ms},3.5\,\rm{ms}]$ using
the method described in \cite{boyle:08}.
For the $(\ell,m)=(2,2)$ mode, we find a total dephasing on the order of $\Delta\phi\sim 1\,\rm{rad}$ and an amplitude difference of about $\sim10\%$
between the waveform obtained at $R=250M_\odot$ and the one obtain at $\scri^+$.
Waveforms extracted at smaller radii naturally yield larger differences to the result at $\scri^+$.
While the amplitude error is rather large, the dephasing is comparable to the dephasing due to numerical error of the orbital evolution of the two neutron stars. 
Since this numerical error is convergent, but the systematic finite radius-error is not,
the finite-radius error becomes a non-neglegible effect as the numerical resolution is increased.
As shown in \cite{reisswig:09, reisswig:10a} for the case of binary black hole mergers, 
extrapolation to infinity using finite-radius data can reduce the errors to a tolerable level in cases where CCE is not available.

Finally, we investigate the influence of the outer boundary when it is \textit{not} causally disconnected from the wave extraction region
and interior evolution.
We compare a setup with a causally connected outer boundary located at $R_{\rm B}=2000M_\odot$ and 
a causally disconnected boundary located at $R_{\rm B}=2800M_\odot$.
The former setup is in causal contact with the interior and wave-extraction region
during the merger and ring-down phases.
We find a difference in GW phase and amplitude, and final spin and mass of about $\sim7\%$.
More details are given in Appendix~\ref{sec:BC}.

By comparing our results with those of \cite{baiotti:10,baiotti:09a}, we conclude that 
the accuracy of the orbital evolution of the two neutron stars is very similar.
The errors in satisfying the Hamiltonian constraint and conserving baryonic mass 
are of comparable size. This is not surprising, since we find little difference between the new cell-centered AMR
scheme compared with the vertex-centered AMR scheme that was also used in \cite{baiotti:10,baiotti:09a}.
Due to our multipatch grids, causally disconnected outer boundaries, and CCE, however, the waveforms that are extracted from our simulations are
more accurate than what has been shown in previous studies.


\section{Summary and Conclusions}
\label{sec:summary}

We have presented a new GR hydrodynamics scheme using multiple Cartesian/curvi-linear grid patches 
and flux-conservative cell-centered adaptive mesh refinement (AMR) to allow for a more 
efficient and accurate spatial discretization of the computational domain.
This is the first study enabling GR hydrodynamic simulations with multipatches and AMR.
Our multipatch scheme consists of a set of curvi-linear spherical ``inflated-cube'' grids with fixed angular resolution and variable radial spacing,
and a central Cartesian grid with AMR.
High-order Lagrange interpolation is used to fill ghost zones at patch boundaries for variables 
that are smooth, and second-order essentially non-oscillatory (ENO) interpolation for variables that contain discontinuities and shocks.

Apart from the successful implementation of multipatches and flux-conservative cell-centered AMR,
we have introduced a number of additional improvements to the publicly available code \texttt{GRHydro}:
(i) We have applied the enhanced piecewise-parabolic method (PPM) to ensure high-order reconstruction at smooth maxima, a property that we have found
to be crucial for cell-centered AMR.
(ii) To speed up the computation, we have applied a multirate Runge-Kutta time integrator that exploits the less restrictive Courant-Friedrich-Lewy (CFL)
condition for the hydrodynamic evolution by switching the the time integration to second order and thus reducing the number of intermediate steps by a factor of two.
Since the hydrodynamic evolution dominates the curvature evolution in terms of computational walltime
when complex microphysics and neutrinos are included, the scheme can yield a speedup of $\gtrsim30\%$ (e.g.~\cite{ott:12b}).

We have presented stable and convergent evolutions for binary neutron star mergers, stellar collapse to a neutron star, neutron star collapse to a black hole,
and evolutions of isolated unperturbed and perturbed neutron stars.
For each test case, due to the more efficient domain discretization, we have been able to enlarge the domain sufficiently so that
the outer boundary is causally disconnected from the interior evolution and wave-extraction zone.
This has allowed us to remove the systematic error that arises from the lack of constraint preserving boundary conditions for the Einstein equations
in the BSSN formulation.
In the case of the binary neutron star merger problem, 
we have found that this error is on the order of a few percent, and thus limits the accuracy of
the simulation and GW extraction.

In addition to enlarging the domain, multipatches have also allowed us to significantly increase the resolution 
in the GW extraction zone compared to previous studies. 
For the neutron star merger problem, we have been able to extract convergent spherical harmonic modes 
of the GW strain $D h$ up to $\ell = 6$. Previous studies have only
considered the dominant $(\ell,m)=(2,2)$ wave mode for this problem.

Furthermore, we have been able to remove the systematic error inherent
in finite-radius wave extraction by application of
Cauchy-characteristic extraction (CCE). This wave-extraction method
computes gauge-invariant radiation at future null infinity $\scri^+$
using boundary data from a worldtube at finite radius.  This method
has previously been applied in simulations of binary black holes and
stellar collapse \cite{reisswig:11ccwave, reisswig:09, reisswig:10a,
  babiuc:11, pollney:10, ott:11a, damour:11}.  Here, we have applied
CCE also to simulations of binary neutron star mergers, neutron star
collapse to a black hole, and isolated excited neutron stars.  We have
found that the error due to finite-radius extraction can be as large
as $10\%$.

Finally, for each test case, we have compared the original vertex-centered AMR 
scheme using original PPM with the new flux-conservative cell-centered AMR scheme using enhanced PPM.
The accuracy has been investigated and compared to results from previous studies.
We have found that simulations of stellar collapse greatly benefit from flux-conservative cell-centered AMR with enhanced PPM
compared to the original vertex-centered AMR scheme with original PPM. Conservation of mass and the satisfaction of the Hamiltonian
constraint are significantly better with the new scheme. 
The isolated neutron star and binary neutron star test cases, on the other hand, are not much affected by the choice of 
cell-centered or vertex-centered AMR. This is mainly due to the choice of grid setup: no matter is
crossing any refinement boundaries so that flux-conservation is not important.
It can become important, however, in the post-merger phase of binary neutron star coalescence, 
especially in cases where a massive accretion torus forms.

The multipatch infrastructure, 
the associated curvature and hydrodynamics evolution codes, and 
all other computer codes used in this paper
will be made (or are already) publicly available via the \texttt{EinsteinToolkit} \cite{einsteintoolkitweb}.


\acknowledgments
We acknowledge helpful discussions with Peter Diener, Frank L\"offler, 
Uschi~C.~T.~Gamma, and members of our Simulating eXtreme
Spacetimes (SXS) collaboration (\url{http://www.black-holes.org}). This
research is partially supported by NSF grant nos.\ AST-0855535,
AST-1212170, PHY-1212460, PHY-1151197, and OCI-0905046,
by the Alfred P. Sloan Foundation, and by the Sherman Fairchild
Foundation.  CR acknowledges support by NASA through Einstein
Postdoctoral Fellowship grant number PF2-130099 awarded by the Chandra
X-ray center, which is operated by the Smithsonian Astrophysical
Observatory for NASA under contract NAS8-03060.  RH acknowledges
support by the Natural Sciences and Engineering Council of Canada.
The simulations were performed on the Caltech compute cluster
\emph{Zwicky} (NSF MRI award No.\ PHY-0960291), on supercomputers of the
NSF XSEDE network under computer time allocation TG-PHY100033, on
machines of the Louisiana Optical Network Initiative under grant
loni\_numrel07, and at the National Energy Research Scientific
Computing Center (NERSC), which is supported by the Office of Science
of the US Department of Energy under contract DE-AC02-05CH11231. 
All figures were generated with the \code{Python}-based \code{matplotlib} package
(\url{http://matplotlib.org/}).


\appendix

\section{Shock-tube Tests}
\label{sec:shock}

\begin{figure}
  \includegraphics[width=1.\linewidth]{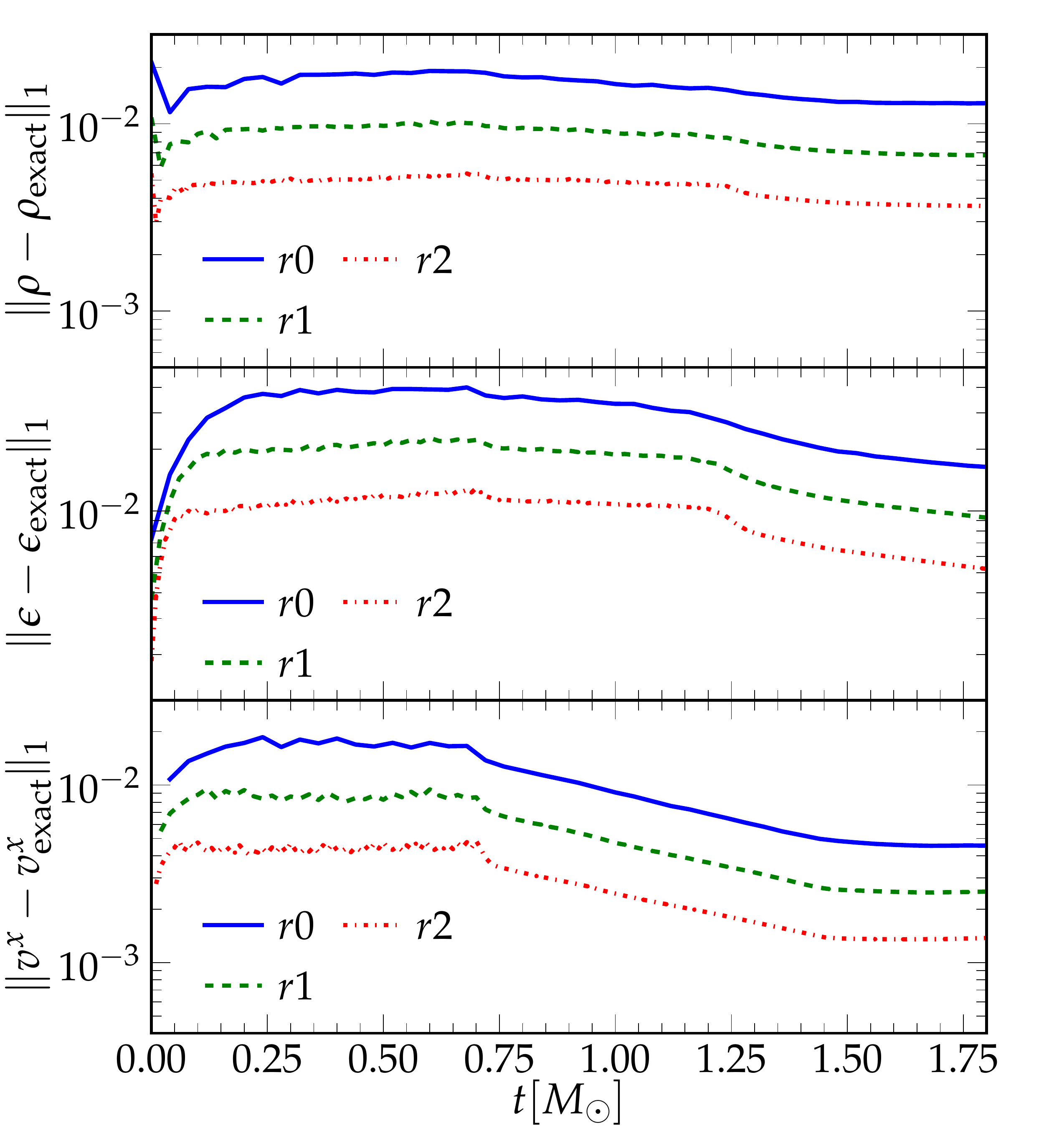}
  \caption{$L_1$-norm of the difference between exact and evolved fluid state for a Sod shock tube problem on low $r0$, medium $r1$, and high $r2$ resolutions.
           As the resolution is increased, the error in primitive density $\rho$ (upper panel), specific internal energy $\epsilon$ (middle panel), and 
           $x$-component of the 3-velocity $v^x$ (lower panel) correctly decrease by a factor of two in accordance with first-order convergence.
    }
  \label{fig:SOD_7p}
\end{figure}

\begin{figure}
  \includegraphics[width=1.\linewidth]{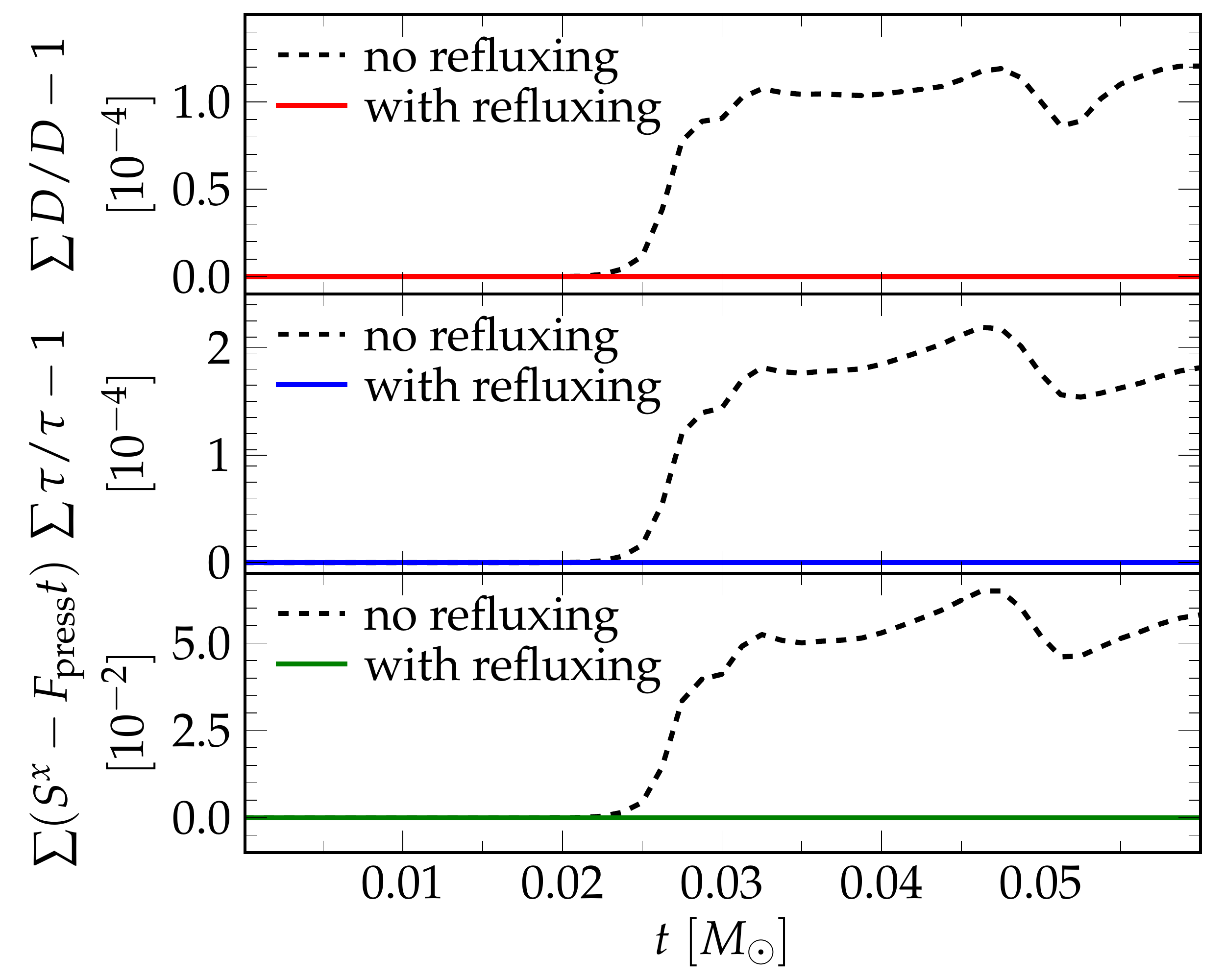}
  \vspace{-0.5cm}
  \caption{Conservation of mass (top panel), energy (middle panel), and momentum (bottom panel) as a function of time for a shock front crossing a refinement boundary.
           The solid (red/blue/green) lines are from a simulation with refluxing, while the dashed (black) curves show the case without refluxing.
           With refluxing, mass, energy, and momentum are exactly conserved (to machine precision). 
           Without refluxing, conservation of mass, energy, and momentum is violated.}
  \label{fig:refluxing}
\end{figure}

We perform a number of basic Sod shock tube and spherical blast wave tests on fixed
backgrounds to ensure correctness and convergence of our scheme at mesh-refinement and inter-patch boundaries.

In this appendix, we restrict our attention to a simple Sod test to show convergence
of the primitive variables across inter-patch boundaries (see Sec.~\ref{sec:hydro-multiblock}), and to demonstrate mass, energy, and momentum conservation
at refinement boundaries when refluxing (see Sec.~\ref{sec:ccamr}) is used.

The Sod shock-tube test consists of setting the initial fluid state according to \cite{sod:78}.
The shock front is located at a position $x_0$.
The background metric is set to the flat space Minkowksi metric.
The tests below use a gamma-law equation of state $P=(\Gamma-1)\rho\epsilon$ with $\Gamma=1.4$.

If not stated otherwise, the tests below use cell-centered AMR with refluxing, ePPM reconstruction,
second-order ENO inter-patch interpolation, RK4 time integration with $\Delta t / \Delta x = 0.4$,
and the HLLE Riemann solver.

\subsection{Inter-patch Interpolation}

In this particular test, we check that
shock fronts are correctly transported across inter-patch boundaries by maintaining convergence, and
without introducing local oscillations at the shock, even in the presence of non-trivial Jacobians
and coordinate transformations.
We setup a multipatch grid consisting of a
central Cartesian grid surrounded by the spherical inflated-cube grids.
The outer boundary extends to $R_{\rm B}=2.5M_\odot$.
The boundary between Cartesian and spherical grids is located at $R_{\rm S}=0.5M_\odot$.
No AMR is employed.
For the coarsest resolution ($r0$), we set the Cartesian and radial resolution to $\Delta x = \Delta r = 0.05$,
and use $(N_\rho,N_\sigma)=(20,20)$ cells per spherical patch per direction.
Medium ($r1$) and high ($r2$) resolutions double and quadruple, respectively, the resolution with respect to
the coarsest resolution.

We set Sod initial data with $x_0=0$ and evolve the system for sufficiently long
so that the shock propagates across inter-patch boundaries.
At each timestep, we compare the evolved fluid state with a solution from an exact special relativistic Riemann solver \cite{marti:03}.

In Fig.~\ref{fig:SOD_7p}, we show the $L_1$-norm of the difference between exact and evolved primitive density $\rho$, specific internal energy $\epsilon$, and
the $x$-component of the 3-velocity $v^x$.
All quantities are plotted for the three resolutions $r0$, $r1$, and $r2$.
As the resolution is increased, the error correctly decreases by a factor of two between successive resolutions, thus indicating first-order convergence.
This is consistent with the ENO operator, which reduces to first-order at shocks.

\subsection{Refluxing}

In this simple test, we check the correctness of our refluxing scheme with a shock front crossing a refinement boundary.
As the shock crosses the boundary, mass, momentum and energy must be conserved to machine precision.

The numerical grid consists of two levels of 2:1 AMR.
The coarse level extends from $x=0$ to $x=1$. The fine level has a refinement half-width of $r=0.1$ and is located at $x=0.4$.
We set the Sod shock front \cite{sod:78} at location $x_0=0.48$. Thus, the shock starts off on the fine grid and propagates
onto the coarse grid.

A measure of conservation of energy and mass is given by the sum of the \textit{conserved} internal energy $\tau$ and the \textit{conserved} density $D$ over the entire simulation domain, 
respectively. Both sums must be constant for all times $t$.
A measure for conservation of momentum is given by the balance between the \textit{conserved} momentum and the pressure force per unit time.
The balance as a sum over the entire simulation domain must be constant as a function of time.
In Fig.~\ref{fig:refluxing}, we show the sums of conserved density, energy, and momentum when refluxing is used (solid lines).
Without refluxing (dashed lines), the conserved mass, energy, and momentum grow significantly at time $t\approx 0.025$ when the shock front crosses
the refinement boundary.

\section{Enhanced PPM Scheme}
\label{sec:ePPM}

\begin{figure}
  \includegraphics[width=1.\linewidth]{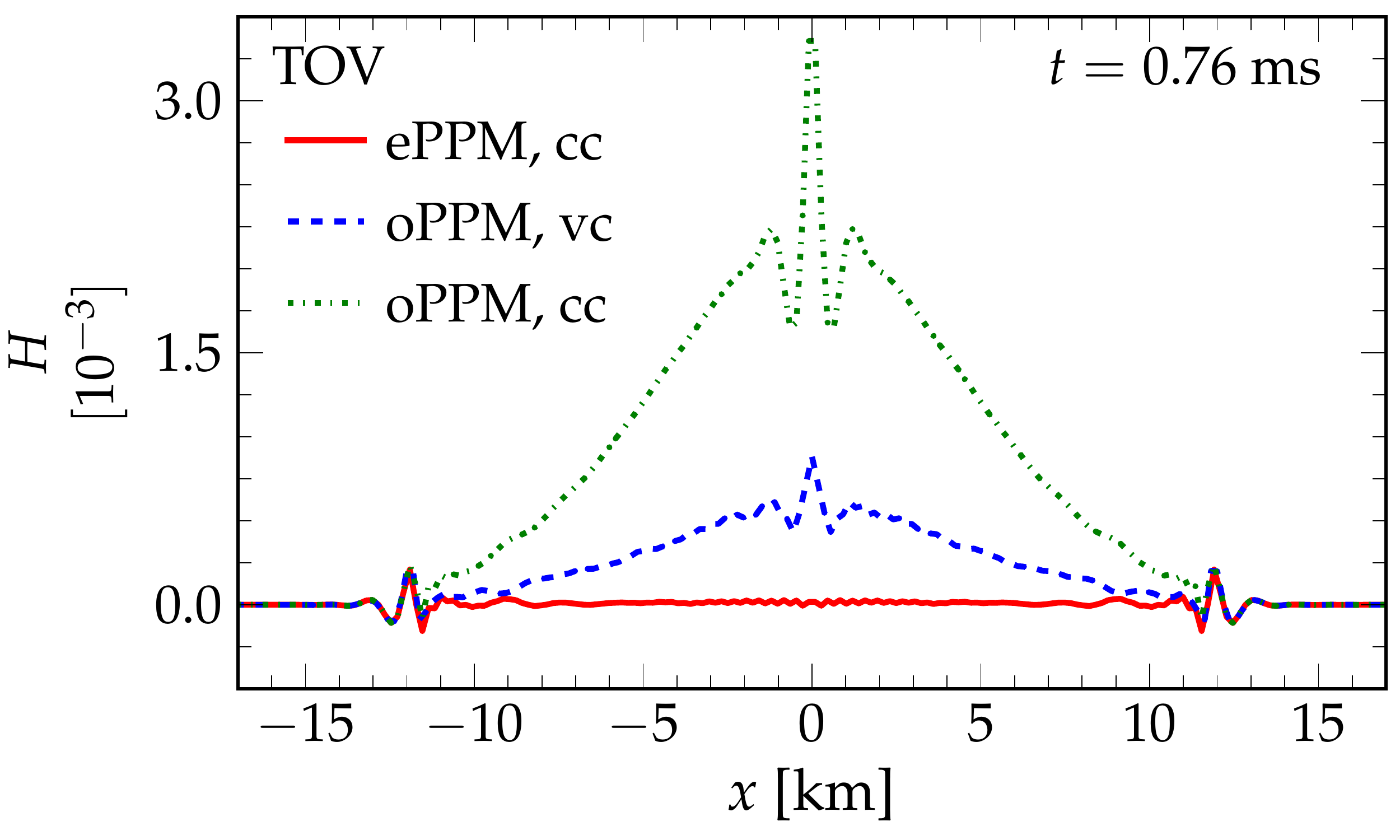}
  \caption{TOV star (from Sec.~\ref{sec:tov}): the effect of original PPM (oPPM) versus enhanced PPM (ePPM) on the Hamiltonian constraint as a function of $x$ at time $t=0.76$ ms.
           on cell-centered (cc) and vertex-centered (vc) AMR grids. The star's radius is $R_e=14.16$ km. The original PPM results in
           large constraint violations on the cell-centered grid. The enhanced PPM clearly outperforms oPPM. For ePPM, the error is 
           dominated by the first-order error at the neutron star surface, where the scheme reduces to first order.}
           \label{fig:tov-constr-1D}
\end{figure}

The PPM scheme seeks to find ``left'' and ``right'' interpolated values, $a_{i,L}$ and $a_{i,R}$ 
at the left and right cell interfaces of a primitive quantity $a_i$ defined on cell centers labeled by $i=0,..,N-1$.
The left and right states are defined on cell interfaces labeled by $a_{i\pm\frac{1}{2}}$.
Rather than assuming a constant value for a cell-averaged quantity within a given cell, 
the PPM scheme uses parabolas to represent cell averages within a given cell.

The \textit{enhanced} PPM reconstruction proceeds in three steps:
(i) Compute an approximation to $a$ at cell interfaces using a high-order interpolation polynomial,
(ii) limit the interpolated cell-interface values obtained in (i) to avoid oscillations near shocks and other discontinuities,
(iii) constrain the parabolic profile so that no new artificial maximum is created within one single cell.
The main difference to the original PPM scheme is in steps (i) and (ii). Both, the limiter and the constraining of the parabolic profiles
is more restrictive in the original PPM scheme, thus reducing the order of accuracy in cases where it is not necessary.

\paragraph{First Step: Interpolation}
We compute an approximation to $a$ at cell interfaces, which, 
assuming a uniform grid, is obtained via fourth order polynomial interpolation
\begin{equation} \label{eq:ppm4}
a_{i+\frac{1}{2}}=\frac{7}{12}(a_{i+1}+a_{i}) - \frac{1}{12}(a_{i-1} + a_{i+2})\,,
\end{equation}
using the cell center values of $a$ from neighboring cells. Ref.~\cite{colella:08}
also suggests to use a sixth-order polynomial. This, however, requires more 
ghost points. In our tests, we find no significant difference between fourth and sixth-order interpolation.
Hence, we stick to the fourth-order interpolant.

\paragraph{Second Step: Limiting}
We require that the values $a_{i+\frac{1}{2}}$ satisfy
\begin{equation} \label{eq:mon}
 \min(a_i, a_{i+1}) \leq a_{i+\frac{1}{2}} \leq \max(a_i, a_{i+1})\,,
\end{equation}
i.e., the interpolated value $a_{i+\frac{1}{2}}$ must lie between adjacent cell values \cite{colella:08}.
This is enforced by the following conditions.
If \eqref{eq:mon} is not satisfied, then we define the second derivatives,
\begin{eqnarray} \label{eq:ppm-derivs}
(D^2a)_{i+\frac{1}{2}} &:=& 3(a_{i} - 2a_{i+\frac{1}{2}} + a_{i+1})\,, \\ 
(D^2a)_{i+\frac{1}{2},L} &:=& (a_{i-1} - 2a_{i} + a_{i+1})\,, \\ 
(D^2a)_{i+\frac{1}{2},R} &:=& (a_{i} - 2a_{i+1} + a_{i+2})\,. 
\end{eqnarray}
If $(D^2a)_{i+\frac{1}{2}}$ and $(D^2a)_{i+\frac{1}{2},L,R}$ all have the same sign $s=\mathrm{sign}((D^2a)_{i+\frac{1}{2}})$, 
we further define
\begin{align}
(D^2a)_{i+\frac{1}{2},\mathrm{lim}} := s \min(& C|(D^2a)_{i+\frac{1}{2},L}|, \nonumber \\
                                             & C|(D^2a)_{i+\frac{1}{2},R}|, \nonumber \\
                                             & |(D^2a)_{i+\frac{1}{2}}|)\,.
\end{align}
where $C$ is a constant that we set according to \cite{colella:08} to $C=1.25$.
Otherwise, if one of the signs is different\footnote{For the specific internal energy $\epsilon$, we also set 
$(D^2a)_{i+\frac{1}{2},\mathrm{lim}}=0$, in cases when $(D^2a)_{i+\frac{1}{2},\mathrm{lim}} > \frac{1}{2}(a_i + a_{i+1})$.
This is different from the procedure in \cite{mccorquodale:11}, but is necessary at very strong contact discontinuities such as
the surface of a neutron star to prevent $\epsilon$ from becoming negative for equations of state that do not allow $\epsilon<0$. 
In practice, this additional limiter has no effect on the measured accuracy.}, we set $(D^2a)_{i+\frac{1}{2},\mathrm{lim}}=0$.
Then, we recompute \eqref{eq:ppm4} by
\begin{equation} \label{eq:monlim}
a_{i+\frac{1}{2}}=\frac{1}{2}(a_i + a_{i+1}) - \frac{1}{3}(D^2a)_{i+\frac{1}{2},\mathrm{lim}}\,.
\end{equation}

\paragraph{Third Step: Constrain Parabolic Profiles}
Here, we apply the refined procedure from \cite{mccorquodale:11}.
We begin by initializing left and right states according to the interpolated (and possibly limited) $a_{i+\frac{1}{2}}$ via
\begin{equation} \label{eq:leftright}
a_{i,R} = a_{i+1,L} = a_{i+\frac{1}{2}}\,,
\end{equation}
so that the Riemann problem is trivial initially.
The conditions below potentially alter $a_{i,R}$ and $a_{i+1,L}$, so that the
Riemann problem becomes non-trivial.

First, we check whether we are at a smooth local maximum.
A condition for local smooth maxima is given by
\begin{eqnarray} \label{eq:cond-max}
(a_{i,L}-a_i)(a_{i} - a_{i,R}) &\leq& 0\,, \qquad \mathrm{or} \nonumber \\
(a_{i-2}-a_i)(a_{i} - a_{i+2}) &\leq& 0\,.
\end{eqnarray}
If \eqref{eq:cond-max} holds, we compute, similar to \eqref{eq:ppm-derivs},
\begin{align}
(D^2a)_i     &= - 12 a_i + 6 (a_{i,L} + a_{i,R})\,, \nonumber \\
(D^2a)_{i,C} &= a_{i-1} - 2 a_{i} + a_{i+1}\,, \nonumber \\
(D^2a)_{i,L} &= a_{i-2} - 2 a_{i-1} + a_{i}\,, \nonumber \\
(D^2a)_{i,R} &= a_{i} - 2 a_{i+1} + a_{i+2}\,.
\end{align}
If $(D^2a)_{i,[C,L,R]}$ all have the same sign $s=\mathrm{sign}((D^2a)_i)$,
we compute
\begin{align}
(D^2a)_{i+\frac{1}{2},\mathrm{lim}} = s \min(& C|(D^2a)_{i+\frac{1}{2},L}|, \nonumber \\
                                             & C|(D^2a)_{i+\frac{1}{2},R}|, \nonumber \\
                                             & C|(D^2a)_{i+\frac{1}{2},C}|, \nonumber \\
                                             & |(D^2a)_{i+\frac{1}{2}}|)\,.
\end{align}
Otherwise, if one of the signs is different, we set $(D^2a)_{i+\frac{1}{2},\mathrm{lim}}=0$.
If 
\begin{equation}
|(D^2a)_i| \leq 10^{-12}\cdot\max(|a_{i-2}|, |a_{i-1}|, |a_{i}|, |a_{i+1}|, |a_{i+2}|)
\end{equation}
then we define and set $\rho_i\equiv0$.
Otherwise, we define
\begin{equation}
\rho_i \equiv \frac{(D^2a)_{i+\frac{1}{2},\mathrm{lim}}}{(D^2a)_i}\,.
\end{equation}
To avoid limiting at small oscillations induced by round-off errors, we do not apply any limiter if $\rho_i \geq 1-10^{-12}$. 
Otherwise, we compute the third derivative according to
\begin{equation}
(D^3a)_{i+\frac{1}{2}} = (D^2a)_{i+1,C} - (D^2a)_{i,C}\,.
\end{equation}
We set
\begin{align}
(D^3a)_i^\mathrm{min} = \min(& (D^3a)_{i-\frac{3}{2}},
                               (D^3a)_{i-\frac{1}{2}}, \nonumber \\
                             & (D^3a)_{i+\frac{1}{2}},
                               (D^3a)_{i+\frac{3}{2}})\,
\end{align}
and
\begin{align}
(D^3a)_i^\mathrm{max} = \max(& (D^3a)_{i-\frac{3}{2}},
                               (D^3a)_{i-\frac{1}{2}}, \nonumber \\
                             & (D^3a)_{i+\frac{1}{2}},
                               (D^3a)_{i+\frac{3}{2}})\,.
\end{align}
Then, we test if 
\begin{align} \label{eq:ppmC3}
C_3\cdot\max(|(D^3a)_i^\mathrm{max}|, & |(D^3a)_i^\mathrm{max}|) \nonumber \\
 & \leq (D^3a)_i^\mathrm{max} - (D^3a)_i^\mathrm{min}\,,
\end{align}
holds. In the expression above, $C_3=0.1$, according to Ref.~\cite{mccorquodale:11}.
If \eqref{eq:ppmC3} does not hold, a limiter is not applied.
Otherwise, we test the following conditions: (i) if $(a_{i,L}-a_i)(a_{i} - a_{i,R}) < 0$, we set
\begin{eqnarray} \label{eq:ppmlt}
a_{i,L} &=& a_i - \rho_i (a_{i}-a_{i,L})\,, \nonumber \\
a_{i,R} &=& a_i + \rho_i (a_{i,R}-a_{i})\,.
\end{eqnarray}
Otherwise, (ii) if $|a_{i}-a_{i,L}| \geq 2|a_{i,R}-a_i|$, we set
\begin{align}
a_{i,L} = a_i - 2(1-\rho_i)(a_{i,R}-a_i)-\rho_i(a_{i}-a_{i,L})\,\,
\end{align}
or (iii) if $|a_{i,R}-a_{i}| \geq  2|a_{i}-a_{i,L}|$, we set
\begin{align}
a_{i,R} = a_i + 2(1-\rho_i)(a_{i}-a_{i,L})+\rho_i(a_{i,R}-a_{i})\,.
\end{align}
In the conditions (i)-(iii) above, we introduce a special treatment for the specific internal energy $\epsilon$.
If $|a_{i}-a_{i,L}|<|a_i|$ or $|a_{i,R}-a_{i}|<|a_i|$,
we set $a_{i,L,R}=a_i$ instead of using the full expressions, respectively.
This is different from the original procedure of Ref.~\cite{mccorquodale:11}.
It essentially reduces the reconstruction of $\epsilon$ to first order in cases when the correction becomes larger
than the value of the reconstructed quantity itself. This is similar to the limiter step further above and
is necessary at very strong contact discontinuities such as the surface of a neutron star.
Without this additional limiter, $\epsilon$ may become ill-conditioned. 
This typically happens when $\epsilon$ is very small and the correction becomes larger than $\epsilon$ itself
potentially leading to negative $\epsilon$.
For some equations of state, $\epsilon < 0$ is ill-defined, causing the HLLE Riemann solver to
fail. 
In practice, this reduction does not affect the overall accuracy of the scheme.
We also note that this special treatment does not \textit{forbid} $\epsilon$ from becoming negative.

Finally, we recompute $a_{i,L}$ ($a_{i,R}$) according to
\begin{eqnarray}
a_{i,L(R)} &=& a_i + (a_{i,L(R)}-a_i) \frac{(D^2a)_{i+\frac{1}{2},\mathrm{lim}}}{(D^2a)_i}\,.  \nonumber \\
\end{eqnarray}
In case the denominator becomes zero in the expression above, we set the last term to zero.

Finally, if \eqref{eq:cond-max} does not hold, we test whether
$|a_{i,R(L)}-a_i| \geq 2|a_{i,L(R)}-a_i|$ holds. In that case, we set
\begin{equation}
a_{i,R(L)} = a_i - 2(a_{i,L(R)} - a_i)\,
\end{equation}
for either $a_{i,L}$ or $a_{i,R}$, respectively.
In the case of reconstructing the specific internal energy $\epsilon$, if $|a_i - 2a_{i,L(R)}| > a_i|$, we
simply set $a_{i,R(L)} = a_i$. This is for the same reason that has been mentioned above already.

After having obtained $a_{i,L}$ and $a_{i,R}$, we apply the ``standard`` flattening procedure discussed in the Appendix of \cite{colella:84}.
This completes the enhanced PPM scheme applied in our code.
Note that Ref.~\cite{mccorquodale:11} (in contrast to \cite{colella:08}) suggests to skip the second step. In our experiments with an 
excited neutron star and a collapsing stellar core, 
however, we find that when skipping this step, the scheme becomes too dissipative.

The enhanced PPM scheme requires four ghost points.
For efficiency reasons, it may be desirable to use only three ghost points, 
since less memory and interprocessor communication is required.
In order to reduce the number of required stencil points to three, 
we use fourth-order polynomial interpolation \eqref{eq:ppm4} instead of sixth-order interpolation \cite{colella:08} in the first step, and
we skip the check \eqref{eq:ppmC3} involving the third derivatives $(D^3a)_i$.
We also use a modified flattening scheme which allows us to use only three ghost points.
This modified flattening scheme is the same as the one presented in the Appendix of \cite{colella:84},
but we drop the maximum in Eqn.~(A.2) of \cite{colella:84}, and directly use $f_i=\tilde{f}_j$.
In our tests, we have found only small differences between the four- and three-point scheme.

In Fig.~\ref{fig:tov-constr-1D}, we show the effect of ePPM compared with oPPM on the Hamiltonian constraint $H$ along the $x$-axis
for the example of an isolated TOV star (Sec.~\ref{sec:tov}) on cell-centered and vertex-centered AMR grids.
Clearly, ePPM results in a significantly lower error compared to oPPM on vertex-centered, and especially on cell-centered AMR grids.

\section{Atmosphere Treatment}
\label{sec:atmo}

In vacuum, obviously,
the equations describing the fluid dynamics break down.
When simulating isolated neutron stars or binary neutron star mergers, a large
fraction of the simulation domain is physically vacuum.
At the surface of the
fluid bodies where a sharp transition to vacuum occurs, 
the Riemann solver breaks down.

As a simple solution to this problem, 
we keep a very low and constant density fluid (the \emph{atmosphere}) in the cells
which would be vacuum otherwise. We also keep track of where the evolution
of the fluid variables fails to produce a physical state and reset these cells
to atmosphere. Typically, there are few such cells, which cluster around the
surface of the star.
The atmosphere density $\rho_{\rm atmo}$ is usually chosen to be $8$ to $10$ orders of magnitudes lower than the 
central density of the fluid body. This ensures that the atmosphere does
not contribute noticeably to the total rest mass and energy in the simulation.

Whether a given fluid cell is set to atmosphere values is decided depending
on the local fluid density. If it drops below atmosphere density
$\rho_{\rm atmo}$, the cell is
set to atmosphere density with zero fluid velocity. 

More specifically, we proceed in the following way.
\begin{enumerate}
 \item 
During each intermediate time step, we set an ``atmosphere'' flag in an atmosphere mask $M_A$ if
$\tau + \Delta t R_\tau < 0$ or $D + \Delta t R_D < 0$, where $R_\tau$ and
$R_D$ are the right-hand sides of the $\tau$ and $D$ equations \eqref{eq:conservation_equations_gr}, respectively
and $\Delta t$ is the temporal timestep size. In addition to
setting the atmosphere flag, we also set all fluid right-hand sides for that
cell to zero, in effect freezing the further evolution of this cell.
In that case, we also skip conversion of conserved to primitive variables of that cell.
 \item
After a full time step, we set all variables of those cells to atmosphere values
that are flagged as atmosphere.
 \item
Finally, we clear the atmosphere mask $M_A$.
\end{enumerate}

Furthermore, we perform the following operations involving atmosphere checks:
\begin{enumerate}
 \item
After reconstruction, we check whether the reconstructed primitive density is below
atmosphere density. If this is the case, we enforce first order reconstruction, i.e.~we 
set left and right cell face $a_{i,L}=a_{i,R}=a_i$ to the cell average $a_i$ for all primitive variables.
 \item 
At the end of conservative to primitive conversion, we check whether the new set of primitive variables
is below atmosphere level for a given cell. If this is the case, we reset that cell to atmosphere level.
\end{enumerate}
In the two cases above, the atmosphere mask is not set.

To limit high-frequency noise in cells slightly above atmosphere level, we 
set cells to atmosphere value if they are within a given tolerance $\delta$ above atmosphere density, i.e.~we
test whether
\begin{equation}
\rho \leq \rho_{\rm atmo} \left(1 + \delta \right)\,.
\end{equation}
In the cases considered here, we set $\delta=0.001$.

The particular treatment of vacuum regions by enforcing a low density atmosphere is 
not ideal and has several drawbacks. If a cell is forced to be not lower than 
a particular minimum density,
small amounts of baryonic mass can be created or removed. This breaks the strictly conservative
nature of our hydrodynamics scheme and can thus lead to small errors. 
As noted in \cite{toro:99}, introducing an artificial atmosphere may also change the local wave structure of the solution.
An artificial low density atmosphere can be avoided by modifying the Riemann solver at those cells adjacent 
to vacuum cells \cite{toro:99}.
In practice, however, 
if the atmosphere level is sufficiently low, the negative influence on the fluid evolution can be neglected.

\section{Scheduling of Ghost-Zone Synchronization}
\label{sec:scheduling}

\begin{table*}[t]
\caption{Required synchronizations for each quantity for the three
synchronization operations. See text for more details.}
\label{tab:synced-quantities}
\begin{ruledtabular}
\begin{tabular}{llll}
Operation              & inter-processor/inter-patch sync.  & prolongation
(buffer zone) & prolongation (regridding) \\
\hline
Quantities             & $\left\{\phi, \tilde\gamma_{ij}, K, \tilde A_{ij},
\tilde\Gamma^i, \alpha, \beta^i, B^i \right\}$ & $\left\{\phi,
\tilde\gamma_{ij}, K, \tilde A_{ij},
\tilde\Gamma^i, \alpha, \beta^i, B^i \right\}$ & $\left\{\phi,
\tilde\gamma_{ij}, K, \tilde A_{ij},
\tilde\Gamma^i, \alpha, \beta^i, B^i \right\}$ \\
 & $\left\{ D, \tau, S_i, Y_e^{\rm con}\right\}$ & $\left\{ D, \tau, S_i, Y_e^{\rm con}, \rho, \epsilon,
v^i, T\right\}$ & $\left\{ D, \tau, S_i, Y_e^{\rm con}, \rho, \epsilon, v^i, Y_e, T, s \right\}$ \\
 & $\left\{M_A\right\}$ & $\left\{M_A\right\}$ & 
\end{tabular}
\end{ruledtabular}
\end{table*}

We find that excessive inter-processor and inter-patch synchronization of
ghost zone information can lead to significant performance drawbacks,
especially on large numbers of processing units ($\gtrsim 1000$).
We have thus optimized our ghost-zone update pattern and reduced the number of
necessary synchronization calls.

We distinguish between three different synchronization update operations:
(i) inter-processor and inter-patch synchronizations performed after each
intermediate time step, and (ii) AMR buffer-zone prolongation performed after
each full time step, and (iii) AMR prolongation after regridding (see
\cite{Schnetter-etal-03b} on the latter two cases for details).

We distinguish between two sets of variables. One set is comprised of the
spacetime variables $\left\{\phi, \tilde\gamma_{ij}, K, \tilde A_{ij},
\tilde\Gamma^i, \alpha, \beta^i, B^i \right\}$ describing the curvature
evolution and gauge (Sec.~\ref{sec:curv}), and the other set is comprised of
variables $\left\{ D, \tau, S_i, \rho, \epsilon, v^i, \tilde v^i, P, W,
Y_e, Y_e^{\rm con}, T, s\right\}$
describing the evolution of the fluid elements (Sec.~\ref{sec:hydro}).
The primitive electron fraction $Y_e$, the conserved electron fraction $Y_e^{\rm con}$,
the temperature $T$, and the specific entropy $s$ are only necessary when 
microphysical finite-temperature equations of state are used.
In addition to these two sets of variables, we
also need to consider the ''pseudo-evolved'' atmosphere mask $M_A$ described in
Appendix~\ref{sec:atmo}.
Thus, in total, we have $24+19+1=44$ evolved components that potentially need to
be synchronized.

As described in Sec.~\ref{sec:curv}, the update terms for the spacetime variables
are computed via finite differences and thus require ghost-zone synchronization
after each intermediate step. In addition, they are also subject to
AMR buffer-zone synchronization via prolongation to obtain valid ghost data from
the coarse grid in the buffer zone.

As described in Sec.~\ref{sec:hydro}, the update terms for the evolved conserved
fluid variables are computed from reconstructed primitive variables at cell
interfaces and thus also require ghost and buffer-zone synchronization in the
same way as the spacetime variables.
The conservative to primitive conversion requires the conserved variables and valid initial
guesses for the primitive variables. Typically, these initial guesses are taken
from the last valid time step on the given cell.
Since cells located in the buffer zone become invalid during time
integration substeps and need to be refilled via buffer-zone prolongation after
a full time step, we also need to synchronize those primitive variables that
are used as initial guesses in the conservative to primitive conversion.
In our case, these are $\rho$, $\epsilon$, $v^i$, and $T$. 
Note that we do not need
to synchronize the \textit{global} primitive velocity $\tilde{v}^i$ since it is
later obtained from a coordinate transformation.

Furthermore, we need to update the atmosphere mask $M_A$ in each
intermediate step via inter-processor and inter-patch synchronization, and 
also via buffer-zone prolongation after each full time step.
This is necessary because the atmosphere mask is only set on cells of the
evolved grid (i.e.~all cells excluding ghost zones).
Operations like conservative to primitive conversion, which depend on the
atmosphere mask, are performed on the entire grid, including ghost zones.
Thus they require a synchronized atmosphere mask.
In addition, the synchronization order of the atmosphere mask is important
during buffer-zone prolongation. 
Before prolongating all other required quantities, we first prolongate the
atmosphere mask.
Immediately afterwards, cells are set to atmosphere values according to the
atmosphere mask. The atmosphere mask itself is cleared (also see
Appendix~\ref{sec:atmo}). This completes the evolution step. 
and all
variables are in their final state for the given evolution step.
Now, it is possible to prolongate also all remaining variables as discussed
above.

Finally, we need to synchronize all variables (except for the atmosphere mask\footnote{The atmosphere mask does not need to be synchronized
because it is not valid during regridding. As explained in Appendix~\ref{sec:atmo}, it is only valid during time integration 
substeps where regridding is not allowed. We clear it in any new grid region.})
via prolongation after regridding.
A subsequent conservative to primitive
conversion ensures that the two conservative and primitive sets of
hydrodynamical variables are consistent with each other.
Even though regridding requires all variables to be synchronized and is thus
rather expensive, fortunately, 
this operation usually does occur only infrequently, say every $64$ iterations, when
moving the fine grids during binary neutron star evolution, and only very
infrequently, say every
couple of thousands of iterations, when adding additional refinement levels
during stellar collapse or neutron star collapse.

In Table~\ref{tab:synced-quantities}, we explicitly list all quantities that
must be updated during one of the three possible synchronization operations.
The most frequent operation, inter-processor and
inter-patch synchronization require the least number of variables to be
updated. Prolongation during regridding, which is the least frequent
synchronization operation, requires the full set of variables (except for the
atmosphere mask $M_A$ which is invalid outside of a full time integration
step).
Also note that the global primitive velocity $\tilde{v}^i$ never needs
to be synchronized because it is obtained from the local primitive velocity
$v^i$ via a coordinate transformation after each synchronization step.
Similarly, the Lorentz factor $W$ and the pressure $P$ are never synchronized since
they are computed in the conservative to primitive routine, which is exectued after
each synchronization operation.

\section{Volume Integration}
\label{sec:vol-int}

Several quantities in our code require volume integration over the entire numerical grid.
For instance, the total baryonic mass is given by
\begin{equation} \label{eq:baryonic-mass}
M_B = \int d^3x\, D(x,y,z)\,
\end{equation}
in terms of the conserved density $D$ in the Cartesian tensor basis\footnote{We remark 
that our code uses the conserved density $D$ in the \textit{local} coordinate basis. Since $D$ is a \textit{densitized} scalar,
\eqref{eq:baryonic-mass} requires an additional Jacobian factor to transform $D$ to the global basis. 
For simplicity of discussion, we omit this here and temporarily assume that $D$ is given in the global basis.}.
In Cartesian coordinates, this can be approximated numerically by
\begin{equation} \label{eq:vol-int-ex}
M_B = \Delta x \Delta y \Delta z \sum_{ijk} D_{ijk}\,,
\end{equation}
where $\Delta x$, $\Delta y$, and $\Delta z$ is the grid spacing and the indices $i,j,k$, in this context, denote grid indices.
In generic curvi-linear coordinates, the global grid spacing is not constant anymore.
In order to compute the volume integral with respect to global coordinates, we make use of the local volume element
\begin{equation}
d^3u = \Delta u \Delta v \Delta w\,,
\end{equation}
where $\Delta u$, $\Delta v$, and $\Delta w$ denotes the local uniform grid spacing,
and we make use of the relation between 
local volume form $d^3u$ and global volume form
\begin{equation} \label{eq:vol-form}
d^3x = d^3u \left|\det{\frac{\p x^i}{\p u^j}}\right|\,.
\end{equation}
The volume form $d^3x$ is introduced as an additional grid function which can be computed once the coordinates and grids are set up.

Next, we need to take into account the non-trivial overlap between neighboring grid patches.
For instance, the spherical boundary of the spherical outer grid (Fig.~\ref{fig:7patch-system}) cuts through cells of the central Cartesian patch, i.e.,
parts of the Cartesian cells reach into the nominal domain of the spherical grid.
Consequently, the volume associated with each of those cells is only a fraction of the volume of the entire cell.
In practice, we set up a weight mask $\mathcal{W}_{ijk}$ defining the contribution of each cell to the total volume. 
A cell fully contained on the nominal grid has a weight of $\mathcal{W}_{ijk}=1$. 
Correspondingly, a cell completely outside of the nominal grid has a weight of $\mathcal{W}_{ijk}=0$.
Cells, whose vertices are not all on the nominal grid, carry a weight $0 < \mathcal{W}_{ijk} < 1$.
In that case, we determine the weight by using 3D Monte-Carlo
integration~\cite[e.g.,][]{numrep} of the volume fraction of the
overlapping regions. 
The weights need to be calculated only once after the grids
have been setup and therefore the cost of Monte Carlo
volume integration is negligible compared to the total cost of 
the simulation. 

For simplicity, we absorb the weight mask into the volume form \eqref{eq:vol-form}, i.e., we effectively store
\begin{equation} \label{eq:vol-form2}
(d^3x)_{ijk} = \Delta u \Delta v \Delta w \left|\det{\frac{\p x^l}{\p u^m}}\right|_{ijk}\, \mathcal{W}_{ijk}\,,
\end{equation}
where the indices $i,j,k$ label grid points and are not subject to the Einstein sum convention.
Similar to the Jacobians introduced for computing global Cartesian derivatives from local finite differences, 
any volume integration needs to take into account \eqref{eq:vol-form2}.
For instance \eqref{eq:vol-int-ex} takes the form
\begin{equation}
M_B = \sum_{ijk} D_{ijk} (d^3x)_{ijk}\,.
\end{equation}

\section{Influence of the Outer Boundary}
\label{sec:BC}

\begin{figure}
  \includegraphics[width=1.\linewidth]{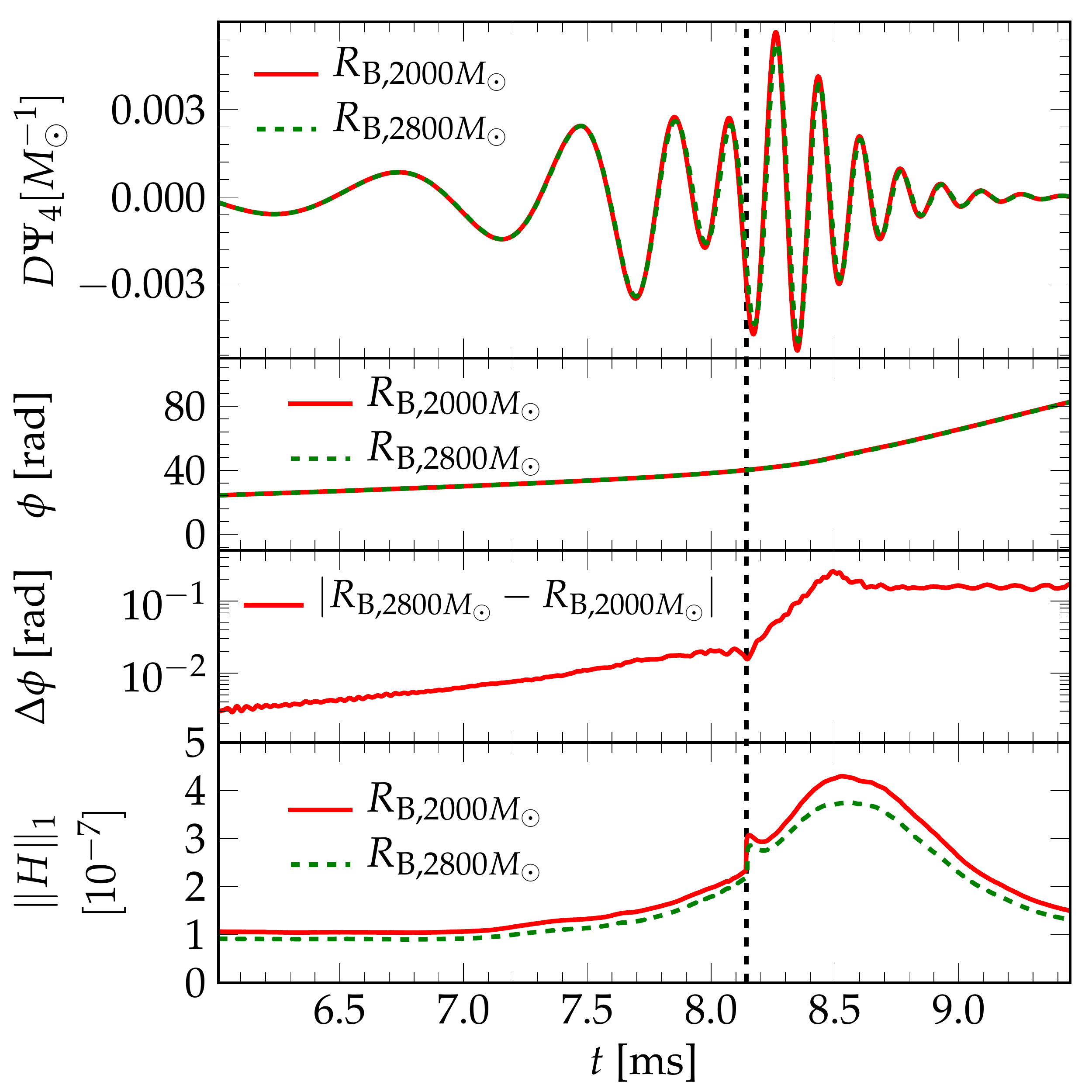}
  \caption{Binary neutron stars: influence of the outer boundary on the accuracy of the wave extraction and evolution.
           The upper panel shows the ``+'' polarization of the Weyl scalar $D \Psi_4$ extracted via CCE for 
           the two setups with different outer boundary locations. At time $t\sim7.5$ ms, when the outer boundary in the setup with $R_{\rm B}=2000M_\odot$
           comes in causal contact with the interior evolution, differences start to become visible for the $R_{\rm B}=2000M_\odot$ setup: the amplitude of $D \Psi_4$ deviates by $\sim 7\%$,
           the phase $\phi$ deviates by $\sim 0.2\,\rm{rad}$, and the $L_1$-norm of the Hamiltonian constraint $\norm{H}_1$ is larger by $\sim15\%$.}
           \label{fig:bns-boundary}
\end{figure}

All GR binary neutron star merger simulations to date employ grids which are too small to allow for causally disconnected
outer boundaries. Since no constraint preserving boundary conditions are known for the BSSN evolution system,
the simulations may be affected by incoming constraint violations.
Thus, it is interesting to investigate the influence of the outer boundary condition on the interior evolution and extracted GWs
of the binary neutron star merger problem considered in Sec.~\ref{sec:bns} when the boundary is \textit{not}
causally disconnected. 

We compare a simulation with outer boundary at $R_{\rm B}=2000M_\odot$ to the simulations in Sec.~\ref{sec:bns}, which use
an outer boundary at $R_{\rm B}=2800M_\odot$.
The setup with $R_{\rm B}=2000M_\odot$ has an outer boundary wich is in causal contact with the interior evolution and the wave-extraction region
during the merger and ring-down phases.
All simulations impose an approximate and non-constraint preserving radiative boundary condition (e.g.~\cite{reisswig:11ccwave}).
We focus on baseline resolution $r1$. We expect the simulations to be very similar at least up to the point when the constraint violations
from the outer boundary reach the wave-extraction region which happens at $t\sim7.5\,\rm{ms}$.

In Fig.~\ref{fig:bns-boundary}, we show the ``+'' polarization of the 
leading order harmonic $(\ell,m)=(2,2)$ mode of the complex Weyl scalar $D \Psi_4$ computed via CCE.
The difference in amplitude are on the order of $\sim7\%$. The effects on the phase are more subtle and not clearly visible 
from a simple inspection of the waveform itself. Therefore, in the two panels below, we plot the phase $\phi$ of the $(\ell,m)=(2,2)$ mode.
The maximum dephasing in the two simulations is $\sim 0.2\,\rm{rad}$ and thus, the systematic dephasing due to the influence from the outer boundary
is only slightly below the one due to the convergent numerical error.
This indicates that when the resolution is further increased, 
the error due to constraint violations from the outer boundary cannot be neglected anymore. 

In the same, figure, we also show the $L_1$-norm\footnote{We show here the $L_1$-norm since it does not require a log-scaling. 
Thus, subtle differences are better visible. We note, however, that the $L_2$-norm $\norm{H}_2$ shows similar differences.} of the Hamiltonian constraint 
$\norm{H}_1$ for the two simulations.
We find that the difference of $\sim15\%$ is smaller than the difference of $\sim25\%$ between the numerical resolutions $r1$ and $r2$,
but not so small that it can be ignored. 

Finally, we also compare mass and spin of the merger remnant, and find 
that the differences are on the order of the
numerical error between resolutions $r1$ and $r2$. 

Overall, we find that causally disconnected outer boundaries have a non-negligible impact on the accuracy of the binary neutron star simulation
presented in Sec.~\ref{sec:bns}. It is thus likely that longer inspiral simulations are even more strongly affected.

\bibliography{bibliography/bh_formation_references,bibliography/gw_references,bibliography/gr_references,bibliography/sn_theory_references,bibliography/grb_references,bibliography/stellarevolution_references,bibliography/methods_references,bibliography/gw_data_analysis_references,bibliography/NSNS_NSBH_references,bibliography/misc_references,bibliography/cs_hpc_references,bibliography/numrel_references,bibliography/publications-schnetter,bibliography/accretion_disk_references,bibliography/ns_references}

\begin{thebibliography}{139}
\expandafter\ifx\csname natexlab\endcsname\relax\def\natexlab#1{#1}\fi
\expandafter\ifx\csname bibnamefont\endcsname\relax
  \def\bibnamefont#1{#1}\fi
\expandafter\ifx\csname bibfnamefont\endcsname\relax
  \def\bibfnamefont#1{#1}\fi
\expandafter\ifx\csname citenamefont\endcsname\relax
  \def\citenamefont#1{#1}\fi
\expandafter\ifx\csname url\endcsname\relax
  \def\url#1{\texttt{#1}}\fi
\expandafter\ifx\csname urlprefix\endcsname\relax\def\urlprefix{URL }\fi
\providecommand{\bibinfo}[2]{#2}
\providecommand{\eprint}[2][]{\url{#2}}

\bibitem[{\citenamefont{{Ott} et~al.}(2012{\natexlab{a}})\citenamefont{{Ott},
  {Abdikamalov}, {O'Connor}, {Reisswig}, {Haas}, {Kalmus}, {Drasco}, {Burrows},
  and {Schnetter}}}]{ott:12a}
\bibinfo{author}{\bibfnamefont{C.~D.} \bibnamefont{{Ott}}},
  \bibinfo{author}{\bibfnamefont{E.}~\bibnamefont{{Abdikamalov}}},
  \bibinfo{author}{\bibfnamefont{E.}~\bibnamefont{{O'Connor}}},
  \bibinfo{author}{\bibfnamefont{C.}~\bibnamefont{{Reisswig}}},
  \bibinfo{author}{\bibfnamefont{R.}~\bibnamefont{{Haas}}},
  \bibinfo{author}{\bibfnamefont{P.}~\bibnamefont{{Kalmus}}},
  \bibinfo{author}{\bibfnamefont{S.}~\bibnamefont{{Drasco}}},
  \bibinfo{author}{\bibfnamefont{A.}~\bibnamefont{{Burrows}}},
  \bibnamefont{and}
  \bibinfo{author}{\bibfnamefont{E.}~\bibnamefont{{Schnetter}}},
  \bibinfo{journal}{\prd} \textbf{\bibinfo{volume}{86}}, \bibinfo{eid}{024026}
  (\bibinfo{year}{2012}{\natexlab{a}}).

\bibitem[{\citenamefont{{Ott} et~al.}(2011)\citenamefont{{Ott}, {Reisswig},
  {Schnetter}, {O'Connor}, {Sperhake}, {L{\"o}ffler}, {Diener}, {Abdikamalov},
  {Hawke}, and {Burrows}}}]{ott:11a}
\bibinfo{author}{\bibfnamefont{C.~D.} \bibnamefont{{Ott}}},
  \bibinfo{author}{\bibfnamefont{C.}~\bibnamefont{{Reisswig}}},
  \bibinfo{author}{\bibfnamefont{E.}~\bibnamefont{{Schnetter}}},
  \bibinfo{author}{\bibfnamefont{E.}~\bibnamefont{{O'Connor}}},
  \bibinfo{author}{\bibfnamefont{U.}~\bibnamefont{{Sperhake}}},
  \bibinfo{author}{\bibfnamefont{F.}~\bibnamefont{{L{\"o}ffler}}},
  \bibinfo{author}{\bibfnamefont{P.}~\bibnamefont{{Diener}}},
  \bibinfo{author}{\bibfnamefont{E.}~\bibnamefont{{Abdikamalov}}},
  \bibinfo{author}{\bibfnamefont{I.}~\bibnamefont{{Hawke}}}, \bibnamefont{and}
  \bibinfo{author}{\bibfnamefont{A.}~\bibnamefont{{Burrows}}},
  \bibinfo{journal}{Phys. Rev. Lett.} \textbf{\bibinfo{volume}{106}},
  \bibinfo{pages}{161103} (\bibinfo{year}{2011}).

\bibitem[{\citenamefont{{Kuroda} et~al.}(2012)\citenamefont{{Kuroda}, {Kotake},
  and {Takiwaki}}}]{kuroda:12}
\bibinfo{author}{\bibfnamefont{T.}~\bibnamefont{{Kuroda}}},
  \bibinfo{author}{\bibfnamefont{K.}~\bibnamefont{{Kotake}}}, \bibnamefont{and}
  \bibinfo{author}{\bibfnamefont{T.}~\bibnamefont{{Takiwaki}}},
  \bibinfo{journal}{\apj} \textbf{\bibinfo{volume}{755}}, \bibinfo{eid}{11}
  (\bibinfo{year}{2012}).

\bibitem[{\citenamefont{{Kuroda} and {Umeda}}(2010)}]{kuroda:10}
\bibinfo{author}{\bibfnamefont{T.}~\bibnamefont{{Kuroda}}} \bibnamefont{and}
  \bibinfo{author}{\bibfnamefont{H.}~\bibnamefont{{Umeda}}},
  \bibinfo{journal}{\apjs} \textbf{\bibinfo{volume}{191}}, \bibinfo{pages}{439}
  (\bibinfo{year}{2010}).

\bibitem[{\citenamefont{{Baiotti} et~al.}(2007)\citenamefont{{Baiotti},
  {Hawke}, and {Rezzolla}}}]{baiotti:07b}
\bibinfo{author}{\bibfnamefont{L.}~\bibnamefont{{Baiotti}}},
  \bibinfo{author}{\bibfnamefont{I.}~\bibnamefont{{Hawke}}}, \bibnamefont{and}
  \bibinfo{author}{\bibfnamefont{L.}~\bibnamefont{{Rezzolla}}},
  \bibinfo{journal}{Class. Quantum Grav.} \textbf{\bibinfo{volume}{24}},
  \bibinfo{pages}{187} (\bibinfo{year}{2007}).

\bibitem[{\citenamefont{{Bernuzzi}
  et~al.}(2012{\natexlab{a}})\citenamefont{{Bernuzzi}, {Nagar}, {Thierfelder},
  and {Bruegmann}}}]{bernuzzi:12a}
\bibinfo{author}{\bibfnamefont{S.}~\bibnamefont{{Bernuzzi}}},
  \bibinfo{author}{\bibfnamefont{A.}~\bibnamefont{{Nagar}}},
  \bibinfo{author}{\bibfnamefont{M.}~\bibnamefont{{Thierfelder}}},
  \bibnamefont{and}
  \bibinfo{author}{\bibfnamefont{B.}~\bibnamefont{{Bruegmann}}},
  \bibinfo{journal}{arXiv:1205.3403}  (\bibinfo{year}{2012}{\natexlab{a}}).

\bibitem[{\citenamefont{{Baiotti} et~al.}(2011)\citenamefont{{Baiotti},
  {Damour}, {Giacomazzo}, {Nagar}, and {Rezzolla}}}]{baiotti:11}
\bibinfo{author}{\bibfnamefont{L.}~\bibnamefont{{Baiotti}}},
  \bibinfo{author}{\bibfnamefont{T.}~\bibnamefont{{Damour}}},
  \bibinfo{author}{\bibfnamefont{B.}~\bibnamefont{{Giacomazzo}}},
  \bibinfo{author}{\bibfnamefont{A.}~\bibnamefont{{Nagar}}}, \bibnamefont{and}
  \bibinfo{author}{\bibfnamefont{L.}~\bibnamefont{{Rezzolla}}},
  \bibinfo{journal}{\prd} \textbf{\bibinfo{volume}{84}},
  \bibinfo{pages}{024017} (\bibinfo{year}{2011}).

\bibitem[{\citenamefont{{Giacomazzo} et~al.}(2011)\citenamefont{{Giacomazzo},
  {Rezzolla}, and {Baiotti}}}]{giacomazzo:11}
\bibinfo{author}{\bibfnamefont{B.}~\bibnamefont{{Giacomazzo}}},
  \bibinfo{author}{\bibfnamefont{L.}~\bibnamefont{{Rezzolla}}},
  \bibnamefont{and}
  \bibinfo{author}{\bibfnamefont{L.}~\bibnamefont{{Baiotti}}},
  \bibinfo{journal}{\prd} \textbf{\bibinfo{volume}{83}},
  \bibinfo{pages}{044014} (\bibinfo{year}{2011}).

\bibitem[{\citenamefont{{Rezzolla} et~al.}(2011)\citenamefont{{Rezzolla},
  {Giacomazzo}, {Baiotti}, {Granot}, {Kouveliotou}, and {Aloy}}}]{rezzolla:11}
\bibinfo{author}{\bibfnamefont{L.}~\bibnamefont{{Rezzolla}}},
  \bibinfo{author}{\bibfnamefont{B.}~\bibnamefont{{Giacomazzo}}},
  \bibinfo{author}{\bibfnamefont{L.}~\bibnamefont{{Baiotti}}},
  \bibinfo{author}{\bibfnamefont{J.}~\bibnamefont{{Granot}}},
  \bibinfo{author}{\bibfnamefont{C.}~\bibnamefont{{Kouveliotou}}},
  \bibnamefont{and} \bibinfo{author}{\bibfnamefont{M.~A.}
  \bibnamefont{{Aloy}}}, \bibinfo{journal}{\apjl}
  \textbf{\bibinfo{volume}{732}}, \bibinfo{pages}{L6} (\bibinfo{year}{2011}).

\bibitem[{\citenamefont{{Etienne} et~al.}(2012)\citenamefont{{Etienne}, {Liu},
  {Paschalidis}, and {Shapiro}}}]{etienne:12}
\bibinfo{author}{\bibfnamefont{Z.~B.} \bibnamefont{{Etienne}}},
  \bibinfo{author}{\bibfnamefont{Y.~T.} \bibnamefont{{Liu}}},
  \bibinfo{author}{\bibfnamefont{V.}~\bibnamefont{{Paschalidis}}},
  \bibnamefont{and} \bibinfo{author}{\bibfnamefont{S.~L.}
  \bibnamefont{{Shapiro}}}, \bibinfo{journal}{\prd}
  \textbf{\bibinfo{volume}{85}}, \bibinfo{pages}{064029}
  (\bibinfo{year}{2012}).

\bibitem[{\citenamefont{{Kiuchi} et~al.}(2012)\citenamefont{{Kiuchi},
  {Sekiguchi}, {Kyutoku}, and {Shibata}}}]{kiuchi:12}
\bibinfo{author}{\bibfnamefont{K.}~\bibnamefont{{Kiuchi}}},
  \bibinfo{author}{\bibfnamefont{Y.}~\bibnamefont{{Sekiguchi}}},
  \bibinfo{author}{\bibfnamefont{K.}~\bibnamefont{{Kyutoku}}},
  \bibnamefont{and}
  \bibinfo{author}{\bibfnamefont{M.}~\bibnamefont{{Shibata}}},
  \bibinfo{journal}{\cqg} \textbf{\bibinfo{volume}{29}},
  \bibinfo{pages}{124003} (\bibinfo{year}{2012}).

\bibitem[{\citenamefont{{East} and {Pretorius}}(2012)}]{east:12}
\bibinfo{author}{\bibfnamefont{W.~E.} \bibnamefont{{East}}} \bibnamefont{and}
  \bibinfo{author}{\bibfnamefont{F.}~\bibnamefont{{Pretorius}}},
  \bibinfo{journal}{\apjl} \textbf{\bibinfo{volume}{760}}, \bibinfo{pages}{L4}
  (\bibinfo{year}{2012}).

\bibitem[{\citenamefont{{Anderson} et~al.}(2008)\citenamefont{{Anderson},
  {Hirschmann}, {Lehner}, {Liebling}, {Motl}, {Neilsen}, {Palenzuela}, and
  {Tohline}}}]{anderson:08b}
\bibinfo{author}{\bibfnamefont{M.}~\bibnamefont{{Anderson}}},
  \bibinfo{author}{\bibfnamefont{E.~W.} \bibnamefont{{Hirschmann}}},
  \bibinfo{author}{\bibfnamefont{L.}~\bibnamefont{{Lehner}}},
  \bibinfo{author}{\bibfnamefont{S.~L.} \bibnamefont{{Liebling}}},
  \bibinfo{author}{\bibfnamefont{P.~M.} \bibnamefont{{Motl}}},
  \bibinfo{author}{\bibfnamefont{D.}~\bibnamefont{{Neilsen}}},
  \bibinfo{author}{\bibfnamefont{C.}~\bibnamefont{{Palenzuela}}},
  \bibnamefont{and} \bibinfo{author}{\bibfnamefont{J.~E.}
  \bibnamefont{{Tohline}}}, \bibinfo{journal}{\prl}
  \textbf{\bibinfo{volume}{100}}, \bibinfo{pages}{191101}
  (\bibinfo{year}{2008}).

\bibitem[{\citenamefont{{Faber} and {Rasio}}(2012)}]{faber:12}
\bibinfo{author}{\bibfnamefont{J.~A.} \bibnamefont{{Faber}}} \bibnamefont{and}
  \bibinfo{author}{\bibfnamefont{F.~A.} \bibnamefont{{Rasio}}},
  \bibinfo{journal}{Liv. Rev. Rel.} \textbf{\bibinfo{volume}{15}},
  \bibinfo{pages}{8} (\bibinfo{year}{2012}).

\bibitem[{\citenamefont{{Ott} et~al.}(2007{\natexlab{a}})\citenamefont{{Ott},
  {Dimmelmeier}, {Marek}, {Janka}, {Hawke}, {Zink}, and
  {Schnetter}}}]{ott:07prl}
\bibinfo{author}{\bibfnamefont{C.~D.} \bibnamefont{{Ott}}},
  \bibinfo{author}{\bibfnamefont{H.}~\bibnamefont{{Dimmelmeier}}},
  \bibinfo{author}{\bibfnamefont{A.}~\bibnamefont{{Marek}}},
  \bibinfo{author}{\bibfnamefont{H.-T.} \bibnamefont{{Janka}}},
  \bibinfo{author}{\bibfnamefont{I.}~\bibnamefont{{Hawke}}},
  \bibinfo{author}{\bibfnamefont{B.}~\bibnamefont{{Zink}}}, \bibnamefont{and}
  \bibinfo{author}{\bibfnamefont{E.}~\bibnamefont{{Schnetter}}},
  \bibinfo{journal}{\prl} \textbf{\bibinfo{volume}{98}},
  \bibinfo{pages}{261101} (\bibinfo{year}{2007}{\natexlab{a}}).

\bibitem[{\citenamefont{{Ott} et~al.}(2007{\natexlab{b}})\citenamefont{{Ott},
  {Dimmelmeier}, {Marek}, {Janka}, {Zink}, {Hawke}, and
  {Schnetter}}}]{ott:07cqg}
\bibinfo{author}{\bibfnamefont{C.~D.} \bibnamefont{{Ott}}},
  \bibinfo{author}{\bibfnamefont{H.}~\bibnamefont{{Dimmelmeier}}},
  \bibinfo{author}{\bibfnamefont{A.}~\bibnamefont{{Marek}}},
  \bibinfo{author}{\bibfnamefont{H.-T.} \bibnamefont{{Janka}}},
  \bibinfo{author}{\bibfnamefont{B.}~\bibnamefont{{Zink}}},
  \bibinfo{author}{\bibfnamefont{I.}~\bibnamefont{{Hawke}}}, \bibnamefont{and}
  \bibinfo{author}{\bibfnamefont{E.}~\bibnamefont{{Schnetter}}},
  \bibinfo{journal}{Class. Quantum Grav.} \textbf{\bibinfo{volume}{24}},
  \bibinfo{pages}{139} (\bibinfo{year}{2007}{\natexlab{b}}).

\bibitem[{\citenamefont{{Scheidegger} et~al.}(2008)\citenamefont{{Scheidegger},
  {Fischer}, {Whitehouse}, and {Liebend{\"o}rfer}}}]{scheidegger:08}
\bibinfo{author}{\bibfnamefont{S.}~\bibnamefont{{Scheidegger}}},
  \bibinfo{author}{\bibfnamefont{T.}~\bibnamefont{{Fischer}}},
  \bibinfo{author}{\bibfnamefont{S.~C.} \bibnamefont{{Whitehouse}}},
  \bibnamefont{and}
  \bibinfo{author}{\bibfnamefont{M.}~\bibnamefont{{Liebend{\"o}rfer}}},
  \bibinfo{journal}{\aap} \textbf{\bibinfo{volume}{490}}, \bibinfo{pages}{231}
  (\bibinfo{year}{2008}).

\bibitem[{\citenamefont{{Scheidegger} et~al.}(2010)\citenamefont{{Scheidegger},
  {K{\"a}ppeli}, {Whitehouse}, {Fischer}, and
  {Liebend{\"o}rfer}}}]{scheidegger:10b}
\bibinfo{author}{\bibfnamefont{S.}~\bibnamefont{{Scheidegger}}},
  \bibinfo{author}{\bibfnamefont{R.}~\bibnamefont{{K{\"a}ppeli}}},
  \bibinfo{author}{\bibfnamefont{S.~C.} \bibnamefont{{Whitehouse}}},
  \bibinfo{author}{\bibfnamefont{T.}~\bibnamefont{{Fischer}}},
  \bibnamefont{and}
  \bibinfo{author}{\bibfnamefont{M.}~\bibnamefont{{Liebend{\"o}rfer}}},
  \bibinfo{journal}{\aap} \textbf{\bibinfo{volume}{514}}, \bibinfo{pages}{A51}
  (\bibinfo{year}{2010}).

\bibitem[{\citenamefont{Winicour}(2009)}]{winicour:09}
\bibinfo{author}{\bibfnamefont{J.}~\bibnamefont{Winicour}},
  \bibinfo{journal}{{Liv. Rev. Rel.}} \textbf{\bibinfo{volume}{12}}
  (\bibinfo{year}{2009}),
  \bibinfo{note}{{http://www.livingreviews.org/lrr-2009-3}}.

\bibitem[{\citenamefont{Bishop et~al.}(1997)\citenamefont{Bishop, G{\'o}mez,
  Lehner, Maharaj, and Winicour}}]{bishop:97b}
\bibinfo{author}{\bibfnamefont{N.~T.} \bibnamefont{Bishop}},
  \bibinfo{author}{\bibfnamefont{R.}~\bibnamefont{G{\'o}mez}},
  \bibinfo{author}{\bibfnamefont{L.}~\bibnamefont{Lehner}},
  \bibinfo{author}{\bibfnamefont{M.}~\bibnamefont{Maharaj}}, \bibnamefont{and}
  \bibinfo{author}{\bibfnamefont{J.}~\bibnamefont{Winicour}},
  \bibinfo{journal}{\prd} \textbf{\bibinfo{volume}{56}}, \bibinfo{pages}{6298}
  (\bibinfo{year}{1997}).

\bibitem[{\citenamefont{{Reisswig} et~al.}(2011)\citenamefont{{Reisswig},
  {Ott}, {Sperhake}, and {Schnetter}}}]{reisswig:11ccwave}
\bibinfo{author}{\bibfnamefont{C.}~\bibnamefont{{Reisswig}}},
  \bibinfo{author}{\bibfnamefont{C.~D.} \bibnamefont{{Ott}}},
  \bibinfo{author}{\bibfnamefont{U.}~\bibnamefont{{Sperhake}}},
  \bibnamefont{and}
  \bibinfo{author}{\bibfnamefont{E.}~\bibnamefont{{Schnetter}}},
  \bibinfo{journal}{\prd} \textbf{\bibinfo{volume}{83}},
  \bibinfo{pages}{064008} (\bibinfo{year}{2011}).

\bibitem[{\citenamefont{Reisswig et~al.}(2009)\citenamefont{Reisswig, Bishop,
  Pollney, and Szilagyi}}]{reisswig:09}
\bibinfo{author}{\bibfnamefont{C.}~\bibnamefont{Reisswig}},
  \bibinfo{author}{\bibfnamefont{N.~T.} \bibnamefont{Bishop}},
  \bibinfo{author}{\bibfnamefont{D.}~\bibnamefont{Pollney}}, \bibnamefont{and}
  \bibinfo{author}{\bibfnamefont{B.}~\bibnamefont{Szilagyi}},
  \bibinfo{journal}{Phys. Rev. Lett.} \textbf{\bibinfo{volume}{103}},
  \bibinfo{pages}{221101} (\bibinfo{year}{2009}).

\bibitem[{\citenamefont{{Reisswig} et~al.}(2010)\citenamefont{{Reisswig},
  {Bishop}, {Pollney}, and {Szil{\'a}gyi}}}]{reisswig:10a}
\bibinfo{author}{\bibfnamefont{C.}~\bibnamefont{{Reisswig}}},
  \bibinfo{author}{\bibfnamefont{N.~T.} \bibnamefont{{Bishop}}},
  \bibinfo{author}{\bibfnamefont{D.}~\bibnamefont{{Pollney}}},
  \bibnamefont{and}
  \bibinfo{author}{\bibfnamefont{B.}~\bibnamefont{{Szil{\'a}gyi}}},
  \bibinfo{journal}{\cqg} \textbf{\bibinfo{volume}{27}},
  \bibinfo{pages}{075014} (\bibinfo{year}{2010}).

\bibitem[{\citenamefont{{Babiuc} et~al.}(2011)\citenamefont{{Babiuc},
  {Szil{\'a}gyi}, {Winicour}, and {Zlochower}}}]{babiuc:11}
\bibinfo{author}{\bibfnamefont{M.~C.} \bibnamefont{{Babiuc}}},
  \bibinfo{author}{\bibfnamefont{B.}~\bibnamefont{{Szil{\'a}gyi}}},
  \bibinfo{author}{\bibfnamefont{J.}~\bibnamefont{{Winicour}}},
  \bibnamefont{and}
  \bibinfo{author}{\bibfnamefont{Y.}~\bibnamefont{{Zlochower}}},
  \bibinfo{journal}{\prd} \textbf{\bibinfo{volume}{84}},
  \bibinfo{pages}{044057} (\bibinfo{year}{2011}).

\bibitem[{\citenamefont{{Winicour}}(2012)}]{winicour:12}
\bibinfo{author}{\bibfnamefont{J.}~\bibnamefont{{Winicour}}},
  \bibinfo{journal}{\cqg} \textbf{\bibinfo{volume}{29}},
  \bibinfo{pages}{113001} (\bibinfo{year}{2012}).

\bibitem[{\citenamefont{Brown et~al.}(2007)\citenamefont{Brown, Sarbach,
  Schnetter, Tiglio, Diener, Hawke, and Pollney}}]{brown:07a}
\bibinfo{author}{\bibfnamefont{D.}~\bibnamefont{Brown}},
  \bibinfo{author}{\bibfnamefont{O.}~\bibnamefont{Sarbach}},
  \bibinfo{author}{\bibfnamefont{E.}~\bibnamefont{Schnetter}},
  \bibinfo{author}{\bibfnamefont{M.}~\bibnamefont{Tiglio}},
  \bibinfo{author}{\bibfnamefont{P.}~\bibnamefont{Diener}},
  \bibinfo{author}{\bibfnamefont{I.}~\bibnamefont{Hawke}}, \bibnamefont{and}
  \bibinfo{author}{\bibfnamefont{D.}~\bibnamefont{Pollney}},
  \bibinfo{journal}{Phys. Rev. D} \textbf{\bibinfo{volume}{76}},
  \bibinfo{pages}{081503(R)} (\bibinfo{year}{2007}).

\bibitem[{\citenamefont{Brown et~al.}(2009)\citenamefont{Brown, Diener,
  Sarbach, Schnetter, and Tiglio}}]{ES-Brown2007b}
\bibinfo{author}{\bibfnamefont{D.}~\bibnamefont{Brown}},
  \bibinfo{author}{\bibfnamefont{P.}~\bibnamefont{Diener}},
  \bibinfo{author}{\bibfnamefont{O.}~\bibnamefont{Sarbach}},
  \bibinfo{author}{\bibfnamefont{E.}~\bibnamefont{Schnetter}},
  \bibnamefont{and} \bibinfo{author}{\bibfnamefont{M.}~\bibnamefont{Tiglio}},
  \bibinfo{journal}{Phys. Rev. D} \textbf{\bibinfo{volume}{79}},
  \bibinfo{pages}{044023} (\bibinfo{year}{2009}).

\bibitem[{\citenamefont{{Zwerger} and {M\"uller}}(1997)}]{zwerger:97}
\bibinfo{author}{\bibfnamefont{T.}~\bibnamefont{{Zwerger}}} \bibnamefont{and}
  \bibinfo{author}{\bibfnamefont{E.}~\bibnamefont{{M\"uller}}},
  \bibinfo{journal}{\aap} \textbf{\bibinfo{volume}{320}}, \bibinfo{pages}{209}
  (\bibinfo{year}{1997}).

\bibitem[{\citenamefont{{Dimmelmeier}
  et~al.}(2002{\natexlab{a}})\citenamefont{{Dimmelmeier}, {Font}, and
  {M{\"u}ller}}}]{dimmelmeier:02}
\bibinfo{author}{\bibfnamefont{H.}~\bibnamefont{{Dimmelmeier}}},
  \bibinfo{author}{\bibfnamefont{J.~A.} \bibnamefont{{Font}}},
  \bibnamefont{and}
  \bibinfo{author}{\bibfnamefont{E.}~\bibnamefont{{M{\"u}ller}}},
  \bibinfo{journal}{\aap} \textbf{\bibinfo{volume}{393}}, \bibinfo{pages}{523}
  (\bibinfo{year}{2002}{\natexlab{a}}).

\bibitem[{\citenamefont{{Dimmelmeier}
  et~al.}(2002{\natexlab{b}})\citenamefont{{Dimmelmeier}, {Font}, and
  {M{\"u}ller}}}]{dimmelmeier:02a}
\bibinfo{author}{\bibfnamefont{H.}~\bibnamefont{{Dimmelmeier}}},
  \bibinfo{author}{\bibfnamefont{J.~A.} \bibnamefont{{Font}}},
  \bibnamefont{and}
  \bibinfo{author}{\bibfnamefont{E.}~\bibnamefont{{M{\"u}ller}}},
  \bibinfo{journal}{\aap} \textbf{\bibinfo{volume}{388}}, \bibinfo{pages}{917}
  (\bibinfo{year}{2002}{\natexlab{b}}).

\bibitem[{\citenamefont{{Dimmelmeier} et~al.}(2007)\citenamefont{{Dimmelmeier},
  {Ott}, {Janka}, {Marek}, and {M{\"u}ller}}}]{dimmelmeier:07}
\bibinfo{author}{\bibfnamefont{H.}~\bibnamefont{{Dimmelmeier}}},
  \bibinfo{author}{\bibfnamefont{C.~D.} \bibnamefont{{Ott}}},
  \bibinfo{author}{\bibfnamefont{H.-T.} \bibnamefont{{Janka}}},
  \bibinfo{author}{\bibfnamefont{A.}~\bibnamefont{{Marek}}}, \bibnamefont{and}
  \bibinfo{author}{\bibfnamefont{E.}~\bibnamefont{{M{\"u}ller}}},
  \bibinfo{journal}{\prl} \textbf{\bibinfo{volume}{98}},
  \bibinfo{pages}{251101} (\bibinfo{year}{2007}).

\bibitem[{\citenamefont{{Abdikamalov} et~al.}(2010)\citenamefont{{Abdikamalov},
  {Ott}, {Rezzolla}, {Dessart}, {Dimmelmeier}, {Marek}, and
  {Janka}}}]{abdikamalov:10}
\bibinfo{author}{\bibfnamefont{E.~B.} \bibnamefont{{Abdikamalov}}},
  \bibinfo{author}{\bibfnamefont{C.~D.} \bibnamefont{{Ott}}},
  \bibinfo{author}{\bibfnamefont{L.}~\bibnamefont{{Rezzolla}}},
  \bibinfo{author}{\bibfnamefont{L.}~\bibnamefont{{Dessart}}},
  \bibinfo{author}{\bibfnamefont{H.}~\bibnamefont{{Dimmelmeier}}},
  \bibinfo{author}{\bibfnamefont{A.}~\bibnamefont{{Marek}}}, \bibnamefont{and}
  \bibinfo{author}{\bibfnamefont{H.}~\bibnamefont{{Janka}}},
  \bibinfo{journal}{\prd} \textbf{\bibinfo{volume}{81}},
  \bibinfo{pages}{044012} (\bibinfo{year}{2010}).

\bibitem[{\citenamefont{{M{\"u}ller} et~al.}(2010)\citenamefont{{M{\"u}ller},
  {Janka}, and {Dimmelmeier}}}]{mueller:10}
\bibinfo{author}{\bibfnamefont{B.}~\bibnamefont{{M{\"u}ller}}},
  \bibinfo{author}{\bibfnamefont{H.-T.} \bibnamefont{{Janka}}},
  \bibnamefont{and}
  \bibinfo{author}{\bibfnamefont{H.}~\bibnamefont{{Dimmelmeier}}},
  \bibinfo{journal}{\apjs} \textbf{\bibinfo{volume}{189}}, \bibinfo{pages}{104}
  (\bibinfo{year}{2010}).

\bibitem[{\citenamefont{{Hanke} et~al.}(2012)\citenamefont{{Hanke}, {Marek},
  {M{\"u}ller}, and {Janka}}}]{hanke:12}
\bibinfo{author}{\bibfnamefont{F.}~\bibnamefont{{Hanke}}},
  \bibinfo{author}{\bibfnamefont{A.}~\bibnamefont{{Marek}}},
  \bibinfo{author}{\bibfnamefont{B.}~\bibnamefont{{M{\"u}ller}}},
  \bibnamefont{and} \bibinfo{author}{\bibfnamefont{H.-T.}
  \bibnamefont{{Janka}}}, \bibinfo{journal}{\apj}
  \textbf{\bibinfo{volume}{755}}, \bibinfo{eid}{138} (\bibinfo{year}{2012}).

\bibitem[{\citenamefont{{Takiwaki} et~al.}(2012)\citenamefont{{Takiwaki},
  {Kotake}, and {Suwa}}}]{takiwaki:12}
\bibinfo{author}{\bibfnamefont{T.}~\bibnamefont{{Takiwaki}}},
  \bibinfo{author}{\bibfnamefont{K.}~\bibnamefont{{Kotake}}}, \bibnamefont{and}
  \bibinfo{author}{\bibfnamefont{Y.}~\bibnamefont{{Suwa}}},
  \bibinfo{journal}{\apj} \textbf{\bibinfo{volume}{749}}, \bibinfo{pages}{98}
  (\bibinfo{year}{2012}).

\bibitem[{\citenamefont{{Dimmelmeier} et~al.}(2006)\citenamefont{{Dimmelmeier},
  {Stergioulas}, and {Font}}}]{dimmelmeier:06}
\bibinfo{author}{\bibfnamefont{H.}~\bibnamefont{{Dimmelmeier}}},
  \bibinfo{author}{\bibfnamefont{N.}~\bibnamefont{{Stergioulas}}},
  \bibnamefont{and} \bibinfo{author}{\bibfnamefont{J.~A.}
  \bibnamefont{{Font}}}, \bibinfo{journal}{\mnras}
  \textbf{\bibinfo{volume}{368}}, \bibinfo{pages}{1609} (\bibinfo{year}{2006}).

\bibitem[{\citenamefont{{Abdikamalov} et~al.}(2009)\citenamefont{{Abdikamalov},
  {Dimmelmeier}, {Rezzolla}, and {Miller}}}]{abdikamalov:09b}
\bibinfo{author}{\bibfnamefont{E.~B.} \bibnamefont{{Abdikamalov}}},
  \bibinfo{author}{\bibfnamefont{H.}~\bibnamefont{{Dimmelmeier}}},
  \bibinfo{author}{\bibfnamefont{L.}~\bibnamefont{{Rezzolla}}},
  \bibnamefont{and} \bibinfo{author}{\bibfnamefont{J.~C.}
  \bibnamefont{{Miller}}}, \bibinfo{journal}{\mnras}
  \textbf{\bibinfo{volume}{392}}, \bibinfo{pages}{52} (\bibinfo{year}{2009}).

\bibitem[{\citenamefont{{Komissarov}}(2006)}]{komissarov:06}
\bibinfo{author}{\bibfnamefont{S.~S.} \bibnamefont{{Komissarov}}},
  \bibinfo{journal}{\mnras} \textbf{\bibinfo{volume}{367}}, \bibinfo{pages}{19}
  (\bibinfo{year}{2006}).

\bibitem[{\citenamefont{{Font} and {Ibanez}}(1998)}]{font:98}
\bibinfo{author}{\bibfnamefont{J.~A.} \bibnamefont{{Font}}} \bibnamefont{and}
  \bibinfo{author}{\bibfnamefont{J.~M.~A.} \bibnamefont{{Ibanez}}},
  \bibinfo{journal}{\apj} \textbf{\bibinfo{volume}{494}}, \bibinfo{pages}{297}
  (\bibinfo{year}{1998}).

\bibitem[{\citenamefont{{Font} and {Daigne}}(2002)}]{font:02}
\bibinfo{author}{\bibfnamefont{J.~A.} \bibnamefont{{Font}}} \bibnamefont{and}
  \bibinfo{author}{\bibfnamefont{F.}~\bibnamefont{{Daigne}}},
  \bibinfo{journal}{\mnras} \textbf{\bibinfo{volume}{334}},
  \bibinfo{pages}{383} (\bibinfo{year}{2002}).

\bibitem[{\citenamefont{{Zanotti} et~al.}(2003)\citenamefont{{Zanotti},
  {Rezzolla}, and {Font}}}]{zanotti:03}
\bibinfo{author}{\bibfnamefont{O.}~\bibnamefont{{Zanotti}}},
  \bibinfo{author}{\bibfnamefont{L.}~\bibnamefont{{Rezzolla}}},
  \bibnamefont{and} \bibinfo{author}{\bibfnamefont{J.~A.}
  \bibnamefont{{Font}}}, \bibinfo{journal}{\mnras}
  \textbf{\bibinfo{volume}{341}}, \bibinfo{pages}{832} (\bibinfo{year}{2003}).

\bibitem[{\citenamefont{{Narayan} et~al.}(2012)\citenamefont{{Narayan},
  {S{\k{a}}dowski}, {Penna}, and {Kulkarni}}}]{narayan:12}
\bibinfo{author}{\bibfnamefont{R.}~\bibnamefont{{Narayan}}},
  \bibinfo{author}{\bibfnamefont{A.}~\bibnamefont{{S{\k{a}}dowski}}},
  \bibinfo{author}{\bibfnamefont{R.~F.} \bibnamefont{{Penna}}},
  \bibnamefont{and} \bibinfo{author}{\bibfnamefont{A.~K.}
  \bibnamefont{{Kulkarni}}}, \bibinfo{journal}{\mnras}
  \textbf{\bibinfo{volume}{426}}, \bibinfo{pages}{3241} (\bibinfo{year}{2012}).

\bibitem[{\citenamefont{Alcubierre and Mendez}(2011)}]{alcubierre:11}
\bibinfo{author}{\bibfnamefont{M.}~\bibnamefont{Alcubierre}} \bibnamefont{and}
  \bibinfo{author}{\bibfnamefont{M.~D.} \bibnamefont{Mendez}},
  \bibinfo{journal}{Gen.Rel.Grav.} \textbf{\bibinfo{volume}{43}},
  \bibinfo{pages}{2769} (\bibinfo{year}{2011}).

\bibitem[{\citenamefont{{Montero} and
  {Cordero-Carri{\'o}n}}(2012)}]{montero:11}
\bibinfo{author}{\bibfnamefont{P.~J.} \bibnamefont{{Montero}}}
  \bibnamefont{and}
  \bibinfo{author}{\bibfnamefont{I.}~\bibnamefont{{Cordero-Carri{\'o}n}}},
  \bibinfo{journal}{\prd} \textbf{\bibinfo{volume}{85}},
  \bibinfo{pages}{124037} (\bibinfo{year}{2012}).

\bibitem[{\citenamefont{{Cordero-Carri{\'o}n} and
  {Cerd{\'a}-Dur{\'a}n}}(2012)}]{cordero-carrion:12}
\bibinfo{author}{\bibfnamefont{I.}~\bibnamefont{{Cordero-Carri{\'o}n}}}
  \bibnamefont{and}
  \bibinfo{author}{\bibfnamefont{P.}~\bibnamefont{{Cerd{\'a}-Dur{\'a}n}}},
  \bibinfo{journal}{arXiv:1211.5930}  (\bibinfo{year}{2012}).

\bibitem[{\citenamefont{{Baumgarte} et~al.}(2012)\citenamefont{{Baumgarte},
  {Montero}, {Cordero-Carri{\'o}n}, and {M{\"u}ller}}}]{baumgarte:12}
\bibinfo{author}{\bibfnamefont{T.~W.} \bibnamefont{{Baumgarte}}},
  \bibinfo{author}{\bibfnamefont{P.~J.} \bibnamefont{{Montero}}},
  \bibinfo{author}{\bibfnamefont{I.}~\bibnamefont{{Cordero-Carri{\'o}n}}},
  \bibnamefont{and}
  \bibinfo{author}{\bibfnamefont{E.}~\bibnamefont{{M{\"u}ller}}},
  \bibinfo{journal}{arXiv:1211.6632}  (\bibinfo{year}{2012}).

\bibitem[{\citenamefont{{Korobkin} et~al.}(2011)\citenamefont{{Korobkin},
  {Abdikamalov}, {Schnetter}, {Stergioulas}, and {Zink}}}]{korobkin:11}
\bibinfo{author}{\bibfnamefont{O.}~\bibnamefont{{Korobkin}}},
  \bibinfo{author}{\bibfnamefont{E.~B.} \bibnamefont{{Abdikamalov}}},
  \bibinfo{author}{\bibfnamefont{E.}~\bibnamefont{{Schnetter}}},
  \bibinfo{author}{\bibfnamefont{N.}~\bibnamefont{{Stergioulas}}},
  \bibnamefont{and} \bibinfo{author}{\bibfnamefont{B.}~\bibnamefont{{Zink}}},
  \bibinfo{journal}{\prd} \textbf{\bibinfo{volume}{83}},
  \bibinfo{pages}{043007} (\bibinfo{year}{2011}).

\bibitem[{\citenamefont{{Korobkin} et~al.}(2012)\citenamefont{{Korobkin},
  {Abdikamalov}, {Stergioulas}, {Schnetter}, {Zink}, {Rosswog}, and
  {Ott}}}]{korobkin:12}
\bibinfo{author}{\bibfnamefont{O.}~\bibnamefont{{Korobkin}}},
  \bibinfo{author}{\bibfnamefont{E.}~\bibnamefont{{Abdikamalov}}},
  \bibinfo{author}{\bibfnamefont{N.}~\bibnamefont{{Stergioulas}}},
  \bibinfo{author}{\bibfnamefont{E.}~\bibnamefont{{Schnetter}}},
  \bibinfo{author}{\bibfnamefont{B.}~\bibnamefont{{Zink}}},
  \bibinfo{author}{\bibfnamefont{S.}~\bibnamefont{{Rosswog}}},
  \bibnamefont{and} \bibinfo{author}{\bibfnamefont{C.~D.} \bibnamefont{{Ott}}},
  \bibinfo{journal}{ArXiv e-prints}  (\bibinfo{year}{2012}).

\bibitem[{\citenamefont{Thornburg}(2004{\natexlab{a}})}]{thornburg:04}
\bibinfo{author}{\bibfnamefont{J.}~\bibnamefont{Thornburg}},
  \bibinfo{journal}{Class. Quant. Grav.} \textbf{\bibinfo{volume}{21}},
  \bibinfo{pages}{743} (\bibinfo{year}{2004}{\natexlab{a}}).

\bibitem[{\citenamefont{Pazos et~al.}(2007)}]{pazos:06}
\bibinfo{author}{\bibfnamefont{E.}~\bibnamefont{Pazos}} \bibnamefont{et~al.},
  \bibinfo{journal}{\cqg} \textbf{\bibinfo{volume}{24}}, \bibinfo{pages}{S341}
  (\bibinfo{year}{2007}).

\bibitem[{\citenamefont{Thornburg}(2004{\natexlab{b}})}]{thornburg:04b}
\bibinfo{author}{\bibfnamefont{J.}~\bibnamefont{Thornburg}},
  \bibinfo{journal}{Class. Quantum Grav.} \textbf{\bibinfo{volume}{21}},
  \bibinfo{pages}{3365} (\bibinfo{year}{2004}{\natexlab{b}}).

\bibitem[{\citenamefont{Dorband et~al.}(2006)\citenamefont{Dorband, Berti,
  Diener, Schnetter, and Tiglio}}]{dorband:06a}
\bibinfo{author}{\bibfnamefont{E.~N.} \bibnamefont{Dorband}},
  \bibinfo{author}{\bibfnamefont{E.}~\bibnamefont{Berti}},
  \bibinfo{author}{\bibfnamefont{P.}~\bibnamefont{Diener}},
  \bibinfo{author}{\bibfnamefont{E.}~\bibnamefont{Schnetter}},
  \bibnamefont{and} \bibinfo{author}{\bibfnamefont{M.}~\bibnamefont{Tiglio}},
  \bibinfo{journal}{Phys. Rev. D} \textbf{\bibinfo{volume}{74}},
  \bibinfo{pages}{084028} (\bibinfo{year}{2006}).

\bibitem[{\citenamefont{Pazos et~al.}(2009)\citenamefont{Pazos, Tiglio, Duez,
  Kidder, and Teukolsky}}]{pazos:09}
\bibinfo{author}{\bibfnamefont{E.}~\bibnamefont{Pazos}},
  \bibinfo{author}{\bibfnamefont{M.}~\bibnamefont{Tiglio}},
  \bibinfo{author}{\bibfnamefont{M.~D.} \bibnamefont{Duez}},
  \bibinfo{author}{\bibfnamefont{L.~E.} \bibnamefont{Kidder}},
  \bibnamefont{and} \bibinfo{author}{\bibfnamefont{S.~A.}
  \bibnamefont{Teukolsky}}, \bibinfo{journal}{\prd}
  \textbf{\bibinfo{volume}{80}}, \bibinfo{pages}{024027}
  (\bibinfo{year}{2009}).

\bibitem[{\citenamefont{{Zink} et~al.}(2008)\citenamefont{{Zink}, {Schnetter},
  and {Tiglio}}}]{zink:08b}
\bibinfo{author}{\bibfnamefont{B.}~\bibnamefont{{Zink}}},
  \bibinfo{author}{\bibfnamefont{E.}~\bibnamefont{{Schnetter}}},
  \bibnamefont{and} \bibinfo{author}{\bibfnamefont{M.}~\bibnamefont{{Tiglio}}},
  \bibinfo{journal}{\prd} \textbf{\bibinfo{volume}{77}},
  \bibinfo{pages}{103015} (\bibinfo{year}{2008}).

\bibitem[{\citenamefont{Ansorg}(2007)}]{ansorg:07}
\bibinfo{author}{\bibfnamefont{M.}~\bibnamefont{Ansorg}},
  \bibinfo{journal}{\cqg} \textbf{\bibinfo{volume}{24}}, \bibinfo{pages}{S1}
  (\bibinfo{year}{2007}).

\bibitem[{\citenamefont{{Pfeiffer} et~al.}(2003)\citenamefont{{Pfeiffer},
  {Kidder}, {Scheel}, and {Teukolsky}}}]{Pfeiffer:03}
\bibinfo{author}{\bibfnamefont{H.~P.} \bibnamefont{{Pfeiffer}}},
  \bibinfo{author}{\bibfnamefont{L.~E.} \bibnamefont{{Kidder}}},
  \bibinfo{author}{\bibfnamefont{M.~A.} \bibnamefont{{Scheel}}},
  \bibnamefont{and} \bibinfo{author}{\bibfnamefont{S.~A.}
  \bibnamefont{{Teukolsky}}}, \bibinfo{journal}{Comp. Phys. Comm.}
  \textbf{\bibinfo{volume}{152}}, \bibinfo{pages}{253} (\bibinfo{year}{2003}).

\bibitem[{\citenamefont{Gourgoulhon et~al.}(2001)\citenamefont{Gourgoulhon,
  Grandclement, Taniguchi, Marck, and Bonazzola}}]{gourgoulhon:00}
\bibinfo{author}{\bibfnamefont{E.}~\bibnamefont{Gourgoulhon}},
  \bibinfo{author}{\bibfnamefont{P.}~\bibnamefont{Grandclement}},
  \bibinfo{author}{\bibfnamefont{K.}~\bibnamefont{Taniguchi}},
  \bibinfo{author}{\bibfnamefont{J.-A.} \bibnamefont{Marck}}, \bibnamefont{and}
  \bibinfo{author}{\bibfnamefont{S.}~\bibnamefont{Bonazzola}},
  \bibinfo{journal}{Phys. Rev. D} \textbf{\bibinfo{volume}{63}},
  \bibinfo{pages}{064029} (\bibinfo{year}{2001}).

\bibitem[{\citenamefont{{Gourgoulhon} et~al.}(2002)\citenamefont{{Gourgoulhon},
  {Grandcl{\'e}ment}, and {Bonazzola}}}]{gourgoulhon:02}
\bibinfo{author}{\bibfnamefont{E.}~\bibnamefont{{Gourgoulhon}}},
  \bibinfo{author}{\bibfnamefont{P.}~\bibnamefont{{Grandcl{\'e}ment}}},
  \bibnamefont{and}
  \bibinfo{author}{\bibfnamefont{S.}~\bibnamefont{{Bonazzola}}},
  \bibinfo{journal}{\prd} \textbf{\bibinfo{volume}{65}},
  \bibinfo{pages}{044020} (\bibinfo{year}{2002}).

\bibitem[{\citenamefont{{Grandcl{\'e}ment}
  et~al.}(2001)\citenamefont{{Grandcl{\'e}ment}, {Bonazzola}, {Gourgoulhon},
  and {Marck}}}]{grandclement:01}
\bibinfo{author}{\bibfnamefont{P.}~\bibnamefont{{Grandcl{\'e}ment}}},
  \bibinfo{author}{\bibfnamefont{S.}~\bibnamefont{{Bonazzola}}},
  \bibinfo{author}{\bibfnamefont{E.}~\bibnamefont{{Gourgoulhon}}},
  \bibnamefont{and} \bibinfo{author}{\bibfnamefont{J.-A.}
  \bibnamefont{{Marck}}}, \bibinfo{journal}{\jcp}
  \textbf{\bibinfo{volume}{170}}, \bibinfo{pages}{231} (\bibinfo{year}{2001}).

\bibitem[{\citenamefont{{Reisswig} et~al.}(2007)\citenamefont{{Reisswig},
  {Bishop}, {Lai}, {Thornburg}, and {Szilagyi}}}]{reisswig:07}
\bibinfo{author}{\bibfnamefont{C.}~\bibnamefont{{Reisswig}}},
  \bibinfo{author}{\bibfnamefont{N.~T.} \bibnamefont{{Bishop}}},
  \bibinfo{author}{\bibfnamefont{C.~W.} \bibnamefont{{Lai}}},
  \bibinfo{author}{\bibfnamefont{J.}~\bibnamefont{{Thornburg}}},
  \bibnamefont{and}
  \bibinfo{author}{\bibfnamefont{B.}~\bibnamefont{{Szilagyi}}},
  \bibinfo{journal}{\cqg} \textbf{\bibinfo{volume}{24}}, \bibinfo{pages}{327}
  (\bibinfo{year}{2007}).

\bibitem[{\citenamefont{{G{\'o}mez} et~al.}(2007)\citenamefont{{G{\'o}mez},
  {Barreto}, and {Frittelli}}}]{gomez:07}
\bibinfo{author}{\bibfnamefont{R.}~\bibnamefont{{G{\'o}mez}}},
  \bibinfo{author}{\bibfnamefont{W.}~\bibnamefont{{Barreto}}},
  \bibnamefont{and}
  \bibinfo{author}{\bibfnamefont{S.}~\bibnamefont{{Frittelli}}},
  \bibinfo{journal}{\prd} \textbf{\bibinfo{volume}{76}},
  \bibinfo{pages}{124029} (\bibinfo{year}{2007}).

\bibitem[{\citenamefont{{Scheel} et~al.}(2009)\citenamefont{{Scheel}, {Boyle},
  {Chu}, {Kidder}, {Matthews}, and {Pfeiffer}}}]{scheel:09}
\bibinfo{author}{\bibfnamefont{M.~A.} \bibnamefont{{Scheel}}},
  \bibinfo{author}{\bibfnamefont{M.}~\bibnamefont{{Boyle}}},
  \bibinfo{author}{\bibfnamefont{T.}~\bibnamefont{{Chu}}},
  \bibinfo{author}{\bibfnamefont{L.~E.} \bibnamefont{{Kidder}}},
  \bibinfo{author}{\bibfnamefont{K.~D.} \bibnamefont{{Matthews}}},
  \bibnamefont{and} \bibinfo{author}{\bibfnamefont{H.~P.}
  \bibnamefont{{Pfeiffer}}}, \bibinfo{journal}{\prd}
  \textbf{\bibinfo{volume}{79}}, \bibinfo{pages}{024003}
  (\bibinfo{year}{2009}).

\bibitem[{\citenamefont{{Szil{\'a}gyi}
  et~al.}(2009)\citenamefont{{Szil{\'a}gyi}, {Lindblom}, and
  {Scheel}}}]{szilagyi:09}
\bibinfo{author}{\bibfnamefont{B.}~\bibnamefont{{Szil{\'a}gyi}}},
  \bibinfo{author}{\bibfnamefont{L.}~\bibnamefont{{Lindblom}}},
  \bibnamefont{and} \bibinfo{author}{\bibfnamefont{M.~A.}
  \bibnamefont{{Scheel}}}, \bibinfo{journal}{\prd}
  \textbf{\bibinfo{volume}{80}}, \bibinfo{pages}{124010}
  (\bibinfo{year}{2009}).

\bibitem[{\citenamefont{{Chu} et~al.}(2009)\citenamefont{{Chu}, {Pfeiffer}, and
  {Scheel}}}]{chu:09}
\bibinfo{author}{\bibfnamefont{T.}~\bibnamefont{{Chu}}},
  \bibinfo{author}{\bibfnamefont{H.~P.} \bibnamefont{{Pfeiffer}}},
  \bibnamefont{and} \bibinfo{author}{\bibfnamefont{M.~A.}
  \bibnamefont{{Scheel}}}, \bibinfo{journal}{\prd}
  \textbf{\bibinfo{volume}{80}}, \bibinfo{pages}{124051}
  (\bibinfo{year}{2009}).

\bibitem[{\citenamefont{{Lovelace} et~al.}(2012)\citenamefont{{Lovelace},
  {Boyle}, {Scheel}, and {Szil{\'a}gyi}}}]{lovelace:12}
\bibinfo{author}{\bibfnamefont{G.}~\bibnamefont{{Lovelace}}},
  \bibinfo{author}{\bibfnamefont{M.}~\bibnamefont{{Boyle}}},
  \bibinfo{author}{\bibfnamefont{M.~A.} \bibnamefont{{Scheel}}},
  \bibnamefont{and}
  \bibinfo{author}{\bibfnamefont{B.}~\bibnamefont{{Szil{\'a}gyi}}},
  \bibinfo{journal}{\cqg} \textbf{\bibinfo{volume}{29}},
  \bibinfo{pages}{045003} (\bibinfo{year}{2012}).

\bibitem[{\citenamefont{{Buchman} et~al.}(2012)\citenamefont{{Buchman},
  {Pfeiffer}, {Scheel}, and {Szilagyi}}}]{buchman:12}
\bibinfo{author}{\bibfnamefont{L.~T.} \bibnamefont{{Buchman}}},
  \bibinfo{author}{\bibfnamefont{H.~P.} \bibnamefont{{Pfeiffer}}},
  \bibinfo{author}{\bibfnamefont{M.~A.} \bibnamefont{{Scheel}}},
  \bibnamefont{and}
  \bibinfo{author}{\bibfnamefont{B.}~\bibnamefont{{Szilagyi}}},
  \bibinfo{journal}{arxiv:1206.3015}  (\bibinfo{year}{2012}).

\bibitem[{\citenamefont{{MacDonald} et~al.}(2012)\citenamefont{{MacDonald},
  {Mroue}, {Pfeiffer}, {Boyle}, {Kidder}, {Scheel}, {Szilagyi}, and
  {Taylor}}}]{macdonald:12}
\bibinfo{author}{\bibfnamefont{I.}~\bibnamefont{{MacDonald}}},
  \bibinfo{author}{\bibfnamefont{A.~H.} \bibnamefont{{Mroue}}},
  \bibinfo{author}{\bibfnamefont{H.~P.} \bibnamefont{{Pfeiffer}}},
  \bibinfo{author}{\bibfnamefont{M.}~\bibnamefont{{Boyle}}},
  \bibinfo{author}{\bibfnamefont{L.~E.} \bibnamefont{{Kidder}}},
  \bibinfo{author}{\bibfnamefont{M.~A.} \bibnamefont{{Scheel}}},
  \bibinfo{author}{\bibfnamefont{B.}~\bibnamefont{{Szilagyi}}},
  \bibnamefont{and} \bibinfo{author}{\bibfnamefont{N.~W.}
  \bibnamefont{{Taylor}}}, \bibinfo{journal}{arXiv:1210.3007}
  (\bibinfo{year}{2012}).

\bibitem[{\citenamefont{{Scheel} et~al.}(2006)\citenamefont{{Scheel},
  {Pfeiffer}, {Lindblom}, {Kidder}, {Rinne}, and {Teukolsky}}}]{scheel:06}
\bibinfo{author}{\bibfnamefont{M.~A.} \bibnamefont{{Scheel}}},
  \bibinfo{author}{\bibfnamefont{H.~P.} \bibnamefont{{Pfeiffer}}},
  \bibinfo{author}{\bibfnamefont{L.}~\bibnamefont{{Lindblom}}},
  \bibinfo{author}{\bibfnamefont{L.~E.} \bibnamefont{{Kidder}}},
  \bibinfo{author}{\bibfnamefont{O.}~\bibnamefont{{Rinne}}}, \bibnamefont{and}
  \bibinfo{author}{\bibfnamefont{S.~A.} \bibnamefont{{Teukolsky}}},
  \bibinfo{journal}{\prd} \textbf{\bibinfo{volume}{74}},
  \bibinfo{pages}{104006} (\bibinfo{year}{2006}).

\bibitem[{\citenamefont{{Foucart} et~al.}(2012)\citenamefont{{Foucart}, {Duez},
  {Kidder}, {Scheel}, {Szilagyi}, and {Teukolsky}}}]{foucart:12}
\bibinfo{author}{\bibfnamefont{F.}~\bibnamefont{{Foucart}}},
  \bibinfo{author}{\bibfnamefont{M.~D.} \bibnamefont{{Duez}}},
  \bibinfo{author}{\bibfnamefont{L.~E.} \bibnamefont{{Kidder}}},
  \bibinfo{author}{\bibfnamefont{M.~A.} \bibnamefont{{Scheel}}},
  \bibinfo{author}{\bibfnamefont{B.}~\bibnamefont{{Szilagyi}}},
  \bibnamefont{and} \bibinfo{author}{\bibfnamefont{S.~A.}
  \bibnamefont{{Teukolsky}}}, \bibinfo{journal}{\prd}
  \textbf{\bibinfo{volume}{85}}, \bibinfo{pages}{044015}
  (\bibinfo{year}{2012}).

\bibitem[{\citenamefont{Foucart et~al.}(2011)\citenamefont{Foucart, Duez,
  Kidder, and Teukolsky}}]{foucart:10}
\bibinfo{author}{\bibfnamefont{F.}~\bibnamefont{Foucart}},
  \bibinfo{author}{\bibfnamefont{M.~D.} \bibnamefont{Duez}},
  \bibinfo{author}{\bibfnamefont{L.~E.} \bibnamefont{Kidder}},
  \bibnamefont{and} \bibinfo{author}{\bibfnamefont{S.~A.}
  \bibnamefont{Teukolsky}}, \bibinfo{journal}{Phys. Rev. D}
  \textbf{\bibinfo{volume}{83}}, \bibinfo{pages}{024005}
  (\bibinfo{year}{2011}).

\bibitem[{\citenamefont{{Duez} et~al.}(2008)\citenamefont{{Duez}, {Foucart},
  {Kidder}, {Pfeiffer}, {Scheel}, and {Teukolsky}}}]{duez:08}
\bibinfo{author}{\bibfnamefont{M.~D.} \bibnamefont{{Duez}}},
  \bibinfo{author}{\bibfnamefont{F.}~\bibnamefont{{Foucart}}},
  \bibinfo{author}{\bibfnamefont{L.~E.} \bibnamefont{{Kidder}}},
  \bibinfo{author}{\bibfnamefont{H.~P.} \bibnamefont{{Pfeiffer}}},
  \bibinfo{author}{\bibfnamefont{M.~A.} \bibnamefont{{Scheel}}},
  \bibnamefont{and} \bibinfo{author}{\bibfnamefont{S.~A.}
  \bibnamefont{{Teukolsky}}}, \bibinfo{journal}{\prd}
  \textbf{\bibinfo{volume}{78}}, \bibinfo{pages}{104015}
  (\bibinfo{year}{2008}).

\bibitem[{\citenamefont{{Foucart} et~al.}(2008)\citenamefont{{Foucart},
  {Kidder}, {Pfeiffer}, and {Teukolsky}}}]{foucart:08}
\bibinfo{author}{\bibfnamefont{F.}~\bibnamefont{{Foucart}}},
  \bibinfo{author}{\bibfnamefont{L.~E.} \bibnamefont{{Kidder}}},
  \bibinfo{author}{\bibfnamefont{H.~P.} \bibnamefont{{Pfeiffer}}},
  \bibnamefont{and} \bibinfo{author}{\bibfnamefont{S.~A.}
  \bibnamefont{{Teukolsky}}}, \bibinfo{journal}{\prd}
  \textbf{\bibinfo{volume}{77}}, \bibinfo{pages}{124051}
  (\bibinfo{year}{2008}).

\bibitem[{\citenamefont{{Pollney} et~al.}(2011)\citenamefont{{Pollney},
  {Reisswig}, {Schnetter}, {Dorband}, and {Diener}}}]{pollney:11}
\bibinfo{author}{\bibfnamefont{D.}~\bibnamefont{{Pollney}}},
  \bibinfo{author}{\bibfnamefont{C.}~\bibnamefont{{Reisswig}}},
  \bibinfo{author}{\bibfnamefont{E.}~\bibnamefont{{Schnetter}}},
  \bibinfo{author}{\bibfnamefont{N.}~\bibnamefont{{Dorband}}},
  \bibnamefont{and} \bibinfo{author}{\bibfnamefont{P.}~\bibnamefont{{Diener}}},
  \bibinfo{journal}{\prd} \textbf{\bibinfo{volume}{83}}, \bibinfo{eid}{044045}
  (\bibinfo{year}{2011}).

\bibitem[{\citenamefont{Pollney et~al.}(2009)\citenamefont{Pollney, Reisswig,
  Dorband, Schnetter, and Diener}}]{pollney:09}
\bibinfo{author}{\bibfnamefont{D.}~\bibnamefont{Pollney}},
  \bibinfo{author}{\bibfnamefont{C.}~\bibnamefont{Reisswig}},
  \bibinfo{author}{\bibfnamefont{N.}~\bibnamefont{Dorband}},
  \bibinfo{author}{\bibfnamefont{E.}~\bibnamefont{Schnetter}},
  \bibnamefont{and} \bibinfo{author}{\bibfnamefont{P.}~\bibnamefont{Diener}},
  \bibinfo{journal}{\prd} \textbf{\bibinfo{volume}{80}},
  \bibinfo{pages}{121502} (\bibinfo{year}{2009}).

\bibitem[{\citenamefont{{Pollney} and {Reisswig}}(2011)}]{pollney:10}
\bibinfo{author}{\bibfnamefont{D.}~\bibnamefont{{Pollney}}} \bibnamefont{and}
  \bibinfo{author}{\bibfnamefont{C.}~\bibnamefont{{Reisswig}}},
  \bibinfo{journal}{\apjl} \textbf{\bibinfo{volume}{732}}, \bibinfo{pages}{L13}
  (\bibinfo{year}{2011}).

\bibitem[{\citenamefont{{Damour} et~al.}(2012)\citenamefont{{Damour}, {Nagar},
  {Pollney}, and {Reisswig}}}]{damour:11}
\bibinfo{author}{\bibfnamefont{T.}~\bibnamefont{{Damour}}},
  \bibinfo{author}{\bibfnamefont{A.}~\bibnamefont{{Nagar}}},
  \bibinfo{author}{\bibfnamefont{D.}~\bibnamefont{{Pollney}}},
  \bibnamefont{and}
  \bibinfo{author}{\bibfnamefont{C.}~\bibnamefont{{Reisswig}}},
  \bibinfo{journal}{\prl} \textbf{\bibinfo{volume}{108}},
  \bibinfo{pages}{131101} (\bibinfo{year}{2012}).

\bibitem[{\citenamefont{Santamaria et~al.}(2010)\citenamefont{Santamaria, Ohme,
  Ajith, Bruegmann, Dorband et~al.}}]{santamaria:10}
\bibinfo{author}{\bibfnamefont{L.}~\bibnamefont{Santamaria}},
  \bibinfo{author}{\bibfnamefont{F.}~\bibnamefont{Ohme}},
  \bibinfo{author}{\bibfnamefont{P.}~\bibnamefont{Ajith}},
  \bibinfo{author}{\bibfnamefont{B.}~\bibnamefont{Bruegmann}},
  \bibinfo{author}{\bibfnamefont{N.}~\bibnamefont{Dorband}},
  \bibnamefont{et~al.}, \bibinfo{journal}{\prd} \textbf{\bibinfo{volume}{82}},
  \bibinfo{pages}{064016} (\bibinfo{year}{2010}).

\bibitem[{\citenamefont{{Bishop} et~al.}(2011)\citenamefont{{Bishop},
  {Pollney}, and {Reisswig}}}]{bishop:11}
\bibinfo{author}{\bibfnamefont{N.}~\bibnamefont{{Bishop}}},
  \bibinfo{author}{\bibfnamefont{D.}~\bibnamefont{{Pollney}}},
  \bibnamefont{and}
  \bibinfo{author}{\bibfnamefont{C.}~\bibnamefont{{Reisswig}}},
  \bibinfo{journal}{\cqg} \textbf{\bibinfo{volume}{28}},
  \bibinfo{pages}{155019} (\bibinfo{year}{2011}).

\bibitem[{\citenamefont{{Ajith} et~al.}(2011)\citenamefont{{Ajith}, {Hannam},
  {Husa}, {Chen}, {Br{\"u}gmann}, {Dorband}, {M{\"u}ller}, {Ohme}, {Pollney},
  {Reisswig} et~al.}}]{ajith:11}
\bibinfo{author}{\bibfnamefont{P.}~\bibnamefont{{Ajith}}},
  \bibinfo{author}{\bibfnamefont{M.}~\bibnamefont{{Hannam}}},
  \bibinfo{author}{\bibfnamefont{S.}~\bibnamefont{{Husa}}},
  \bibinfo{author}{\bibfnamefont{Y.}~\bibnamefont{{Chen}}},
  \bibinfo{author}{\bibfnamefont{B.}~\bibnamefont{{Br{\"u}gmann}}},
  \bibinfo{author}{\bibfnamefont{N.}~\bibnamefont{{Dorband}}},
  \bibinfo{author}{\bibfnamefont{D.}~\bibnamefont{{M{\"u}ller}}},
  \bibinfo{author}{\bibfnamefont{F.}~\bibnamefont{{Ohme}}},
  \bibinfo{author}{\bibfnamefont{D.}~\bibnamefont{{Pollney}}},
  \bibinfo{author}{\bibfnamefont{C.}~\bibnamefont{{Reisswig}}},
  \bibnamefont{et~al.}, \bibinfo{journal}{\prl} \textbf{\bibinfo{volume}{106}},
  \bibinfo{pages}{241101} (\bibinfo{year}{2011}).

\bibitem[{\citenamefont{Goodale et~al.}(2003)\citenamefont{Goodale, Allen,
  Lanfermann, Mass{\'o}, Radke, Seidel, and Shalf}}]{goodale:03}
\bibinfo{author}{\bibfnamefont{T.}~\bibnamefont{Goodale}},
  \bibinfo{author}{\bibfnamefont{G.}~\bibnamefont{Allen}},
  \bibinfo{author}{\bibfnamefont{G.}~\bibnamefont{Lanfermann}},
  \bibinfo{author}{\bibfnamefont{J.}~\bibnamefont{Mass{\'o}}},
  \bibinfo{author}{\bibfnamefont{T.}~\bibnamefont{Radke}},
  \bibinfo{author}{\bibfnamefont{E.}~\bibnamefont{Seidel}}, \bibnamefont{and}
  \bibinfo{author}{\bibfnamefont{J.}~\bibnamefont{Shalf}}, in
  \emph{\bibinfo{booktitle}{Vector and Parallel Processing -- VECPAR'2002, 5th
  International Conference, Lecture Notes in Computer Science}}
  (\bibinfo{publisher}{Springer}, \bibinfo{address}{Berlin},
  \bibinfo{year}{2003}).

\bibitem[{\citenamefont{Schnetter et~al.}(2004)\citenamefont{Schnetter, Hawley,
  and Hawke}}]{Schnetter-etal-03b}
\bibinfo{author}{\bibfnamefont{E.}~\bibnamefont{Schnetter}},
  \bibinfo{author}{\bibfnamefont{S.}~\bibnamefont{Hawley}}, \bibnamefont{and}
  \bibinfo{author}{\bibfnamefont{I.}~\bibnamefont{Hawke}},
  \bibinfo{journal}{Class. Quantum Grav.} \textbf{\bibinfo{volume}{21}},
  \bibinfo{pages}{1465} (\bibinfo{year}{2004}).

\bibitem[{\citenamefont{Schnetter et~al.}(2006)\citenamefont{Schnetter, Diener,
  Dorband, and Tiglio}}]{ES-Schnetter2006a}
\bibinfo{author}{\bibfnamefont{E.}~\bibnamefont{Schnetter}},
  \bibinfo{author}{\bibfnamefont{P.}~\bibnamefont{Diener}},
  \bibinfo{author}{\bibfnamefont{E.~N.} \bibnamefont{Dorband}},
  \bibnamefont{and} \bibinfo{author}{\bibfnamefont{M.}~\bibnamefont{Tiglio}},
  \bibinfo{journal}{Class. Quantum Grav.} \textbf{\bibinfo{volume}{23}},
  \bibinfo{pages}{S553} (\bibinfo{year}{2006}).

\bibitem[{\citenamefont{{L{\"o}ffler} et~al.}(2012)\citenamefont{{L{\"o}ffler},
  {Faber}, {Bentivegna}, {Bode}, {Diener}, {Haas}, {Hinder}, {Mundim}, {Ott},
  {Schnetter} et~al.}}]{et:12}
\bibinfo{author}{\bibfnamefont{F.}~\bibnamefont{{L{\"o}ffler}}},
  \bibinfo{author}{\bibfnamefont{J.}~\bibnamefont{{Faber}}},
  \bibinfo{author}{\bibfnamefont{E.}~\bibnamefont{{Bentivegna}}},
  \bibinfo{author}{\bibfnamefont{T.}~\bibnamefont{{Bode}}},
  \bibinfo{author}{\bibfnamefont{P.}~\bibnamefont{{Diener}}},
  \bibinfo{author}{\bibfnamefont{R.}~\bibnamefont{{Haas}}},
  \bibinfo{author}{\bibfnamefont{I.}~\bibnamefont{{Hinder}}},
  \bibinfo{author}{\bibfnamefont{B.~C.} \bibnamefont{{Mundim}}},
  \bibinfo{author}{\bibfnamefont{C.~D.} \bibnamefont{{Ott}}},
  \bibinfo{author}{\bibfnamefont{E.}~\bibnamefont{{Schnetter}}},
  \bibnamefont{et~al.}, \bibinfo{journal}{Class. Quantum Grav.}
  \textbf{\bibinfo{volume}{29}}, \bibinfo{pages}{115001}
  (\bibinfo{year}{2012}).

\bibitem[{\citenamefont{Berger and Oliger}(1984)}]{Berger1984}
\bibinfo{author}{\bibfnamefont{M.~J.} \bibnamefont{Berger}} \bibnamefont{and}
  \bibinfo{author}{\bibfnamefont{J.}~\bibnamefont{Oliger}},
  \bibinfo{journal}{\jcp} \textbf{\bibinfo{volume}{53}}, \bibinfo{pages}{484}
  (\bibinfo{year}{1984}), ISSN \bibinfo{issn}{0021-9991}.

\bibitem[{\citenamefont{{East} et~al.}(2012)\citenamefont{{East}, {Pretorius},
  and {Stephens}}}]{east:11}
\bibinfo{author}{\bibfnamefont{W.~E.} \bibnamefont{{East}}},
  \bibinfo{author}{\bibfnamefont{F.}~\bibnamefont{{Pretorius}}},
  \bibnamefont{and} \bibinfo{author}{\bibfnamefont{B.~C.}
  \bibnamefont{{Stephens}}}, \bibinfo{journal}{\prd}
  \textbf{\bibinfo{volume}{85}}, \bibinfo{pages}{124010}
  (\bibinfo{year}{2012}).

\bibitem[{\citenamefont{{Colella} and {Sekora}}(2008)}]{colella:08}
\bibinfo{author}{\bibfnamefont{P.}~\bibnamefont{{Colella}}} \bibnamefont{and}
  \bibinfo{author}{\bibfnamefont{M.~D.} \bibnamefont{{Sekora}}},
  \bibinfo{journal}{\jcp} \textbf{\bibinfo{volume}{227}}, \bibinfo{pages}{7069}
  (\bibinfo{year}{2008}).

\bibitem[{\citenamefont{{McCorquodale} and {Colella}}(2011)}]{mccorquodale:11}
\bibinfo{author}{\bibfnamefont{P.}~\bibnamefont{{McCorquodale}}}
  \bibnamefont{and}
  \bibinfo{author}{\bibfnamefont{P.}~\bibnamefont{{Colella}}},
  \bibinfo{journal}{Comm. Appl. Math. Comp. Sci.} \textbf{\bibinfo{volume}{6}},
  \bibinfo{pages}{1} (\bibinfo{year}{2011}).

\bibitem[{\citenamefont{{Schlegel} et~al.}(2009)\citenamefont{{Schlegel},
  {Knoth}, {Arnold}, and {Wolke}}}]{schlegel:09}
\bibinfo{author}{\bibfnamefont{M.}~\bibnamefont{{Schlegel}}},
  \bibinfo{author}{\bibfnamefont{O.}~\bibnamefont{{Knoth}}},
  \bibinfo{author}{\bibfnamefont{M.}~\bibnamefont{{Arnold}}}, \bibnamefont{and}
  \bibinfo{author}{\bibfnamefont{R.}~\bibnamefont{{Wolke}}},
  \bibinfo{journal}{J. Comp. Appl. Math.} \textbf{\bibinfo{volume}{226}},
  \bibinfo{pages}{345} (\bibinfo{year}{2009}).

\bibitem[{\citenamefont{{Constantinescu} and
  {Sandu}}(2007)}]{constantinescu:07}
\bibinfo{author}{\bibfnamefont{E.}~\bibnamefont{{Constantinescu}}}
  \bibnamefont{and} \bibinfo{author}{\bibfnamefont{A.}~\bibnamefont{{Sandu}}},
  \bibinfo{journal}{SIAM J. Sci. Comput.} \textbf{\bibinfo{volume}{33}},
  \bibinfo{pages}{239} (\bibinfo{year}{2007}).

\bibitem[{\citenamefont{{Ott} et~al.}(2012{\natexlab{b}})\citenamefont{{Ott},
  {Abdikamalov}, {Moesta}, {Haas}, {Drasco}, {O'Connor}, {Reisswig}, {Meakin},
  and {Schnetter}}}]{ott:12b}
\bibinfo{author}{\bibfnamefont{C.~D.} \bibnamefont{{Ott}}},
  \bibinfo{author}{\bibfnamefont{E.}~\bibnamefont{{Abdikamalov}}},
  \bibinfo{author}{\bibfnamefont{P.}~\bibnamefont{{Moesta}}},
  \bibinfo{author}{\bibfnamefont{R.}~\bibnamefont{{Haas}}},
  \bibinfo{author}{\bibfnamefont{S.}~\bibnamefont{{Drasco}}},
  \bibinfo{author}{\bibfnamefont{E.}~\bibnamefont{{O'Connor}}},
  \bibinfo{author}{\bibfnamefont{C.}~\bibnamefont{{Reisswig}}},
  \bibinfo{author}{\bibfnamefont{C.}~\bibnamefont{{Meakin}}}, \bibnamefont{and}
  \bibinfo{author}{\bibfnamefont{E.}~\bibnamefont{{Schnetter}}},
  \bibinfo{journal}{Submitted to the Astrophys.~J., arXiv:1210.6674}
  (\bibinfo{year}{2012}{\natexlab{b}}).

\bibitem[{ein()}]{einsteintoolkitweb}
\bibinfo{note}{EinsteinToolkit: A Community Toolkit for Numerical Relativity},
  \urlprefix\url{http://www.einsteintoolkit.org}.

\bibitem[{\citenamefont{Baiotti et~al.}(2005)\citenamefont{Baiotti, Hawke,
  Montero, L{\"o}ffler, Rezzolla, Stergioulas, Font, and Seidel}}]{baiotti:05}
\bibinfo{author}{\bibfnamefont{B.}~\bibnamefont{Baiotti}},
  \bibinfo{author}{\bibfnamefont{I.}~\bibnamefont{Hawke}},
  \bibinfo{author}{\bibfnamefont{P.~J.} \bibnamefont{Montero}},
  \bibinfo{author}{\bibfnamefont{F.}~\bibnamefont{L{\"o}ffler}},
  \bibinfo{author}{\bibfnamefont{L.}~\bibnamefont{Rezzolla}},
  \bibinfo{author}{\bibfnamefont{N.}~\bibnamefont{Stergioulas}},
  \bibinfo{author}{\bibfnamefont{J.~A.} \bibnamefont{Font}}, \bibnamefont{and}
  \bibinfo{author}{\bibfnamefont{E.}~\bibnamefont{Seidel}},
  \bibinfo{journal}{Phys. Rev. D} \textbf{\bibinfo{volume}{71}},
  \bibinfo{pages}{024035} (\bibinfo{year}{2005}).

\bibitem[{\citenamefont{York}(1983)}]{york:83}
\bibinfo{author}{\bibfnamefont{J.~W.} \bibnamefont{York}, \bibfnamefont{Jr.}},
  in \emph{\bibinfo{booktitle}{Gravitational radiation}}, edited by
  \bibinfo{editor}{\bibfnamefont{N.}~\bibnamefont{Deruelle}} \bibnamefont{and}
  \bibinfo{editor}{\bibfnamefont{T.}~\bibnamefont{Piran}}
  (\bibinfo{publisher}{North-Holland Publishing Company},
  \bibinfo{year}{1983}), pp. \bibinfo{pages}{175--201}.

\bibitem[{\citenamefont{Banyuls et~al.}(1997)\citenamefont{Banyuls, Font,
  Ib{\'a}{\~nez}, Mart\'{\i}, and Miralles}}]{banyuls:97}
\bibinfo{author}{\bibfnamefont{F.}~\bibnamefont{Banyuls}},
  \bibinfo{author}{\bibfnamefont{J.~A.} \bibnamefont{Font}},
  \bibinfo{author}{\bibfnamefont{J.~M.} \bibnamefont{Ib{\'a}{\~nez}}},
  \bibinfo{author}{\bibfnamefont{J.~M.} \bibnamefont{Mart\'{\i}}},
  \bibnamefont{and} \bibinfo{author}{\bibfnamefont{J.~A.}
  \bibnamefont{Miralles}}, \bibinfo{journal}{\apj}
  \textbf{\bibinfo{volume}{476}}, \bibinfo{pages}{221} (\bibinfo{year}{1997}).

\bibitem[{\citenamefont{{Font}}(2008)}]{font:08}
\bibinfo{author}{\bibfnamefont{J.~A.} \bibnamefont{{Font}}},
  \bibinfo{journal}{Liv. Rev. Rel.} \textbf{\bibinfo{volume}{11}},
  \bibinfo{pages}{7} (\bibinfo{year}{2008}).

\bibitem[{\citenamefont{{Einfeldt}}(1988)}]{HLLE:88}
\bibinfo{author}{\bibfnamefont{B.}~\bibnamefont{{Einfeldt}}}, in
  \emph{\bibinfo{booktitle}{Shock tubes and waves; Proceedings of the Sixteenth
  International Symposium, Aachen, Germany, July 26--31, 1987. VCH Verlag,
  Weinheim, Germany}} (\bibinfo{year}{1988}), p. \bibinfo{pages}{671}.

\bibitem[{\citenamefont{{Colella} and {Woodward}}(1984)}]{colella:84}
\bibinfo{author}{\bibfnamefont{P.}~\bibnamefont{{Colella}}} \bibnamefont{and}
  \bibinfo{author}{\bibfnamefont{P.~R.} \bibnamefont{{Woodward}}},
  \bibinfo{journal}{J. Comp. Phys.} \textbf{\bibinfo{volume}{54}},
  \bibinfo{pages}{174} (\bibinfo{year}{1984}).

\bibitem[{\citenamefont{Hyman}(1976)}]{Hyman-1976-Courant-MOL-report}
\bibinfo{author}{\bibfnamefont{J.~M.} \bibnamefont{Hyman}},
  \bibinfo{type}{Tech. Rep.}, \bibinfo{institution}{ERDA Mathematics and
  Computing Laboratory, Courant Institute of Mathematical Sciences, New York
  University} (\bibinfo{year}{1976}).

\bibitem[{\citenamefont{Nakamura et~al.}(1987)\citenamefont{Nakamura, Oohara,
  and Kojima}}]{nakamura:87}
\bibinfo{author}{\bibfnamefont{T.}~\bibnamefont{Nakamura}},
  \bibinfo{author}{\bibfnamefont{K.}~\bibnamefont{Oohara}}, \bibnamefont{and}
  \bibinfo{author}{\bibfnamefont{Y.}~\bibnamefont{Kojima}},
  \bibinfo{journal}{Prog. Theor. Phys. Suppl.} \textbf{\bibinfo{volume}{90}},
  \bibinfo{pages}{1} (\bibinfo{year}{1987}).

\bibitem[{\citenamefont{Shibata and Nakamura}(1995)}]{shibata:95}
\bibinfo{author}{\bibfnamefont{M.}~\bibnamefont{Shibata}} \bibnamefont{and}
  \bibinfo{author}{\bibfnamefont{T.}~\bibnamefont{Nakamura}},
  \bibinfo{journal}{Phys. Rev. D} \textbf{\bibinfo{volume}{52}},
  \bibinfo{pages}{5428} (\bibinfo{year}{1995}).

\bibitem[{\citenamefont{Baumgarte and Shapiro}(1999)}]{baumgarte:99}
\bibinfo{author}{\bibfnamefont{T.~W.} \bibnamefont{Baumgarte}}
  \bibnamefont{and} \bibinfo{author}{\bibfnamefont{S.~L.}
  \bibnamefont{Shapiro}}, \bibinfo{journal}{Phys. Rev. D}
  \textbf{\bibinfo{volume}{59}}, \bibinfo{pages}{024007}
  (\bibinfo{year}{1999}).

\bibitem[{\citenamefont{Alcubierre et~al.}(2000)\citenamefont{Alcubierre,
  Br\"{u}gmann, Dramlitsch, Font, Papadopoulos, Seidel, Stergioulas, and
  Takahashi}}]{alcubierre:00}
\bibinfo{author}{\bibfnamefont{M.}~\bibnamefont{Alcubierre}},
  \bibinfo{author}{\bibfnamefont{B.}~\bibnamefont{Br\"{u}gmann}},
  \bibinfo{author}{\bibfnamefont{T.}~\bibnamefont{Dramlitsch}},
  \bibinfo{author}{\bibfnamefont{J.~A.} \bibnamefont{Font}},
  \bibinfo{author}{\bibfnamefont{P.}~\bibnamefont{Papadopoulos}},
  \bibinfo{author}{\bibfnamefont{E.}~\bibnamefont{Seidel}},
  \bibinfo{author}{\bibfnamefont{N.}~\bibnamefont{Stergioulas}},
  \bibnamefont{and}
  \bibinfo{author}{\bibfnamefont{R.}~\bibnamefont{Takahashi}},
  \bibinfo{journal}{Phys. Rev. D} \textbf{\bibinfo{volume}{62}},
  \bibinfo{pages}{044034} (\bibinfo{year}{2000}).

\bibitem[{\citenamefont{{Alcubierre}}(2008)}]{alcubierre:08}
\bibinfo{author}{\bibfnamefont{M.}~\bibnamefont{{Alcubierre}}},
  \emph{\bibinfo{title}{{Introduction to 3+1 Numerical Relativity}}}
  (\bibinfo{publisher}{Oxford University Press}, \bibinfo{year}{2008}).

\bibitem[{\citenamefont{Bona et~al.}(1995)\citenamefont{Bona, Mass\'o, Seidel,
  and Stela}}]{bona:95}
\bibinfo{author}{\bibfnamefont{C.}~\bibnamefont{Bona}},
  \bibinfo{author}{\bibfnamefont{J.}~\bibnamefont{Mass\'o}},
  \bibinfo{author}{\bibfnamefont{E.}~\bibnamefont{Seidel}}, \bibnamefont{and}
  \bibinfo{author}{\bibfnamefont{J.}~\bibnamefont{Stela}},
  \bibinfo{journal}{\prl} \textbf{\bibinfo{volume}{75}}, \bibinfo{pages}{600}
  (\bibinfo{year}{1995}).

\bibitem[{\citenamefont{Alcubierre et~al.}(2003)\citenamefont{Alcubierre,
  Br\"ugmann, Diener, Koppitz, Pollney, Seidel, and
  Takahashi}}]{alcubierre:03a}
\bibinfo{author}{\bibfnamefont{M.}~\bibnamefont{Alcubierre}},
  \bibinfo{author}{\bibfnamefont{B.}~\bibnamefont{Br\"ugmann}},
  \bibinfo{author}{\bibfnamefont{P.}~\bibnamefont{Diener}},
  \bibinfo{author}{\bibfnamefont{M.}~\bibnamefont{Koppitz}},
  \bibinfo{author}{\bibfnamefont{D.}~\bibnamefont{Pollney}},
  \bibinfo{author}{\bibfnamefont{E.}~\bibnamefont{Seidel}}, \bibnamefont{and}
  \bibinfo{author}{\bibfnamefont{R.}~\bibnamefont{Takahashi}},
  \bibinfo{journal}{Phys. Rev. D} \textbf{\bibinfo{volume}{67}},
  \bibinfo{pages}{084023} (\bibinfo{year}{2003}).

\bibitem[{\citenamefont{Schnetter}(2010)}]{ES-Schnetter2010a}
\bibinfo{author}{\bibfnamefont{E.}~\bibnamefont{Schnetter}},
  \bibinfo{journal}{Class. Quantum Grav.} \textbf{\bibinfo{volume}{27}},
  \bibinfo{pages}{167001} (\bibinfo{year}{2010}).

\bibitem[{\citenamefont{M{\"u}ller and Br{\"u}gmann}(2010)}]{Muller:2009jx}
\bibinfo{author}{\bibfnamefont{D.}~\bibnamefont{M{\"u}ller}} \bibnamefont{and}
  \bibinfo{author}{\bibfnamefont{B.}~\bibnamefont{Br{\"u}gmann}},
  \bibinfo{journal}{Class. Quantum Grav.} \textbf{\bibinfo{volume}{27}},
  \bibinfo{pages}{114008} (\bibinfo{year}{2010}).

\bibitem[{\citenamefont{Diener et~al.}(2007)\citenamefont{Diener, Dorband,
  Schnetter, and Tiglio}}]{diener:05}
\bibinfo{author}{\bibfnamefont{P.}~\bibnamefont{Diener}},
  \bibinfo{author}{\bibfnamefont{E.~N.} \bibnamefont{Dorband}},
  \bibinfo{author}{\bibfnamefont{E.}~\bibnamefont{Schnetter}},
  \bibnamefont{and} \bibinfo{author}{\bibfnamefont{M.}~\bibnamefont{Tiglio}},
  \bibinfo{journal}{J. Sci. Comput.} \textbf{\bibinfo{volume}{32}},
  \bibinfo{pages}{109} (\bibinfo{year}{2007}).

\bibitem[{\citenamefont{Shu}(1998)}]{shu:98}
\bibinfo{author}{\bibfnamefont{C.-W.} \bibnamefont{Shu}},
  \bibinfo{journal}{Lecture Notes in Mathematics}
  \textbf{\bibinfo{volume}{1697}}, \bibinfo{pages}{325} (\bibinfo{year}{1998}).

\bibitem[{\citenamefont{Berger and Colella}(1989)}]{berger:89}
\bibinfo{author}{\bibfnamefont{M.~J.} \bibnamefont{Berger}} \bibnamefont{and}
  \bibinfo{author}{\bibfnamefont{P.}~\bibnamefont{Colella}},
  \bibinfo{journal}{J. Comp. Phys.} \textbf{\bibinfo{volume}{82}},
  \bibinfo{pages}{64} (\bibinfo{year}{1989}).

\bibitem[{\citenamefont{Press et~al.}(2007)\citenamefont{Press, Teukolsky,
  Vetterling, and Flannery}}]{numrep}
\bibinfo{author}{\bibfnamefont{W.~H.} \bibnamefont{Press}},
  \bibinfo{author}{\bibfnamefont{S.~A.} \bibnamefont{Teukolsky}},
  \bibinfo{author}{\bibfnamefont{W.~T.} \bibnamefont{Vetterling}},
  \bibnamefont{and} \bibinfo{author}{\bibfnamefont{B.~P.}
  \bibnamefont{Flannery}}, \emph{\bibinfo{title}{Numerical Recipes, 3rd.
  edition}} (\bibinfo{publisher}{Cambridge University Press},
  \bibinfo{address}{Cambridge, U. K.}, \bibinfo{year}{2007}).

\bibitem[{\citenamefont{{Thorne}}(1980)}]{thorne:80}
\bibinfo{author}{\bibfnamefont{K.~S.} \bibnamefont{{Thorne}}},
  \bibinfo{journal}{Rev. Mod. Phys.} \textbf{\bibinfo{volume}{52}},
  \bibinfo{pages}{299} (\bibinfo{year}{1980}).

\bibitem[{\citenamefont{{Shibata} and {Sekiguchi}}(2004)}]{shibata:04}
\bibinfo{author}{\bibfnamefont{M.}~\bibnamefont{{Shibata}}} \bibnamefont{and}
  \bibinfo{author}{\bibfnamefont{Y.}~\bibnamefont{{Sekiguchi}}},
  \bibinfo{journal}{Phys. Rev. D} \textbf{\bibinfo{volume}{69}},
  \bibinfo{pages}{084024} (\bibinfo{year}{2004}).

\bibitem[{\citenamefont{{Dimmelmeier} et~al.}(2008)\citenamefont{{Dimmelmeier},
  {Ott}, {Marek}, and {Janka}}}]{dimmelmeier:08}
\bibinfo{author}{\bibfnamefont{H.}~\bibnamefont{{Dimmelmeier}}},
  \bibinfo{author}{\bibfnamefont{C.~D.} \bibnamefont{{Ott}}},
  \bibinfo{author}{\bibfnamefont{A.}~\bibnamefont{{Marek}}}, \bibnamefont{and}
  \bibinfo{author}{\bibfnamefont{H.-T.} \bibnamefont{{Janka}}},
  \bibinfo{journal}{\prd} \textbf{\bibinfo{volume}{78}},
  \bibinfo{pages}{064056} (\bibinfo{year}{2008}).

\bibitem[{\citenamefont{{Nagar} and {Rezzolla}}(2005)}]{nagar:05}
\bibinfo{author}{\bibfnamefont{A.}~\bibnamefont{{Nagar}}} \bibnamefont{and}
  \bibinfo{author}{\bibfnamefont{L.}~\bibnamefont{{Rezzolla}}},
  \bibinfo{journal}{Class. Quantum Grav.} \textbf{\bibinfo{volume}{22}},
  \bibinfo{pages}{167} (\bibinfo{year}{2005}).

\bibitem[{\citenamefont{Penrose}(1963)}]{penrose:63}
\bibinfo{author}{\bibfnamefont{R.}~\bibnamefont{Penrose}},
  \bibinfo{journal}{\prl} \textbf{\bibinfo{volume}{10}}, \bibinfo{pages}{66}
  (\bibinfo{year}{1963}).

\bibitem[{\citenamefont{Newman and Penrose}(1962)}]{newman:62}
\bibinfo{author}{\bibfnamefont{E.~T.} \bibnamefont{Newman}} \bibnamefont{and}
  \bibinfo{author}{\bibfnamefont{R.}~\bibnamefont{Penrose}},
  \bibinfo{journal}{J. Math. Phys.} \textbf{\bibinfo{volume}{3}},
  \bibinfo{pages}{566} (\bibinfo{year}{1962}).

\bibitem[{\citenamefont{{Reisswig} et~al.}(2012)\citenamefont{{Reisswig},
  {Bishop}, and {Pollney}}}]{reisswig:12null}
\bibinfo{author}{\bibfnamefont{C.}~\bibnamefont{{Reisswig}}},
  \bibinfo{author}{\bibfnamefont{N.~T.} \bibnamefont{{Bishop}}},
  \bibnamefont{and}
  \bibinfo{author}{\bibfnamefont{D.}~\bibnamefont{{Pollney}}},
  \bibinfo{journal}{arXiv:1208.3891}  (\bibinfo{year}{2012}).

\bibitem[{\citenamefont{{Reisswig} and {Pollney}}(2011)}]{reisswig:11}
\bibinfo{author}{\bibfnamefont{C.}~\bibnamefont{{Reisswig}}} \bibnamefont{and}
  \bibinfo{author}{\bibfnamefont{D.}~\bibnamefont{{Pollney}}},
  \bibinfo{journal}{\cqg} \textbf{\bibinfo{volume}{28}},
  \bibinfo{pages}{195015} (\bibinfo{year}{2011}).

\bibitem[{\citenamefont{Lousto and Zlochower}(2007)}]{lousto:07a}
\bibinfo{author}{\bibfnamefont{C.~O.} \bibnamefont{Lousto}} \bibnamefont{and}
  \bibinfo{author}{\bibfnamefont{Y.}~\bibnamefont{Zlochower}},
  \bibinfo{journal}{\prd} \textbf{\bibinfo{volume}{76}},
  \bibinfo{pages}{041502} (\bibinfo{year}{2007}).

\bibitem[{\citenamefont{{Ruiz} et~al.}(2008)\citenamefont{{Ruiz}, {Alcubierre},
  {N{\'u}{\~n}ez}, and {Takahashi}}}]{ruiz:08}
\bibinfo{author}{\bibfnamefont{M.}~\bibnamefont{{Ruiz}}},
  \bibinfo{author}{\bibfnamefont{M.}~\bibnamefont{{Alcubierre}}},
  \bibinfo{author}{\bibfnamefont{D.}~\bibnamefont{{N{\'u}{\~n}ez}}},
  \bibnamefont{and}
  \bibinfo{author}{\bibfnamefont{R.}~\bibnamefont{{Takahashi}}},
  \bibinfo{journal}{Gen. Rel. Grav.} \textbf{\bibinfo{volume}{40}},
  \bibinfo{pages}{2467} (\bibinfo{year}{2008}).

\bibitem[{\citenamefont{Dreyer et~al.}(2003)\citenamefont{Dreyer, Krishnan,
  Schnetter, and Shoemaker}}]{dreyer:02}
\bibinfo{author}{\bibfnamefont{O.}~\bibnamefont{Dreyer}},
  \bibinfo{author}{\bibfnamefont{B.}~\bibnamefont{Krishnan}},
  \bibinfo{author}{\bibfnamefont{E.}~\bibnamefont{Schnetter}},
  \bibnamefont{and}
  \bibinfo{author}{\bibfnamefont{E.}~\bibnamefont{Shoemaker}},
  \bibinfo{journal}{Phys. Rev. D} \textbf{\bibinfo{volume}{67}},
  \bibinfo{pages}{024018} (\bibinfo{year}{2003}).

\bibitem[{\citenamefont{{Tolman}}(1939)}]{tolman:39}
\bibinfo{author}{\bibfnamefont{R.~C.} \bibnamefont{{Tolman}}},
  \bibinfo{journal}{Phys. Rev.} \textbf{\bibinfo{volume}{55}},
  \bibinfo{pages}{364} (\bibinfo{year}{1939}).

\bibitem[{\citenamefont{{Oppenheimer} and {Volkoff}}(1939)}]{oppenheimer:39}
\bibinfo{author}{\bibfnamefont{J.~R.} \bibnamefont{{Oppenheimer}}}
  \bibnamefont{and} \bibinfo{author}{\bibfnamefont{G.~M.}
  \bibnamefont{{Volkoff}}}, \bibinfo{journal}{Phys. Rev.}
  \textbf{\bibinfo{volume}{55}}, \bibinfo{pages}{374} (\bibinfo{year}{1939}).

\bibitem[{\citenamefont{{Baiotti}
  et~al.}(2009{\natexlab{a}})\citenamefont{{Baiotti}, {Bernuzzi}, {Corvino},
  {de Pietri}, and {Nagar}}}]{baiotti:09}
\bibinfo{author}{\bibfnamefont{L.}~\bibnamefont{{Baiotti}}},
  \bibinfo{author}{\bibfnamefont{S.}~\bibnamefont{{Bernuzzi}}},
  \bibinfo{author}{\bibfnamefont{G.}~\bibnamefont{{Corvino}}},
  \bibinfo{author}{\bibfnamefont{R.}~\bibnamefont{{de Pietri}}},
  \bibnamefont{and} \bibinfo{author}{\bibfnamefont{A.}~\bibnamefont{{Nagar}}},
  \bibinfo{journal}{\prd} \textbf{\bibinfo{volume}{79}},
  \bibinfo{pages}{024002} (\bibinfo{year}{2009}{\natexlab{a}}).

\bibitem[{\citenamefont{{Komatsu}
  et~al.}(1989{\natexlab{a}})\citenamefont{{Komatsu}, {Eriguchi}, and
  {Hachisu}}}]{komatsu:89a}
\bibinfo{author}{\bibfnamefont{H.}~\bibnamefont{{Komatsu}}},
  \bibinfo{author}{\bibfnamefont{Y.}~\bibnamefont{{Eriguchi}}},
  \bibnamefont{and}
  \bibinfo{author}{\bibfnamefont{I.}~\bibnamefont{{Hachisu}}},
  \bibinfo{journal}{\mnras} \textbf{\bibinfo{volume}{237}},
  \bibinfo{pages}{355} (\bibinfo{year}{1989}{\natexlab{a}}).

\bibitem[{\citenamefont{{Komatsu}
  et~al.}(1989{\natexlab{b}})\citenamefont{{Komatsu}, {Eriguchi}, and
  {Hachisu}}}]{komatsu:89b}
\bibinfo{author}{\bibfnamefont{H.}~\bibnamefont{{Komatsu}}},
  \bibinfo{author}{\bibfnamefont{Y.}~\bibnamefont{{Eriguchi}}},
  \bibnamefont{and}
  \bibinfo{author}{\bibfnamefont{I.}~\bibnamefont{{Hachisu}}},
  \bibinfo{journal}{\mnras} \textbf{\bibinfo{volume}{239}},
  \bibinfo{pages}{153} (\bibinfo{year}{1989}{\natexlab{b}}).

\bibitem[{\citenamefont{{Janka} et~al.}(1993)\citenamefont{{Janka}, {Zwerger},
  and {M\"onchmeyer}}}]{janka:93}
\bibinfo{author}{\bibfnamefont{H.-T.} \bibnamefont{{Janka}}},
  \bibinfo{author}{\bibfnamefont{T.}~\bibnamefont{{Zwerger}}},
  \bibnamefont{and}
  \bibinfo{author}{\bibfnamefont{R.}~\bibnamefont{{M\"onchmeyer}}},
  \bibinfo{journal}{\aap} \textbf{\bibinfo{volume}{268}}, \bibinfo{pages}{360}
  (\bibinfo{year}{1993}).

\bibitem[{\citenamefont{{Berti} et~al.}(2009)\citenamefont{{Berti}, {Cardoso},
  and {Starinets}}}]{berti:09}
\bibinfo{author}{\bibfnamefont{E.}~\bibnamefont{{Berti}}},
  \bibinfo{author}{\bibfnamefont{V.}~\bibnamefont{{Cardoso}}},
  \bibnamefont{and} \bibinfo{author}{\bibfnamefont{A.~O.}
  \bibnamefont{{Starinets}}}, \bibinfo{journal}{\cqg}
  \textbf{\bibinfo{volume}{26}}, \bibinfo{pages}{163001}
  (\bibinfo{year}{2009}).

\bibitem[{\citenamefont{{Thierfelder} et~al.}(2011)\citenamefont{{Thierfelder},
  {Bernuzzi}, and {Br{\"u}gmann}}}]{thierfelder:11}
\bibinfo{author}{\bibfnamefont{M.}~\bibnamefont{{Thierfelder}}},
  \bibinfo{author}{\bibfnamefont{S.}~\bibnamefont{{Bernuzzi}}},
  \bibnamefont{and}
  \bibinfo{author}{\bibfnamefont{B.}~\bibnamefont{{Br{\"u}gmann}}},
  \bibinfo{journal}{\prd} \textbf{\bibinfo{volume}{84}},
  \bibinfo{pages}{044012} (\bibinfo{year}{2011}).

\bibitem[{\citenamefont{{Bernuzzi}
  et~al.}(2012{\natexlab{b}})\citenamefont{{Bernuzzi}, {Thierfelder}, and
  {Br{\"u}gmann}}}]{bernuzzi:12b}
\bibinfo{author}{\bibfnamefont{S.}~\bibnamefont{{Bernuzzi}}},
  \bibinfo{author}{\bibfnamefont{M.}~\bibnamefont{{Thierfelder}}},
  \bibnamefont{and}
  \bibinfo{author}{\bibfnamefont{B.}~\bibnamefont{{Br{\"u}gmann}}},
  \bibinfo{journal}{\prd} \textbf{\bibinfo{volume}{85}},
  \bibinfo{pages}{104030} (\bibinfo{year}{2012}{\natexlab{b}}).

\bibitem[{\citenamefont{{Gold} et~al.}(2011)\citenamefont{{Gold}, {Bernuzzi},
  {Thierfelder}, {Bruegmann}, and {Pretorius}}}]{gold:11}
\bibinfo{author}{\bibfnamefont{R.}~\bibnamefont{{Gold}}},
  \bibinfo{author}{\bibfnamefont{S.}~\bibnamefont{{Bernuzzi}}},
  \bibinfo{author}{\bibfnamefont{M.}~\bibnamefont{{Thierfelder}}},
  \bibinfo{author}{\bibfnamefont{B.}~\bibnamefont{{Bruegmann}}},
  \bibnamefont{and}
  \bibinfo{author}{\bibfnamefont{F.}~\bibnamefont{{Pretorius}}},
  \bibinfo{journal}{arXiv:1109.5128}  (\bibinfo{year}{2011}).

\bibitem[{LORENE()}]{LORENE:web}
LORENE, \emph{\bibinfo{title}{{LORENE}: {L}angage {O}bjet pour la
  {RE}lativit\'e {N}um\'eriqu{E}}},
  \urlprefix\url{http://www.lorene.obspm.fr/}.

\bibitem[{\citenamefont{{Baiotti} et~al.}(2010)\citenamefont{{Baiotti},
  {Shibata}, and {Yamamoto}}}]{baiotti:10}
\bibinfo{author}{\bibfnamefont{L.}~\bibnamefont{{Baiotti}}},
  \bibinfo{author}{\bibfnamefont{M.}~\bibnamefont{{Shibata}}},
  \bibnamefont{and}
  \bibinfo{author}{\bibfnamefont{T.}~\bibnamefont{{Yamamoto}}},
  \bibinfo{journal}{\prd} \textbf{\bibinfo{volume}{82}},
  \bibinfo{pages}{064015} (\bibinfo{year}{2010}).

\bibitem[{\citenamefont{{Baiotti}
  et~al.}(2009{\natexlab{b}})\citenamefont{{Baiotti}, {Giacomazzo}, and
  {Rezzolla}}}]{baiotti:09a}
\bibinfo{author}{\bibfnamefont{L.}~\bibnamefont{{Baiotti}}},
  \bibinfo{author}{\bibfnamefont{B.}~\bibnamefont{{Giacomazzo}}},
  \bibnamefont{and}
  \bibinfo{author}{\bibfnamefont{L.}~\bibnamefont{{Rezzolla}}},
  \bibinfo{journal}{Classical and Quantum Gravity}
  \textbf{\bibinfo{volume}{26}}, \bibinfo{pages}{114005}
  (\bibinfo{year}{2009}{\natexlab{b}}).

\bibitem[{\citenamefont{Boyle et~al.}(2008)\citenamefont{Boyle, Buonanno,
  Kidder, Mroue, Pan, Pfeiffer, and Scheel}}]{boyle:08}
\bibinfo{author}{\bibfnamefont{M.}~\bibnamefont{Boyle}},
  \bibinfo{author}{\bibfnamefont{A.}~\bibnamefont{Buonanno}},
  \bibinfo{author}{\bibfnamefont{L.~E.} \bibnamefont{Kidder}},
  \bibinfo{author}{\bibfnamefont{A.~H.} \bibnamefont{Mroue}},
  \bibinfo{author}{\bibfnamefont{Y.}~\bibnamefont{Pan}},
  \bibinfo{author}{\bibfnamefont{H.~P.} \bibnamefont{Pfeiffer}},
  \bibnamefont{and} \bibinfo{author}{\bibfnamefont{M.~A.}
  \bibnamefont{Scheel}}, \bibinfo{journal}{\prd} \textbf{\bibinfo{volume}{78}},
  \bibinfo{pages}{104020} (\bibinfo{year}{2008}).

\bibitem[{\citenamefont{{Sod}}(1978)}]{sod:78}
\bibinfo{author}{\bibfnamefont{G.~A.} \bibnamefont{{Sod}}},
  \bibinfo{journal}{Journal of Computational Physics}
  \textbf{\bibinfo{volume}{27}}, \bibinfo{pages}{1} (\bibinfo{year}{1978}).

\bibitem[{\citenamefont{{Mart{\'{\i}}} and {M{\"u}ller}}(2003)}]{marti:03}
\bibinfo{author}{\bibfnamefont{J.~M.} \bibnamefont{{Mart{\'{\i}}}}}
  \bibnamefont{and}
  \bibinfo{author}{\bibfnamefont{E.}~\bibnamefont{{M{\"u}ller}}},
  \bibinfo{journal}{Liv. Rev. Rel.} \textbf{\bibinfo{volume}{6}},
  \bibinfo{pages}{7} (\bibinfo{year}{2003}).

\bibitem[{\citenamefont{Toro}(1999)}]{toro:99}
\bibinfo{author}{\bibfnamefont{E.~F.} \bibnamefont{Toro}},
  \emph{\bibinfo{title}{{R}iemann {S}olvers and {N}umerical {M}ethods for
  {F}luid {D}ynamics}} (\bibinfo{publisher}{Springer},
  \bibinfo{address}{Berlin}, \bibinfo{year}{1999}).

\end{thebibliography}

\end{document}